\documentclass[singlecolumn]{svjour3}          % twocolumn
\usepackage{steinmetz}
\usepackage{balance}
\usepackage{etex}

\usepackage[usenames,dvipsnames]{pstricks}
\usepackage{pst-grad} % For gradients
\usepackage{pst-plot} % For axes

\usepackage{amsmath}
\usepackage{amssymb}
\usepackage{epsfig}
\usepackage{psfrag}
\usepackage{tikz}
\usepackage{array}
\usepackage{graphicx}
\usepackage{multirow}
\usepackage{framed}
\usepackage{float}
\usepackage{caption}
\usepackage{balance}
\usepackage{subfig}
\usepackage{tabularx}
\usepackage{url}
\begin{document}

\title{Do we need two forms of feedback in the Rate Control Protocol (RCP)?}
\author{Abuthahir \and Gaurav~Raina \and Thomas Voice}%
% \thanks{Abuthahir and G. Raina are with Department of Electrical Engineering, IIT Madras, Chennai  600036, India (e-mail: ee12d207@ee.iitm.ac.in; gaurav@ee.iitm.ac.in).}}

\institute{A. Abuthahir \and G. Raina \at
              Department of Electrical Engineering, IIT Madras, Chennai  600036, India \\
              \email{ee12d207@ee.iitm.ac.in;gaurav@ee.iitm.ac.in}           %  \\
 \and
           T. Voice \at
              Independent Researcher
           \email{tdvoice@gmail.com}           %  \\
}

\date{Received: \today}

\maketitle
\begin{abstract} 

% \begin{document}
% 
% \begin{frontmatter}
% 
% \title{ }
% %\title{Local Hopf Bifurcation Analysis of a Rate Control Protocol\,(RCP)}
% %\title{Impact of queue size feedback on the performance of a Rate Control Protocol\,(RCP)}
% 
% 
% \author[iitm]{Abuthahir\corref{mycorrespondingauthor}}
% \ead{ee12d207@ee.iitm.ac.in}
% 
% \author[apple]{Thomas Voice}
% \ead{tdvoice@gmail.com}
% 
% \author[iitm]{Gaurav Raina}
% \ead{gaurav@ee.iitm.ac.in}
% 
% \address[iitm]{Department of Electrical Engineering, Indian Institute of Technology Madras, Chennai-600036, India}
% \address[apple]{Independent Researcher}
% %\address[apple]{Statistical Laboratory, University of Cambridge, Cambridge, UK}
% \cortext[mycorrespondingauthor]{Corresponding author at: Department of Electrical Engineering, Indian Institute of Technology Madras, Chennai-600036, India. }
% 
% \begin{abstract}

% Rate Control Protocol (RCP) uses explicit feedback from routers to manage it's flow and congestion control algorithms.
There is considerable interest in the networking community in explicit congestion control as it may allow the design of a fair, stable, low loss, low delay, and high utilization network. The Rate Control Protocol (RCP) is an example of such a congestion control protocol. The current design of RCP suggests that it should employ two forms of feedback; i.e. rate mismatch and queue size, in order to manage its flow control algorithms. An outstanding design question in RCP is whether the presence of queue size feedback is useful or not, given feedback based on rate mismatch. In this paper, we address this question using tools from control and bifurcation theory. We linearize the actual non-linear system and analyze the local asymptotic stability, robust stability and rate of convergence of both the design choices, i.e., with and without queue size feedback. But such analyses do not offer clear design recommendations on whether the queue feedback is useful or not. This motivates a bifurcation-theoretic analysis where we have to take non-linear terms into consideration, which helps to learn additional dynamical properties of the RCP system. In particular, we proceed to analyze two non-linear properties, namely, the type of Hopf bifurcation and the asymptotic stability of the bifurcating limit cycles. Analytical results reveal that the presence of queue feedback in RCP can induce a sub-critical Hopf bifurcation, which can lead to undesirable system behavior. Whereas, in the absence of queue feedback, the Hopf bifurcation is always super-critical where the bifurcating limit cycles are stable and of small amplitude. The analysis is corroborated by numerical computations and some packet-level simulations as well. Based on our work, the suggestion for RCP is to only include feedback based on rate mismatch in the design of the protocol.            
\end{abstract}

\section{Introduction}
\label{sec:introduction}

In recent times, most service systems provide some form of delay-related information like waiting times, queue size to inform their users of the congestion level of the system. These feedback aims to control the congestion and improve system performance. However, if such feedback is time-delayed, it will have serious impacts on the behavior and dynamics of the underlying system; for example, see \cite{allon2011,novitzkysiads2019,pender2018,sharma2001}. The presence of feedback delays makes the system infinite-dimensional, and may pose numerous theoretical and practical challenges. In general, the stability of a closed-loop system is sensitive to feedback delays, which normally necessitates a detailed stability analysis. The initial, and in fact very common, style of analysis for non-linear time delayed systems is to first linearize the equation and then study the stability properties of the linearized system. Local stability analysis retains only the linear component and ignores all higher order terms of the nonlinear system before addressing the issue of stability. However, the feedback delays of a nonlinear dynamical system may result in various complex dynamics like bifurcation, chaos, etc. So, it looks appealing to have an analytical methodology which may allow us to capture the impact of some nonlinear terms while performing a Taylor expansion of the nonlinear system about its equilibrium. Local bifurcation theory is one such methodology \cite{hassard1981}. For example, see \cite{dubeycnsns2019,liaonld2015,novitzkysiads2019,ranchaonld2018} for some stability and bifurcation analysis of dynamical systems with feedback delays. Moreover, without an understanding of the dynamics of the system in the unstable regime, choosing an operating point close to the boundary of the stable region could be risky. A comprehensive understanding of local bifurcation phenomena may help yield insights into the behavior of the system in the unstable regime. This paper employs both linear systems theory and non-linear techniques to investigate how the feedback of queue size can impact the system dynamics in the setting of congestion control protocols for the Internet. We consider protocols where end-systems use feedback, which is time-delayed, from routers to adjust their rates. There is a considerable interest in analyzing the stability and dynamical properties of fluid models for Internet congestion control algorithms \cite{khoshcsf2019,peicnsns2019,peiijbc2018,tangnld2017,voice2009maxminrcp,xunld2019}. In this study, we focus on a well-known explicit congestion control protocol called the Rate Control Protocol (RCP) \cite{balakrishnan2007stability,dukkipatircpac,krv2009}.

%  There is a considerable interest in analyzing the dynamics of Internet congestion control algorithms \cite{gentilecnsns2014,peicnsns2019,raina2005,voice2009maxminrcp}.
% 
% The goal of this paper is to investigate how the feedback of queue size at network routers can impact the system dynamics in the setting of Internet congestion control protocols. 

The most widely implemented congestion control algorithm in the Internet today is the Transmission Control Protocol (TCP). Despite the tremendous success of TCP, it is now well acknowledged that its performance would degrade in high bandwidth-delay environments \cite{katabi2002}. In the TCP protocol, endpoints implicitly estimate congestion from noisy information, which is essentially the single bit of feedback provided by a dropped or marked packet. There is a continued interest in the development of explicit congestion control algorithms that rely on explicit feedback from routers~\cite{baretto2015rcp,he2017,jose2016xcc,katabi2002,krv2009,lei15,liu2016xcc,Wydrowskimaxnet,Zhangjetmax}. Rate Control Protocol (RCP) is an explicit congestion control protocol that aims to reduce the flow completion time by quickly assigning the flows their fair rate. Moreover, RCP continues to receive attention not only in the currently used host-centric~(IP-based) networks \cite{baretto2015rcp,krbookchap,sharma17,sun2012rcp}, but also in the future data-centric networking architectures like Named Data Networking~(NDN) \cite{lei15,mahdian2016xcc,zhong2017}. See \cite{zhangndn2010} for an overview of Named Data Networking. In NDN, there is no IP address, and all data are named with unique names. Moreover, the data can be fetched from multiple sources via multiple paths which makes the implicit signaling mechanism unreliable in NDN \cite{ren2016xcc}. Therefore, researchers focus on employing rate-based RCP-style algorithms in NDN. In both the networking architectures, the motivation for using RCP lies in its advantage of quickly assigning a fair rate for all the flows.

RCP computes the fair rate using two forms of feedback: rate mismatch and queue size. An important question that has not been fully addressed in the design of RCP is whether we really need two forms of feedback. Currently, regardless of networking architecture, RCP uses both rate mismatch and queue size feedback.  
% Therefore, addressing the design question of whether the queue feedback is useful, and may invoke further interests from the networks that deploy RCP-based algorithms. 
In this paper, we focus on a variant of the RCP dynamical system which was introduced in~\cite{krv2009}. We consider the RCP model where all the flows have a common feedback delay, operating over a single bottleneck link.

In general, the congestion control algorithms will always have to contend with feedback delays: the presence of propagation delays at the very least, which makes stability an important concern. 
% The initial, and in fact very common, style of analysis for non-linear time delayed systems is to first linearize the equation and then study the stability properties of the linearized system.
% 
% In general, the feedback delays of closed loop systems may introduce instabilities, which normally necessitates a stability analysis. 
We derive necessary and sufficient conditions for local asymptotic stability of RCP. We highlight that the RCP which uses both rate mismatch and queue size feedback readily loses local stability via a \textit{Hopf bifurcation} \cite{hassard1981}. Apart from ensuring stability, another important design objective is to make sure that the system converges quickly to a stable equilibrium. We study the convergence rate of RCP in the presence and absence of queue size feedback. The rate of convergence analysis can help understand any trade-off among system parameters and could guide us in tuning the parameter values for faster convergence. But, it does not provide any insights on which design choice is desirable. Similarly, one can devise conditions for robust stability, for each of these design choices, but such conditions do not offer clear design recommendations on whether the queue feedback is beneficial or not. Therefore, based on the linear analysis we are unable to distinguish between the two different design choices. This provides motivation for non-linear analysis to study some additional dynamical properties of the RCP system, with and without the queue size feedback. We employ a bifurcation theoretic style of analysis where we study the dynamics of the system as it transits from a stable to an unstable regime. We are concerned with the loss of local stability occurring via a Hopf bifurcation leading to the onset of limit cycles, as a parameter crosses a critical value. The birth of limit cycles is not disastrous, as long they are \textit{stable} and of \textit{small} amplitude. Thus, key concerns about the direction and the stability of the limit cycles bifurcating from the steady state are also addressed. From a bifurcation theoretic perspective, we would like our algorithms to always produce {stable} limit cycles of {small} amplitude. We opt for the method of Poincar{\`e} normal forms and the center manifold theorem (see \cite{hassard1981} for details) to analyze the nature of Hopf bifurcation.
% In addition to ensuring stability, it is also crucial to make sure that any loss of stability produces only stable and small amplitude limit cycles. 
%  This paper addresses the above design question of RCP by investigating the local stability and Hopf bifurcation for the cases with and without queue feedback.

We need to decide which parameter will be used to violate the stability condition and hence act as the bifurcation parameter. We prefer not to use any of the system parameters as the bifurcation parameter, as varying it may change the value of the system equilibrium. Therefore, a non-dimensional exogenous parameter is used to induce instability. This has various advantages. We need not be concerned with the dimensions of the parameter, and as it is common for both the design choices, we can compare the results fairly.
% 
% In \cite{raina2005}, the Hopf bifurcation properties of a class of non-linear delay differential equations are studied using Poincar{\`e} normal forms and the center manifold theorem \cite{hassard1981}. 
% Therefore, based on the linear analysis we are unable to distinguish between the two different design choices. This provides motivation for non-linear analysis to study some additional dynamical properties of the RCP system, with and without the queue size feedback. Then, the next natural step is to investigate the dynamical behavior of the system as it transits from a stable to an unstable regime. We explore the impact of loss of local stability for both the design options, i.e., with and without queue size feedback.

Using the analytical results from \cite{raina2005}, we perform the requisite calculations to investigate the nature of Hopf bifurcation both in the presence and absence of queue feedback. We establish that, in the absence of queue feedback, the system always undergoes a super-critical Hopf bifurcation and leads to the emergence of stable limit cycles of small amplitude. However, if the queue feedback is included in the protocol, the Hopf bifurcation would be sub-critical, at high link utilization. A sub-critical Hopf bifurcation is often undesirable for real engineering systems as it would give rise either to the onset of large amplitude limit cycles or to unstable limit cycles \cite{strogatz2018}. In essence, the insights of our analyses suggest the removal of the queuing term from the definition of RCP. Some of the theoretical insights are validated with bifurcation diagrams, numerical computations, and packet-level simulations.

The rest of the paper is organized as follows: In Section 2, we outline the non-linear fluid model of RCP. In Section 3, we derive the necessary and sufficient conditions to ensure local asymptotic stability. The rate of convergence and robust stability analysis are outlined in Sections 4 and 5. In section 6, we conduct a local Hopf bifurcation analysis. In section 7, we present some packet-level simulations. In Section 8, we conclude with the summary of our contributions and offer some avenues for further research.
%For our analysis, we consider single bottleneck link and single delay setting.

\section{Model Description}
%In this section, we outline the model that represents the behavior of RCP. 
The closed loop feedback systems with delays are often modeled as delay differential equations. The non-linear fluid model of a proportionally fair variant of RCP \cite{krv2009} is governed by the following equation 
\begin{equation}
\label{eq:smallbuf_model}
\dot{R}_{j}(t)=\dfrac{aR_{j}(t)}{C_{j}\overline{T_{j}}(t)}\Big(C_{j}-y_{j}(t)-b_{j}C_{j}p_{j}\big(y_{j}(t)\big) \Big),
\end{equation}
where
\begin{equation}
y_{j}(t) = \sum\limits_{r:j\in r}x_{r}\big(t-T_{rj}\big) \label{eq:aggr}
\end{equation}
is the aggregate load arriving at link $j$ via all the routes passing through link $j$. We write $j \in r$ to indicate that the route $r$ passes through the link $j$. $R_{j}(t)$ is the fair rate that RCP calculates for all flows passing through link $j$, $x_{r}(t)$ is the flow rate on route $r$, $p_{j}(y_{j})$ is the mean queue size at link $j$ when the arriving load is $y_{j}$, $C_{j}$ is the capacity of link $j$, $a$ and $b_{j}$ are non-negative protocol parameters. Here, $\overline{T}_{j}(t)$ is the average round trip delay of packets passing through link $j$ given by
\begin{equation}
\overline{T}_{j}(t) = \dfrac{\sum\limits_{r:j\in r}x_{r}(t)T_{r}}{\sum\limits_{r:j\in r}x_{r}(t)},\label{eq:avgrtt}
\end{equation}
where
% \begin{equation}
% T_{r}=T_{rj} + T_{jr}\nonumber
% \end{equation}
$T_{r}=T_{rj}+T_{jr}$ represents the sum of the propagation delay on route $r$ from source to link $j$ and the return delay from link $j$ to source. The control equations at RCP router periodically calculates a common fair rate to be used by all flows traversing the bottleneck link. RCP communicates this fair rate via packet headers to destination, which then informs the source through acknowledgement packets. The rate information is then used by the sources to adjust their data rates, and thereby controls the network congestion. We assume that the queuing delay can be ignored relative to the propagation delay. 
% Here, we assume that the queuing delay is negligible as compared to the propagation delay, which conforms with our assumption of small buffers.
The flow rate $x_{r}(t)$ is given by~\cite{krv2009} 
\begin{equation}
 x_{r}(t) = \left( \sum\limits_{j\in r}R_{j}(t-T_{jr})^{-1}\right)^{-1}.\label{eq:florate}
\end{equation}
% Now we will model the mean queue size term as follows. Consider the workload arriving at link $j$ of capacity $C_{j}$, over a time interval $\tau$ is Gaussian with mean $y_{j}\tau$ and variance $y_{j}\tau\sigma_{j}^2$, where $\sigma_{j}^2$ represents the traffic variability at link $j$. For Poisson traffic, $\sigma_{j}=1$. The workload at the queue reflects Brownian motion, with mean under its stationary distribution of 
The mean queue size $p_{j}(y_{j})$ is approximated as follows~\cite{krv2009}
\begin{equation}
p_{j}(y_{j}) = \dfrac{y_{j}\sigma_{j}^2}{2\big(C_{j}-y_{j}\big)},\label{eq:stat_dist}
\end{equation}
where $\sigma_{j}^2$ represents the traffic variability at link $j$.
We can interpret that the rate equation \eqref{eq:smallbuf_model} contains two forms of feedback: rate mismatch term $C_{j}-y_{j}(t)$, and a term based on the mean queue size. Note that the equations \eqref{eq:aggr} and \eqref{eq:florate} make proper allowance for the propagation delays, and average round-trip time \eqref{eq:avgrtt} scales the rate of adaption \eqref{eq:smallbuf_model} at the bottleneck link.

\newcommand{\equil}[1]{#1^{\star}}
\section{Local stability analysis}
In this section, we derive conditions for local asymptotic stability, and highlight that the system loses local stability via a Hopf bifurcation. For our  analysis, we consider the network with single
bottleneck link of capacity $C$, carrying flows with a same feedback delay $\tau$. We assume $\sigma_{j}^2 = 1$, which corresponds to the Poisson arrival of packets of constant size.
We use an exogenous non-dimensional parameter $\kappa$, to push the system just into the locally unstable regime. Then the model is governed by the following delay differential equation 
\begin{equation}
\label{eq:simple_rate_B}
\frac{d}{dt}R(t) = \frac{\kappa aR(t)}{C\tau}\biggl(C - y(t) - bCp\Bigl(y(t)\Bigr)\biggr),
\end{equation}
where 
\begin{equation*}
 y(t) = R\left(t-\tau\right),\quad p(y) = y\Big/\Big(2\big(C - y\big)\Big).
\end{equation*}
%\subsection{Existence of a Hopf bifurcation}
\subsection{Without queue feedback}
To model RCP which uses only rate mismatch feedback, $b$ is set to zero in~\eqref{eq:simple_rate_B}. We know that the target link utilization depends on the value of the parameter $b$. Now, to aim for a particular target link utilization, say a fraction $\gamma$ of the actual link capacity, $C$ is replaced with $\gamma C$. Then, the system model is given by 
% \begin{equation*}
%   \frac{d}{dt}\, R(t) = \frac{ \kappa aR(t)}{C\, \tau}\Bigl(C - R(t-\tau)\Bigr).  \\
%   \label{eq:noqfeedbackK}
% \end{equation*}
% With the introduction of non-dimensional parameter $\kappa$, the system is now represented as
 \begin{equation}
   \frac{d}{dt}\, R(t) = \frac{ \kappa aR(t)}{\gamma C\, \tau}\Bigl(\gamma C - R(t-\tau)\Bigr).  \\
   \label{eq:noqfeedbackKappa}
 \end{equation}
For local stability, we only need to consider the linearized form of  \eqref{eq:noqfeedbackKappa}.		
The equilibrium of \eqref{eq:noqfeedbackKappa} is  $\equil{R}=\gamma C$. Linearizing~\eqref{eq:noqfeedbackKappa} about $\equil{R}$, we obtain
\begin{equation}
\frac{d}{dt}u(t) = -\frac{\kappa a}{\tau}u\left(t-\tau\right). \label{eq:finLinNoqB}
\end{equation}
It is to be noted that the local stability of fixed point of \eqref{eq:noqfeedbackKappa} is given by the stability of the trivial fixed point ($u=0$) of \eqref{eq:finLinNoqB}. 
% The characteristic equation for \eqref{eq:finLinNoqB} is
% %We now consider the case where $b = 0$. Substituting $b=0$ in (\ref{eq:simple_rate_B}), we obtain the characteristic equation 
% \begin{equation}
% \lambda  + \frac{\kappa a}{\tau}e^{-\lambda\tau}  = 0. \label{eq:charac_without_q}
% \end{equation}
From~\cite{raina2005}, the necessary and sufficient condition for local asymptotic stability of \eqref{eq:finLinNoqB} can be written as
\begin{equation}
 \left(\frac{\kappa a}{\tau}\right) \times \tau < \frac{\pi}{2}.
\end{equation}
For $\kappa =1$, we get the necessary and sufficient condition for local stability of \eqref{eq:noqfeedbackKappa} as
\begin{equation}
 a < \frac{\pi}{2}
  \label{eq:nscondwithoutq}
\end{equation}
and the first local Hopf bifurcation occurs at $a = \pi/2.$

%\begin{tabular}{c} $\leftarrow$ super-critical $\rightarrow$  \\ Hopf  \end{tabular}

\subsection{With queue feedback} 
\begin{figure}[hbtp!]
\centering
\psfrag{x}{\hspace{-16mm}Protocol parameter, $a$}
\psfrag{y}{\hspace{-16mm}Protocol parameter, $b$}
\psfrag{unstble}{\footnotesize{}}
\psfrag{0.79}{\small{$\pi/4$}}
\psfrag{1.57}{\small{$\pi/2$}}
\psfrag{1.18}{\small{$3\pi/8$}}
\psfrag{stbbbbbbbbbbbbbbbb}{\footnotesize{Stable region}}
%\psfrag{Hopfboundary}{\footnotesize{$\leftarrow$Hopf boundary}}
\psfrag{Hopf}{\footnotesize{Hopf}}
\psfrag{Hopfboundary}{\footnotesize{}}
\includegraphics[width=3.55in,height=2.35in]{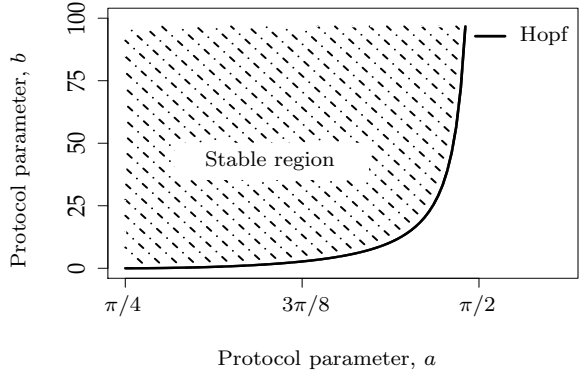}

\vspace{-2mm}
\caption{Stability chart for the RCP system which uses both rate mismatch and queue size feedback.}
%\vspace{-2mm}
\label{fig:ab_stable}
\end{figure}

In this subsection, we analyze the local stability of RCP which uses both rate mismatch and queue size feedback. The equilibrium of~\eqref{eq:simple_rate_B} is
\begin{equation}
 \equil{R} = C\left(\dfrac{b+4-\sqrt{b^{2} + 8b}\,}{4}\right).
\end{equation}
%$\equil{R} = C\left(b+4-\sqrt{b^{2} + 8b}\,\right)\Big/4.$
Let $R(t)=\equil{R}+u(t)$, and linearizing~\eqref{eq:simple_rate_B} about the equilibrium we get
\begin{equation}
\frac{d}{dt}u(t) = -\frac{\kappa \tilde{a}}{\tau} u\left(t-\tau\right), \label{eq:finLinB}
\end{equation}
where
\begin{align}
 \tilde{a} &= a\,\bigl(1+\equil{\rho}\,\bigr),\\
 \equil{\rho}&=\frac{\equil{R}}{C}=\Bigg(\dfrac{b+4-\sqrt{b^{2} + 8b}\,}{4}\Bigg) \label{eq:rhointermsofb}
\end{align}
is the equilibrium link utilization.
% Looking for exponential solutions, the characteristic equation for (\ref{eq:finLinB}) is
% \begin{equation}
% \label{eq:charaa}
% \lambda  + \frac{\kappa\tilde{a}}{\tau} e^{-\lambda\tau} = 0,
% \end{equation}
% where $\kappa$, $\tilde{a}$ and $\tau>0$.
% For the system to be stable, all the roots of the characteristic equation should lie in the left half of the complex plane. As the system parameter varies, the stable fixed point loses its stability when its corresponding eigenvalues cross the imaginary axis. Therefore, the condition for the crossover defines the bounds on the system parameters to maintain stability. To find the critical condition, where this crossover occurs, we substitute $\lambda = \pm i\omega$, $\omega>0$ in (\ref{eq:charaa}). 
% Equating the real and imaginary parts, we obtain 
% \begin{equation}
% \kappa\tilde{a}\cos(\omega\tau) = 0, \label{eq:real0}
% \end{equation}
% \begin{equation}
%  \ \  \kappa\tilde{a}\sin(\omega\tau) = \omega\tau. \label{eq:img0}
% \end{equation}
% Solving~(\ref{eq:real0}) and~(\ref{eq:img0}), we get
% $$\omega\tau=\frac{(2n+1)\pi}{2},\qquad n=0,1,2,\cdots.$$
% We only treat the case $n=0$, which gives 
% \begin{equation}
%  \omega_0=\frac{\pi}{2\tau}.
% \end{equation}
% 
Using results from \cite{raina2005}, we obtain the necessary and sufficient condition for local stability of \eqref{eq:simple_rate_B} as
%$(\kappa \tilde{a}/\tau)\times\tau < \pi/2$, that is
\begin{equation}
\label{eq:stableBns}
 \kappa a\big(1+\equil{\rho}\big)< \frac{\pi}{2},
\end{equation}
and the Hopf bifurcation occurs  at $\kappa a\big(1+\equil{\rho}\big) = \pi/2$.
Therefore, using \eqref{eq:rhointermsofb}, we can write the critical value of the bifurcation
parameter, at which the system loses local stability, as
\begin{equation}
\label{eq:kappa_cval}
 \kappa_c = \frac{2\pi}{a\Big(b+8-\sqrt{b^{2} + 8b}\Big)}.
\end{equation}
For $\kappa=\kappa_c=1$, we obtain the necessary and sufficient condition for local stability of \eqref{eq:simple_rate_B} as
\begin{equation}
\label{eq:stableBns}
 \frac{a\Big(b+8-\sqrt{b^{2} + 8b}\,\Big)}{4}< \frac{\pi}{2}.
\end{equation}
See  Figure \ref{fig:ab_stable} for the graphical representation of \eqref{eq:stableBns}.

From \eqref{eq:nscondwithoutq} and \eqref{eq:stableBns}, we can note that the removal of queue size feedback increases the range of parameter $a$ for which the system is stable. Thus, the necessary and sufficient conditions for stability enable us to determine the stability region in the parameter space, and could guide us in tuning the parameter values to ensure stable performance.

%%%%%%%%%%%%%%%%%%%%%%%%%%%%%%%%%%%%%%%%%%%%%%%%%%%%%%%%%%%%%%%%%%%%%%%%%%%%%%%%%%%%%%%%%%%%%%%
\section{Rate of convergence}
% Local stability analysis revealed the nature of instabilities that may occur on violating those conditions. Now, within the stable regime, it is important to study how fast the system converges to equilibrium, when perturbed. Especially, in the real network environments, these perturbations are more frequent due to the random arrival and departure of flows.
% Rate of convergence is a key performance metric which must be considered for a good design of a congestion control algorithm. 
%We now consider some convergence properties of the system in the stable regime. 
Rate of convergence is a key performance metric which must be considered for a congestion control algorithm. The impact of queue feedback on the convergence rate can be studied by conducting the rate of convergence analysis in the presence and absence of queue size feedback.

\subsection{Without queue feedback}
In this subsection, following the style of analysis outlined in \cite{brauer79}, rate of convergence analysis is performed for the RCP which uses only rate mismatch feedback. The analytical results enables us to investigate the impact of protocol parameters on the convergence rate. Here, we consider $\kappa=1$, to get back the original system. Now the characteristic equation can be written as 
\begin{equation}
\label{eq:ce_withoutq}
\lambda + \left(\frac{a}{\tau}\right)e^{-\lambda \tau} = 0.
\end{equation}
Substituting $\lambda\tau=x-\sigma\tau$ in \eqref{eq:ce_withoutq} yields
\begin{equation}\label{eq:newce_withoutq}
 (\sigma\tau-x)e^{x}-ae^{\sigma\tau}=0.
\end{equation} 
Here $\sigma$ is considered to be the supremum of the solutions of \eqref{eq:newce_withoutq} over $(0,\infty)$ which guarantees that all the characteristic roots lie on the open left half of the complex plane. Let $-\alpha<0$ be the largest real part of all the roots of \eqref{eq:newce_withoutq}. Then the rate at which the system approaches stable equilibrium is given by $\sigma=(\alpha/\tau)$. A necessary and sufficient condition for all the eigenvalues of \eqref{eq:newce_withoutq} to lie on the left half-plane, stated in \cite{hayes50} is,
\begin{align}
\sigma\tau&<1, \label{eq:nscroc1} \\
\sigma\tau&<ae^{\sigma\tau},\label{eq:nscroc2} \\
ae^{\sigma\tau}&<\frac{u}{\sin(u)}, \label{eq:nscroc3}
\end{align}
where $u$ is the solution of the equation
\begin{equation}
\label{eq:uroc}
 u=\sigma\tau \tan(u),
\end{equation}
in $0<u<\pi$, with $u=\pi/2$ if $\sigma=0$. Consider the following function
\begin{equation}
\label{eq:gofu}
 g(u)=\dfrac{u}{\sin(u)}e^{-u/\tan(u)},
\end{equation}
 which increases monotonically in the interval $u\in (0,\pi)$, with $g(0)=1/e$, $g(\pi/2)=\pi/2$ and $\lim_{u \rightarrow \pi}\ g(u)=\infty$. Now, using \eqref{eq:uroc} and \eqref{eq:gofu}, the inequality \eqref{eq:nscroc3} can be rewritten as
 \begin{equation}
  \label{eq:nscro3_1}
  a<g(u).
 \end{equation}
As $\sigma$ increases, $u$ decreases, and hence $g(u)$ is decreasing function of $\sigma$. Therefore, the maximum value of $\sigma$ that satisfies \eqref{eq:nscro3_1} can be obtained by solving its corresponding equality. By a similar argument, the L.H.S of \eqref{eq:nscroc1} and \eqref{eq:nscroc2} increases with increase in $\sigma$. Thus, the maximum $\sigma$ satisfying the inequalities \eqref{eq:nscroc1}, \eqref{eq:nscroc2} can be determined by solving the corresponding equalities. It is to be noted that if the solution does not exists for any of these equalities, then there is no restriction on the value of $\sigma$. Now, the results can be summarized as follows:\\
Let $\sigma_1$, $\sigma_2$, $\sigma_3$ be the solutions of
\begin{align}
 \sigma \tau&=1,\label{eq:eqlty1} \\
 \sigma \tau e^{-\sigma\tau}&=a, \label{eq:eqlty2}\\
 u&=\sigma\tau \tan(u),\ \  g(u)=a, \label{eq:eqlty3} 
 \end{align}
respectively. Consider $\sigma_i=\infty,$ for $i=1,\ 2,\ 3$ if there is no solution exists for the corresponding equality. Then the convergence rate $\sigma$ is given by
\begin{equation}
 \sigma=\min[\sigma_1,\sigma_2,\sigma_3].
\end{equation}
Now, the next step is to analyze the dependence of convergence rate on protocol parameter $a$, for $\tau>0$. The function $\sigma\tau e^{-\sigma\tau}$ has maximum of $1/e$ at $\sigma\tau=1$. Similarly the function $g(u)$ has minimum of $1/e$ at $u=0$. Let $a^{*}=1/e$, then there is no solution for \eqref{eq:eqlty2} if $a>a^{*}$, and for \eqref{eq:eqlty3} if $a<a^{*}$. Let $\sigma_2$ be the solution of \eqref{eq:eqlty2} on $0< a\leq a^{*}$. Similarly, consider $\sigma_3$ be the solution of \eqref{eq:eqlty3} on $a\ >\ a^{*}$. At $a=0$, it is obvious that the rate of convergence $\sigma=0$.
\paragraph{Case 1\ :\ $a \in (0,a^*)$} Differentiating \eqref{eq:eqlty2} with respect to $a$ gives
\begin{equation}
 \frac{d\sigma}{da}=\frac{e^{\sigma\tau}}{\tau(1-a e^{\sigma\tau})} \label{eq:sigmadash}.
\end{equation}
Using \eqref{eq:eqlty2}, the derivative \eqref{eq:sigmadash} can be written as 
\begin{equation}
 \frac{d\sigma}{da}=\frac{e^{\sigma\tau}}{\tau(1-\sigma\tau)}. \label{eq:sigmadash1}
\end{equation}
From \eqref{eq:sigmadash1}, it can deduced that $d\sigma_2 / da$ $>0$ if $\sigma_2\tau<1$. Hence $\sigma_2<\sigma_1$ for $a\in (0,a^*)$.
\paragraph{Case 2\ :\ $a=a^*$} Substituting $a=a^*=1/e$ in \eqref{eq:eqlty2} yields
\begin{equation}
 \sigma_2 \tau e^{-\sigma_2\tau}=a=1/e.
\end{equation}
It is known that the function $\sigma_2\tau e^{-\sigma_2\tau}$ reaches maximum of $1/e$ at $\sigma_2\tau=1$, thus $\sigma_2=\sigma_1=1/\tau$ at $a=a^*$.
\paragraph{Case 3\ :\ $a>a^*$} For $a\ >\ a^*$, using \eqref{eq:eqlty3}, the following holds
\begin{equation}
 g(u)=\dfrac{u}{\sin(u)}e^{-u/\tan(u)}>1/e. \label{eq:case4ineqlty}
\end{equation}
For $u \in (0,\pi)$ and $u/\sin(u) > 1$, \eqref{eq:case4ineqlty} can be written as 
\begin{equation}
 e^{-u/\tan(u)}>1/e \label{case4ineqlty1}.
\end{equation}
From \eqref{case4ineqlty1}, it can be deduced that $u/\tan(u)<1$, and hence $\sigma_3\tau<1$, and $\sigma_3<\sigma_1$.

To summarize, the convergence rate $\sigma$ is given by
\begin{align}
 \sigma&=\min[\sigma_1,\sigma_2]=\sigma_2 \quad a \in (0,a^*],\\
 &=\min[\sigma_1,\sigma_3]=\sigma_3 \quad a>a^*,
\end{align}
where $a^*=1/e$. 
Fig. \ref{fig:roc} shows the variation of convergence rate with protocol parameter $a$ for various values of $\tau$. It can observed that the rate of convergence increases with $a$ for $a<1/e$, and decreases when $a>1/e$. The convergence rate is maximum at $a=(1/e)$, so the optimal value of the protocol parameter for fast convergence is $a=(1/e)$. Also, the rate of convergence decreases with the increase in RTT (see Fig. \ref{fig:roc}).
\newcommand{\wdth}{\textwidth}
\begin{figure}
\psfrag{R}{\hspace{-1.8cm} \small Rate of convergence, $\sigma$}
\psfrag{a}{{\hspace{-0.8cm} \small Parameter, $a$}}
\psfrag{tau=1}{\small{$\tau=1$}}
\psfrag{tau=2}{\small{$\tau=2$}}
\psfrag{unstable}{\small }
\psfrag{roc}{\small Hopf condition}
\psfrag{0}{\small{$0$}}
\psfrag{0.0000000}{\hspace{0.4cm}\small{$0$}}
\psfrag{0.3678794}[][]{\small{$1/e$}}
\psfrag{0.7853982}{\hspace{0.3cm}\small{$\pi/4$}}
\psfrag{1.5707963}{\hspace{0.3cm}\small{$\pi/2$}}
\psfrag{0.0}{\hspace{0.0cm}\small{$0$}}
\psfrag{0.5}{\hspace{0.0cm}\small{$0.5$}}
\psfrag{1.0}[][]{\small{$1$}}
\psfrag{1.57}{\small{$\pi/2$}}
\includegraphics[width=0.75\wdth]{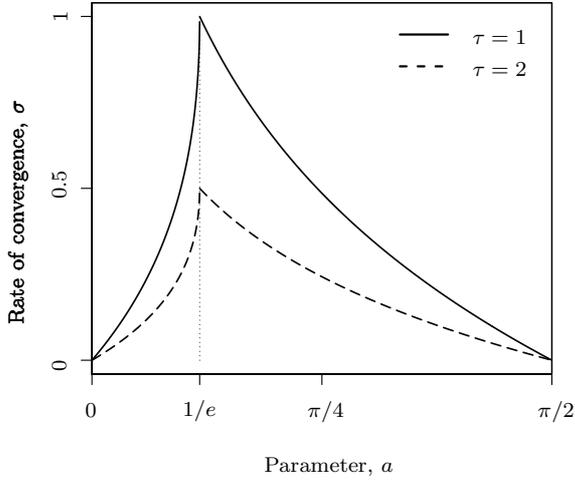}
\caption{Rate of convergence to equilibrium for RCP without queue feedback. The rate of convergence increases with $a$ and reaches maxima of $1/\tau$ at $a=1/e$, and then decreases for $a>1/e$.}
\label{fig:roc}
\end{figure}
\subsection{With queue feedback}
\begin{figure}
\psfrag{b}{\hspace{-1.8cm} rate of convergence $\sigma$}
\psfrag{a}{{\hspace{-0.8cm}parameter $a$}}
\psfrag{tau=1}{\small{$\tau=1$}}
\psfrag{tau=2}{\small{$\tau=2$}}
\psfrag{R}{\hspace{-1.8cm} Rate of convergence, $\sigma$}
\psfrag{b=0}{\hspace{-1mm}\small{$b=0$}}
\psfrag{b=0.736}{\hspace{-1mm}\small{$b=0.736$}}
\psfrag{b=0.4}{\hspace{-1mm}\small{$b=0.4$}}
\psfrag{b=0.005}{\hspace{-1mm}\small{$b=0.005$}}
\psfrag{0}{\small{$0$}}
\psfrag{time}{\hspace{-0.0cm}  \small Time}
\psfrag{Rate}{\hspace{-0.5cm}  \small Rate, $R(t)$}
\psfrag{0}{\small{$0$}}
\psfrag{5}{\hspace{0cm}\small{$5$}}
\psfrag{10}{\hspace{0cm}\small{$10$}}
\psfrag{15}{\hspace{0cm}\small{$15$}}
\psfrag{20}{\small{$20$}}
\psfrag{0}{\small{$0$}}
\psfrag{25}{\hspace{0cm}\small{$25$}}
\psfrag{50}{\small{$50$}}
\psfrag{0.0000000}{\hspace{0.50cm}\small{$0$}}
\psfrag{0.5}{\hspace{0.0cm}\small{$0.5$}}
\psfrag{1.0}[][]{\small{$1$}}
\psfrag{1.57}{\small{$\pi/2$}}
\psfrag{0.3678794}[][]{\small{$1/e$}}
\psfrag{0.7853982}{\hspace{0.3cm}\small{$\pi/4$}}
\psfrag{1.5707963}{\hspace{0.3cm}\small{$\pi/2$}}
\includegraphics[width=0.75\wdth]{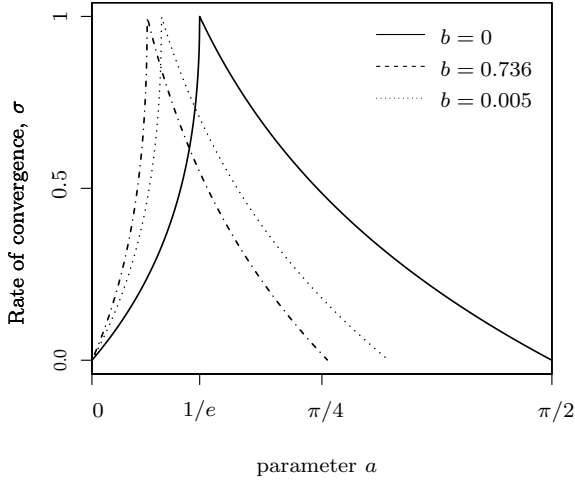}
\caption{Rate of convergence for RCP with and without queue size feedback.}
\label{fig:rcp_1d_roc_Vsaandb_consolidated}
\end{figure}

The characteristic equation of RCP which uses both rate mismatch and queue size feedback is 
\begin{equation}
\label{eq:ce_withq}
\lambda + \left(\frac{\tilde{a}}{\tau}\right)e^{-\lambda \tau} = 0.
\end{equation}
where $\tilde{a}=a(1+\rho^*)$. As outlined in the previous subsection, we consider $\sigma_1$, $\sigma_2$, $\sigma_3$ be the solutions of
\begin{align}
 \sigma \tau&=1,\label{eq:withqeqlty1} \\
 \sigma \tau e^{-\sigma\tau}&=\tilde{a}, \label{eq:withqeqlty2}\\
 u&=\sigma\tau \tan(u),\ \  g(u)=\tilde{a}, \label{eq:withqeqlty3} 
 \end{align}
respectively. Following the style of analysis outlined in the previous subsection, we can obtain the rate of convergence for various values of $a$
 and $b$. The results are shown in Fig. \ref{fig:rcp_1d_roc_Vsaandb_consolidated}.
%%%%%%%%%%%%%%%%%%%%%%%%%%%%%%%%%%%%%%%%%%%%%%%%%%%

\section{Robust stability analysis}
\subsection{Without queue feedback}
The characteristic equation for RCP without queue feedback is given by 
\begin{equation}
\label{eq:robust_ce_withoutq}
\lambda + \bar{a} e^{-\lambda \tau} = 0,
\end{equation}
where $\bar{a} = a/\tau$.

For systems with characteristic equation of the form \eqref{eq:robust_ce_withoutq}, a sufficient condition for robust stability is \cite{kharitonov99}
% Note that this is of the form 
% \begin{equation}
%  p(\lambda) + q(\lambda) e^{-\lambda \tau} = 0,
% \end{equation}
% where $p(\lambda) = \lambda$ and $q(\lambda) = a/\tau $. Here, deg($p$) $>$ deg($q$), and $p(\lambda)$ is a stable polynomial. Therefore, the necessary and sufficient condition for the system to robust stable independent of the delay is \cite{kharitonov99}
% \begin{equation}
%  \left|p(j\omega)\right| >  \left|q(j\omega)\right| \quad \quad \forall \  \omega > 0.
% \end{equation}
% This simplifies to 
% \begin{equation}
%  \frac{a}{\omega\tau} < 1  \quad \quad \forall \ \omega > 0.
% \end{equation}
% Substituting the value of $\omega$, we get
\begin{equation}
 \bar{a} \tau < 1.
 \label{eq:rscondnnoqfb}
\end{equation}
Substituting the value of $\bar{a}$ in \eqref{eq:rscondnnoqfb}, we get a sufficient condition for robust stability of RCP which uses only rate mismatch feedback as 
\begin{equation}
 a < 1.
\end{equation}

\subsection{With queue feedback}
The corresponding characteristic equation is given by 
\begin{equation}
\label{eq:robust_ce_withq}
\lambda + \bar{a} e^{-\lambda \tau} = 0,
\end{equation}
where $\bar{a}={a(1 + \rho^*)}/{\tau}$.
As explained in the previous subsection, the necessary and sufficient condition for robust stability is
\begin{equation}
 a (1 + \rho^*) < 1.
\end{equation}
Therefore, based on the results of stability and convergence analysis, we are unable to deduce whether the queue size feedback is beneficial or not. Then, the next natural step is to investigate the dynamical behavior of the system as it transits from a stable to an unstable regime. In the next section, we explore the impact of loss of local stability for both the design options, i.e., with and without queue size feedback.

%Clearly, this is not true. Hence, the system cannot be robust stable independent of 
%%%%%%%%%%%%%%%%%%%%%%%%%%%%%%%%%%%%%%%%%%%%%%%%%%%%%
\section{Local Hopf bifurcation analysis}
Apart from deriving stability conditions, it is also important to analyze the consequences of violating those conditions so that we feel comfortable in operating the system close to the edge of stable regime. To that end, in this section, we conduct a local Hopf bifurcation analysis.
A Hopf bifurcation gives rise to the emergence of limit cycles in system dynamics. A major concern in local Hopf bifurcation analysis is to determine the type of the bifurcation: will the bifurcation be super-critical or sub-critical?.
%Thus, the next natural step would be to investigate the nature of the bifurcating limit cycles.
For now, we will only be concerned with the first Hopf bifurcation. In contrary to local stability analysis which uses only linear terms, the Hopf bifurcation analysis uses both linear and non-linear terms to analyze the impact of queue size feedback on the system dynamics.

%, which are outlined in the Appendix.
\subsection{Without queue feedback}
% In ~\cite{voice2009maxminrcp}, it is shown that the RCP with feedback based on only rate mismatch always undergoes a super-critical Hopf bifurcation, as the Hopf condition is violated. It highlighted that the system \eqref{eq:noqfeedbackKappa} cannot produce a sub-critical Hopf bifurcation.
The Taylor series expansion of \eqref{eq:noqfeedbackKappa} about its equilibrium is given by 
\begin{equation} 
\frac{d}{dt}u(t) = \kappa(\xi_y u(t-\tau) + \xi_{xy}u(t)u(t-\tau)) 
\label{eq:nlwithoutq}
\end{equation}
where 
\begin{equation}
\xi_y=-\frac{\kappa a}{\tau}, \quad \quad  \xi_{xy}=-\frac{\kappa a}{ \gamma C \tau}.
\end{equation}
In \cite{voice2009maxminrcp}, it has been shown that \eqref{eq:nlwithoutq} undergoes first local Hopf bifurcation at $\kappa=\kappa_c$, where $\kappa_c a=\pi/2.$ If the Hopf condition is just violated, the system would lose local stability via a super-critical Hopf bifurcation and the amplitude of the bifurcating limit cycles will be proportional to 
\begin{equation}
 R^{*}\sqrt{\frac{20\pi(\kappa-\kappa_c)}{3\pi-2}}.
\end{equation}
Here $R^*$ denotes the equilibrium of \eqref{eq:nlwithoutq}. It is also highlighted in \cite{voice2009maxminrcp} that equation \eqref{eq:nlwithoutq} cannot undergo a sub-critical Hopf bifurcation. Therefore, the type of Hopf bifurcation for RCP which uses only rate mismatch feedback is always super-critical, and does not depend on any of the system parameters.

% By comparing \eqref{eq:nlwithoutq} with \eqref{eq:gen_eqn}, we can write the expression for $\mu_2$ of \eqref{eq:nlwithoutq} as 
% \begin{equation} 
%  \mu_2 = \frac{\xi^2_{xy}}{5\pi \xi_y^2}(3\pi-2)=\frac{3\pi-2}{5\pi {\gamma}^2 C^2} > 0, 
%  \label{eq:myu2rcpnoq}
% \end{equation}
% which implies that the type of Hopf bifurcation is always super-critical, and does not depend on any of the system parameters. 
% % Here, we can observe that the sign of $\mu_2$ for \eqref{eq:myu2rcpnoq}  does not depend on any of the system parameters. Therefore, the RCP which uses only rate mismatch feedback, always undergoes a super-critical Hopf bifurcation, as the stability condition is violated.
% This substantiates our previous findings in \cite{voice2009maxminrcp}, where we analyzed a max-min fair variant of RCP, and showed that the RCP without queue feedback would always give rise to super-critical Hopf bifurcation.

\subsection{With queue feedback}
We now consider the RCP model which uses both rate mismatch and queue feedback, and perform the requisite calculations to determine the type of Hopf bifurcation.
% Consider the following non-linear delay differential equation
% \begin{equation}
% \label{eq:gen_non_eq0}
% \dfrac{d}{dt}x(t) = \kappa g\big(x(t),x(t-\tau)\big) = \kappa g\big(x, y\big),
% \end{equation}
% where $g$ has a unique equilibrium and $\kappa$, $\tau > 0$. We can observe that equation \eqref{eq:simple_rate_B} is of the 
% form (\ref{eq:gen_non_eq0}).
% 
The analysis relies on the linear, quadratic and cubic terms in the Taylor series expansion, whose non-zero coefficients of \eqref{eq:simple_rate_B} are tabulated in TABLE \ref{tab:tayexp}. 
\begin{table}[h!]
\caption{Coefficients of linear and higher order terms in the Taylor series expansion of~\eqref{eq:simple_rate_B}.}\label{tab:tayexp}
\renewcommand{\arraystretch}{2.25}
\centering
\begin{tabular}{ll}
\hline
\hline
$\text{Coefficients} \qquad \qquad \qquad \qquad \qquad$  & $\text{Expressions} $\\
\hline
\vspace{-5mm}\\
$\xi_{y}$\ \   & $ -\dfrac{a\bigl( 1  + \rho^* \bigr)}{\tau} $\\  
$\xi_{xy}$\ \   & $ -\dfrac{a(1+\rho^*)}{C\tau\rho^*} $\\
$\xi_{yy}$ \ & $ -\dfrac{a}{C\tau(1-\rho^*)}$\\
$\xi_{xyy}$ \ & $ -\dfrac{a}{C^2\tau\rho^*(1-\rho^*)}$\\
$\xi_{yyy}$ \ & $ -\dfrac{a}{C^2\tau(1-\rho^*)^2}$\\
\vspace{-5mm}\\
\hline
\hline
\end{tabular}
\\
%\caption{Coefficients of linear and higher order terms in the Taylor series expansion of~(\ref{eq:simplified_withoutQ_model}).}\label{tab:tayexp}
%
\end{table}
On writing the Taylor series expansion of~\eqref{eq:simple_rate_B} about the equilibrium up to the third order terms, we get
\begin{equation}
\begin{aligned}
 \frac{d}{dt}u(t) =&\,\,  \kappa(\xi_yu(t-\tau) + \xi_{xy}u(t)u(t-\tau) + \xi_{yy}u^2(t-\tau)+ \xi_{xyy}u(t)u^2(t-\tau)   \\
		     & + \xi_{yyy}u^3(t-\tau)). 
\end{aligned}
\label{eq:rcpwithqnleqn}
\end{equation}\\
% The definitions of Taylor series coefficients are as follows
% \begin{alignat*}
% \xi\xi_i&=g^*_i,&\xi_{ii}&=\dfrac{1}{2}g^*_{ii},&\xi_{iii}&=\dfrac{1}{6}g^*_{iii} \quad \forall \  i \in \{x,y\}\\
% \xi_{xy}&=g^*_{xy},\ & \xi_{xyy}&=\dfrac{1}{2}g^*_{xyy},\ & \xi_{xxy}&=\dfrac{1}{2}g^*_{xxy},
% \end{alignat*}
% where $g^*$ denotes the value of $g$ at the equilibrium.
To analyze the type of Hopf bifurcation in RCP, we employ the following result (obtained in~\cite{raina2005}) about the local instability in a non-linear delay differential equation. Following the analysis in \cite{raina2005}, we now recapitulate the result as follows.\\ \\
\textbf{Result:} \textit{For the following delay differential equation}
\begin{align}
 \frac{d}{dt}u(t) =&\,\, \eta(- bu(t-\tau) + \xi_{xx}u^2(t) + \xi_{xy}u(t)u(t-\tau)+ \xi_{yy}u^2(t-\tau) + \xi_{xxx}u^3(t) \notag \\
 &+ \xi_{xxy}u^2(t)u(t-\tau) + \xi_{xyy}u(t)u^2(t-\tau) + \xi_{yyy}u^3(t-\tau) ),
\label{eq:gen_eqn}
 \end{align}
\textit{where $\eta,\  \tau,\ b >0$.}\\
 
(i) \textit{The necessary and sufficient condition for local stability is}
\begin{equation}
 \eta b \tau < \pi/2
\end{equation}
\textit{and treating $\kappa$ as the bifurcation parameter, the first Hopf bifurcation occurs at $\eta = \eta_c$, where}
\begin{equation}
 \eta_c b\tau = \pi/2.
\end{equation}

(ii) \textit{If the first Hopf condition is just violated, the Hopf bifurcation is super-critical if $\mu_2 > 0$ and sub-critical if $\mu_2 < 0$, where}
\begin{align}
 \mu_2 &= \frac{1}{\pi	b}\Biggl( \xi^2_{xx}\frac{4(\pi-9)}{5b} + \xi^2_{xy}\frac{3\pi-2}{5b} + \xi^2_{yy}\frac{2(11\pi-4)}{5b} + \xi_{xx}\xi_{xy}\frac{(7\pi-18)}{5b} \notag \\ 
 &  + \xi_{xx}\xi_{yy}\frac{2(7\pi-18)}{5b}  +  \xi_{xy}\xi_{yy}\frac{(7\pi-18)}{5b}   -6\xi_{xxx} + \pi\xi_{xxy} -2\xi_{xyy} + 3\pi\xi_{yyy}  \Biggr). 
 \label{eq:myu2gr}
\end{align}

% (iii) \textit{The asymptotic form of the bifurcating periodic solutions is given by}
% \begin{align}
%  u(t)=&\sqrt{\dfrac{4(\eta-\eta_c)}{\mu_2}} \cos\left(\dfrac{\pi t}{2\tau}\right) + \left(\dfrac{\eta-\eta_c}{\mu_2}\right)\Bigg(\dfrac{2\xi_{xx}}{5b}\bigg(2\sin{\left(\dfrac{\pi t}{\tau}\right)}-\cos{\left(\dfrac{\pi t }{\tau}\right)} + 5 \bigg) \nonumber \\
%  &-\dfrac{2\xi_{xy}}{5b}\bigg(\sin{\left(\dfrac{\pi t}{\tau}\right)}+2\cos{\left(\dfrac{\pi t }{\tau}\right)}\bigg)+\dfrac{2\xi_{yy}}{5b}\bigg(\cos{\left(\dfrac{\pi t}{\tau}\right)}-2\sin{\left(\dfrac{\pi t }{\tau}\right)} + 5 \bigg)\Bigg) \nonumber \\
%  &+\mathcal{O}(\epsilon^3) 
%  \label{eq:raina2005amp}
% \end{align}
% \textit{for $0 \leq t \leq \mathcal{P}(\epsilon)$, where $\epsilon=\sqrt{{\left(\eta-\eta_c\right)}/{\mu_2}}$.}\\
% 
% (iv) \textit{The period of the bifurcation solutions is}
% \begin{align}
%  \mathcal{P(\epsilon)}=& \, 4\tau \Bigg( 1+\dfrac{\left(\eta-\eta_c\right)}{\mu_2}\times \dfrac{1}{5\pi b}\bigg( \dfrac{36}{b}\xi_{xx}^2 + \dfrac{2}{b} \xi_{xy}^2 + \dfrac{8}{b}\xi_{yy}^2 + \dfrac{18}{b}\xi_{xx}\xi_{xy} + \dfrac{36}{b}\xi_{xx}\xi_{yy} \nonumber \\
%  & + \dfrac{18}{b}\xi_{xy}\xi_{yy} + 30 \xi_{xxx} + 10 \xi_{xyy}\bigg) + \mathcal{O}\big(\epsilon^{4}\big)
%  \Bigg).
%  \label{eq:raina2005period}
% \end{align}
In \cite{raina2005}, the analytical tools employed to study the type of Hopf bifurcation are the Poincar{\'e} normal forms and the center manifold theorem \cite{hassard1981}.

By comparing \eqref{eq:rcpwithqnleqn} with \eqref{eq:gen_eqn}, and using \eqref{eq:myu2gr}, we obtain $\mu_2$ of \eqref{eq:rcpwithqnleqn} as
\begin{equation}
 \mu_2 = \, \xi^2_{xy}\frac{3\pi-2}{5\pi\xi_{y}^2} + \xi^2_{yy}\frac{2(11\pi-4)}{5\pi\xi_{y}^2}     
 +  \xi_{xy}\xi_{yy}\frac{(7\pi-18)}{5\pi\xi_{y}^2}  
 +\dfrac{2\xi_{xyy}}{\pi\xi_y} - \dfrac{3\xi_{yyy}}{\xi_y}. 
  \label{eq:myu2rcp1}
 \end{equation}
In \cite{raina2005}, the Hopf bifurcation properties of a non-linear equation with a single discrete delay was studied in some detail. The analysis allowed us to ascertain that some non-linear terms always produced a Hopf bifurcation of certain type. This enabled us to identify the impact of some non-linear terms on the nature of Hopf bifurcation. So leading from the previous work \cite{raina2005}, it might be natural to ask if we may develop a similar understanding for the non-linear delay equation of RCP. Such analysis can help us identify which non-linear terms may be desirable, in the sense that they always produced a super-critical Hopf.

\begin{figure}[hbtp!]
\centering
\psfrag{x}{\textbf{$\xi_{xy}$}}
\psfrag{y}{\textbf{$\xi_{yy}$}}
\psfrag{0}{\small{$0$}}
\psfrag{5}{\small{$5$}}
\psfrag{-5}{\small{$-5$}}
\psfrag{100}{\small{$100$}}
\psfrag{50}{\small{$50$}}
\psfrag{mu2}[b1][1][1] [90]{\textbf{$\mu_2$}}
%\psfrag{Hopfboundary}{\footnotesize{$\leftarrow$Hopf boundary}}
\includegraphics[trim=0cm 0cm 0cm 2.2cm, clip=true, width=5in,height=2in]{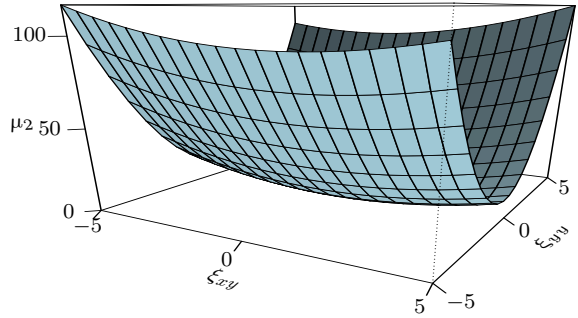}
\vspace{-2mm}
\caption{Effect of the quadratic terms $\xi_{xy}$ and $\xi_{yy}$ on the nature of Hopf bifurcation. Observe that $\mu_2>0$ which implies that the Hopf bifurcation is super-critical.}
%\vspace{-2mm}
\label{fig:mu2_rcp_quadratics}
\end{figure}

\begin{figure}[hbtp!]
\centering
\psfrag{x}{$\xi_{xyy}$}
\psfrag{y}{$\xi_{yyy}$}
\psfrag{0}{\small{$0$}}
\psfrag{2}{\small{$2$}}
\psfrag{-2}{\hspace{-1mm}\small{$-2$}}
\psfrag{-1}{\vspace{1mm}\hspace{-1mm}\small{$-1$}}
\psfrag{-4}{\vspace{-5mm}\hspace{-1mm}\small{$-4$}}
\psfrag{mu2}[b1][1][1] [90]{\textbf{$\mu_2$}}
%\psfrag{Hopfboundary}{\footnotesize{$\leftarrow$Hopf boundary}}
\includegraphics[trim=0cm 0cm 0cm 2.2cm, clip=true, width=5in,height=2in]{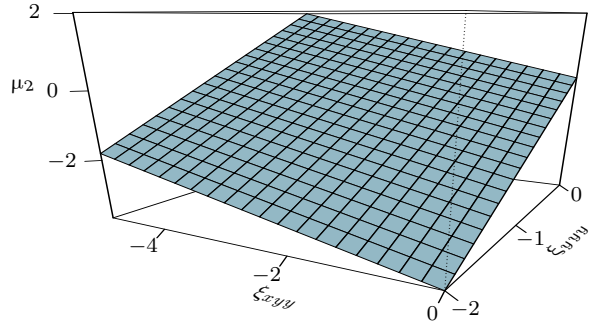}
\vspace{-2mm}
\caption{Effect of the cubic terms $\xi_{xyy}$ and $\xi_{yyy}$ on the nature of Hopf bifurcation. The Hopf bifurcation is \textit{super-critical} if $\mu_2>0$, and is \textit{sub-critical} if $\mu_2<0$. Observe that the sign of $\mu_2$ depends on the values of both the cubic terms.}
%\vspace{-2mm}
\label{fig:mu2_rcp_cubics}
\end{figure}

We first consider each of the non-linear terms in isolation, and analyze its impact on the type of Hopf bifurcation. If we consider $\xi_{xy} u(t)u(t-\tau)$ or $\xi_{yy} u^{2}(t-\tau)$ in isolation, then we get 
\begin{equation}
 \mu_2 = \xi^2_{xy}\frac{3\pi-2}{5\xi_{y}^2} > 0 \, \, \text{and}\, \, \mu_2 =  \xi^2_{yy}\frac{2(11\pi-4)}{5\xi_{y}^2} > 0
\end{equation}
respectively. Therefore, the type of Hopf bifurcation is always super-critical in both the cases. Now, we consider the combination of quadratic terms i.e. both $\xi_{xy} u(t)u(t-\tau)$ and $\xi_{yy} u^{2}(t-\tau)$. Then the value of $\mu_2$ is
\begin{equation}
 \mu_2 = \, \xi^2_{xy}\frac{3\pi-2}{5\pi\xi_{y}^2} + \xi^2_{yy}\frac{2(11\pi-4)}{5\pi\xi_{y}^2}     
 +  \xi_{xy}\xi_{yy}\frac{(7\pi-18)}{5\pi\xi_{y}^2}. 
  \label{eq:myu2rcpquadratic}
 \end{equation}
Figure \ref{fig:mu2_rcp_quadratics} shows the value of $\mu_2$ for various values of $\xi_{xy}$ and $\xi_{yy}$. From Figure \ref{fig:mu2_rcp_quadratics}, we can observe that  $\mu_2 > 0$, which implies that the system undergoes a super-critical Hopf bifurcation.
Whereas, if cubic terms $\xi_{xyy} u(t) u^2(t-\tau)$ and $\xi_{yyy} u^3(t-\tau)$  are considered in isolation, we get
\begin{equation}
 \mu_2 = \frac{2\xi_{xyy}}{\pi\xi_{y}} \, \, \text{and}\, \, \mu_2 =  \frac{-3\xi_{yyy}}{\xi_{y}}
\end{equation}
respectively. In our case, from Table \ref{tab:tayexp}, we get  $\xi_y$, $\xi_{xyy}$ and $\xi_{yyy}$ $<0$. This implies that the cubic terms $\xi_{xyy} u(t) u^2(t-\tau)$ and $\xi_{yyy} u^3(t-\tau)$ induces super-critical and sub-critical Hopf respectively. In the case where we have both cubic terms, then the criticality of Hopf bifurcation depends on the magnitude of the cubic coefficients 
(see Figure \ref{fig:mu2_rcp_cubics}).
In summary, the quadratic terms and the cubic term $\xi_{xyy} u(t) u^2(t-\tau)$ induce super-critical Hopf. Whereas, the cubic term $\xi_{yyy} u^3(t-\tau)$ induces sub-critical Hopf. We now analyze what happens when we have both quadratic terms and cubic terms, using the expressions of its corresponding coefficients from Table \ref{tab:tayexp}.

\begin{figure}[h!]
  \centering
  \psfrag{0.0}{\scriptsize\hspace{1.5pt}$0$}
  \psfrag{0.5}{\hspace{-1pt}\scriptsize $0.5$}
  \psfrag{0.6621}{\hspace{-3pt}\scriptsize $0.6621$}
  \psfrag{1.0}{\hspace{-1pt}\scriptsize $1.0$}
  \psfrag{sub}{\footnotesize \hspace{-1pt} \begin{tabular}{c} $\leftarrow$ sub-critical $\rightarrow$ \\ Hopf \end{tabular} }
  \psfrag{super}{\footnotesize \hspace{-23pt}  \begin{tabular}{c} $\leftarrow$ super-critical $\rightarrow$  \\ Hopf  \end{tabular}}
  \psfrag{y}[b][t]{ {Numerator of $\mu_2$}}
  \psfrag{rho}[c]{ {Equilibrium utilization, $\rho^*$}}
  \psfrag{10}{\hspace{-2pt}\scriptsize $10$}
  \psfrag{0}{\scriptsize $0$}
  \psfrag{-10}{\hspace{-5pt}\scriptsize $-10$}
  \psfrag{-20}{\hspace{-5pt}\scriptsize $-20$}
  \psfrag{-30}{\hspace{-5pt}\scriptsize $-30$}
 \includegraphics[scale=0.27,angle=-90]{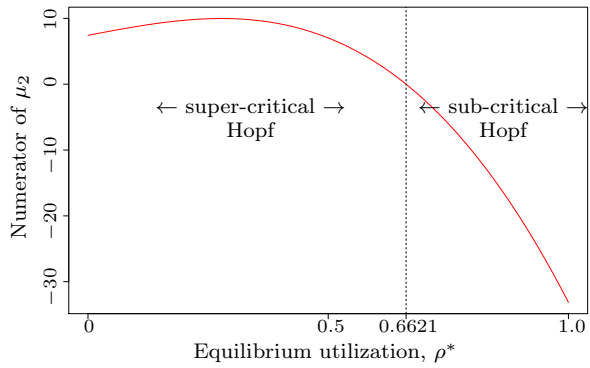}
 \caption{Variation in the numerator of $\mu_2$ of RCP with queue feedback, as the equilibrium utilization ($\rho^*$) changes. Observe that $\mu_2$ turns negative for $\rho^*>0.6621$.
 Hence the Hopf bifurcation is super-critical for $\rho^*<0.6621$ and sub-critical for $\rho^*>0.6621$.}
 \label{fig:u2plot}
\end{figure}

% \begin{align}
%  u(t)=&\sqrt{\dfrac{4(\eta-\eta_c)}{\mu_2}} \cos\left(\dfrac{\pi t}{2\tau}\right) + \left(\dfrac{\eta-\eta_c}{\mu_2}\right)\Bigg(\dfrac{2\xi_{xx}}{5b}\bigg(2\sin{\left(\dfrac{\pi t}{\tau}\right)}-\cos{\left(\dfrac{\pi t }{\tau}\right)} + 5 \bigg) \nonumber \\
%  &-\dfrac{2\xi_{xy}}{5b}\bigg(\sin{\left(\dfrac{\pi t}{\tau}\right)}+2\cos{\left(\dfrac{\pi t }{\tau}\right)}\bigg)+\dfrac{2\xi_{yy}}{5b}\bigg(\cos{\left(\dfrac{\pi t}{\tau}\right)}-2\sin{\left(\dfrac{\pi t }{\tau}\right)} + 5 \bigg)\Bigg) \nonumber \\
%  &+\mathcal{O}(\epsilon^3) 
%  \label{eq:raina2005amp}
% \end{align}
% for $0 \leq t \leq \mathcal{P}(\epsilon)$, where $\epsilon=\sqrt{{\left(\eta-\eta_c\right)}/{\mu_2}}$, $\eta_c b\tau = \pi/2$ and the period of the bifurcation solutions is
% \begin{align}
%  \mathcal{P(\epsilon}=&4\tau \Bigg( 1+\dfrac{\left(\eta-\eta_c\right)}{\mu_2}\times \dfrac{1}{5\pi b}\bigg( \dfrac{36}{b}\xi_{xx}^2 + \dfrac{2}{b} \xi_{xy}^2 + \dfrac{8}{b}\xi_{yy}^2 + \dfrac{18}{b}\xi_{xx}\xi_{xy} + \dfrac{36}{b}\xi_{xx}\xi_{yy} \nonumber \\
%  & + \dfrac{18}{b}\xi_{xy}\xi_{yy} + 30 \xi_{xxx} + 10 \xi_{xyy}\bigg) + \mathcal{O}(\epsilon^{4})
%  \Bigg)
%  \label{eq:raina2005period}
% \end{align}
After substituting the values from Table \ref{tab:tayexp} in~\eqref{eq:myu2rcp1}, and simplifying, we obtain the expression for $\mu_2$ of \eqref{eq:rcpwithqnleqn} as
\begin{align}
 \mu_2 =&\,\, \frac{1}{C^2 5\pi{\rho^*}^2\left(1-{\rho^*}^2\right)^2}\Bigg( (3\pi-2){\rho^*}^4 - (22\pi - 8){\rho^*}^3 - (4-\pi){\rho^*}^2  \notag \\
 & +  (7\pi-8)\rho^* + (3\pi-2)\Bigg).
     \label{eq:myu2rcp2}
\end{align}
% \begin{equation}
%  \mu_2 =(3\pi-2)(1+{\rho^*}^4)-(22\pi - 8){\rho^*}^3-(4-\pi){\rho^*}^2 +(7\pi-8)\rho^*
% \end{equation}
% where
% \begin{align}
%  \rho^* = \frac{R^*}{C}&= \frac{\left(b+4\right)-\sqrt{8b+b^2} }{4}.\label{eq:rhob}
% \end{align}
To analyze the type of the Hopf bifurcation for \eqref{eq:rcpwithqnleqn}, we need to find the sign of its corresponding $\mu_2$. From \eqref{eq:myu2rcp2}, we can deduce that only numerator terms determine the sign of $\mu_2$. Hence we plot the variation in the numerator of $\mu_2$ as the equilibrium utilization $\rho^*$ is varied from $0$ to $1$; see Figure \ref{fig:u2plot}. The equilibrium link utilization depends on the value of the parameter $b$. 
We can observe from Figure \ref{fig:u2plot} that $\mu_2$ reaches zero at $\rho^*=0.6621$, and hence the type of Hopf bifurcation changes from super-critical to sub-critical after $\rho^*=0.6621$.

It is also noteworthy that the sign of $\mu_2$ does not depend on the link capacity ($C$) and round-trip time ($\tau$).
%%%%%%%%%%%%%%%%%%%%%%%%%%%%%%  expressions for amplitude and period    %%%%%%%%%%%%%%%%%%%%%%%%%%%%%%%%%%
%  Similarly, we can write the expressions for amplitude and period as\\
% \begin{align}
%  u(t)=&\sqrt{\dfrac{4(\eta-\eta_c)}{\mu_2}} \cos\left(\dfrac{\pi t}{2\tau}\right) + \left(\dfrac{\eta-\eta_c}{\mu_2}\right)\Bigg(\dfrac{-2\xi_{xy}}{5b}\bigg(\sin{\left(\dfrac{\pi t}{\tau}\right)}+2\cos{\left(\dfrac{\pi t }{\tau}\right)}\bigg) \nonumber \\
%  &+\dfrac{2\xi_{yy}}{5b}\bigg(\cos{\left(\dfrac{\pi t}{\tau}\right)}-2\sin{\left(\dfrac{\pi t }{\tau}\right)} + 5 \bigg)\Bigg) +\mathcal{O}(\epsilon^3) 
%  \label{eq:rcp1damp}
% \end{align}
% \begin{equation}
%  \mathcal{P(\epsilon)}=4\tau \Bigg( 1+\dfrac{\left(\eta-\eta_c\right)}{\mu_2}\times \dfrac{1}{5\pi b}\bigg( \dfrac{2}{b} \xi_{xy}^2 + \dfrac{8}{b}\xi_{yy}^2 + \dfrac{18}{b}\xi_{xy}\xi_{yy} + 10 \xi_{xyy}\bigg) + \mathcal{O}\big(\epsilon^{4}\big)
%  \Bigg).
%  \label{eq:rcp1dperiod}
% \end{equation}

We now validate the analytical results using some numerical examples.

\begin{figure}[h!]
\centering
\psfrag{R}[][][2]{Amplitude of oscillation}
\psfrag{kappa}[][][2]{Bifurcation parameter, $\kappa$}
\psfrag{0.95}[][][2]{$0.95$}
\psfrag{1.00}[][][2]{\scriptsize $1.00$}
\psfrag{1.01}[][][2]{\scriptsize $1.01$}
\psfrag{1.02}[][][2]{\scriptsize $1.02$}
\psfrag{1.03}[][][2]{\scriptsize $1.03$}
\psfrag{1.04}[][][2]{\scriptsize $1.04$}
\psfrag{1.05}[][][2]{\scriptsize $1.05$}
\psfrag{0}[][][2]{\scriptsize $0$}
\psfrag{4}[][][2]{\scriptsize $4$}
\psfrag{8}[][][2]{\scriptsize $8$}
\includegraphics[scale = 0.9,trim=0cm 0cm 0cm 1.7cm, clip=true,width=3.25in,height=2.25in]{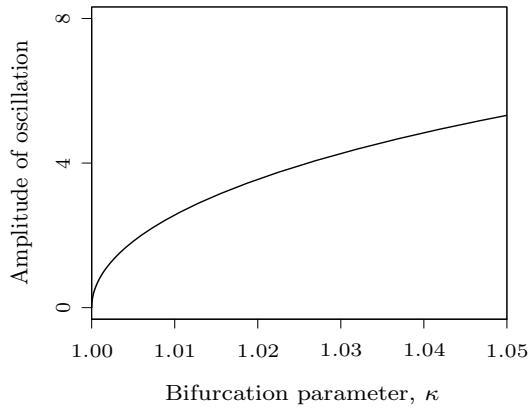}
\vspace{-2mm}
\caption{Bifurcation diagram highlighting that the system undergoes a super-critical Hopf bifurcation at $\kappa=1$. The parameter values used are  $a=1.01$, $\tau=100$, $C=10$ and $b=0.736$ ($\rho^*=0.55$).}
%\vspace{-5mm}
\label{fig:supercriticalplot}
\vspace{-2mm}
\end{figure}

\begin{figure}[h!]
\centering
\psfrag{R}[][][2]{ Amplitude of oscillation}
\psfrag{kappa}[][][2]{ Bifurcation parameter, $\kappa$}
\psfrag{0.95}[][][2]{\scriptsize $0.95$}
\psfrag{0.96}[][][2]{\scriptsize $0.96$}
\psfrag{0.97}[][][2]{\scriptsize $0.97$}
\psfrag{0.98}[][][2]{\scriptsize $0.98$}
\psfrag{0.99}[][][2]{\scriptsize $0.99$}
\psfrag{1.00}[][][2]{\scriptsize $1.00$}
\psfrag{1.01}[][][2]{\scriptsize $1.01$}
\psfrag{1.02}[][][2]{\scriptsize $1.02$}
\psfrag{1.03}[][][2]{\scriptsize $1.03$}
\psfrag{1.04}[][][2]{\scriptsize $1.04$}
\psfrag{1.05}[][][2]{\scriptsize $1.05$}
\psfrag{0}[][][2]{\scriptsize $0$}
\psfrag{10}[][][2]{\scriptsize $10$}
\psfrag{20}[][][2]{$20$}
\psfrag{30}[][][2]{$30$}
\psfrag{40}[][][2]{$40$}
\psfrag{4}[][][2]{$4$}
\psfrag{8}[][][2]{$8$}
\psfrag{5}[][][2]{\scriptsize $5$}
\psfrag{60}[][][2]{$60$}
\psfrag{104}{\small{$\times 10^{-4}$}}
\includegraphics[scale = 0.9,trim=0cm 0cm 0cm 1.7cm, clip=true,width=3.25in,height=2.25in]{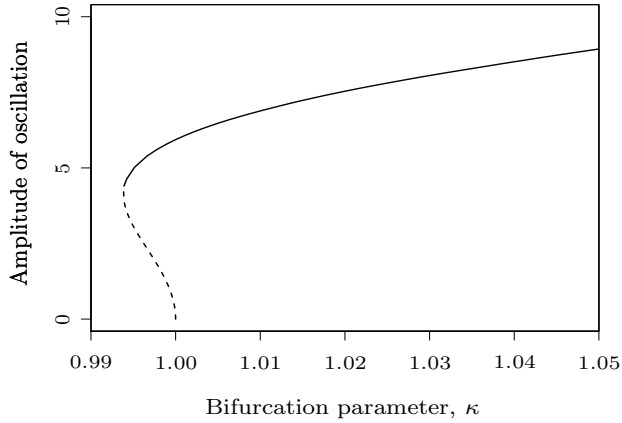}
\vspace{-2mm}
\caption{Bifurcation diagram showing the existence of a sub-critical Hopf for $a=0.924$, $\tau=100$, $C=10$ and $b=0.257$ ($\rho^*=0.70$). The solid and dashed lines denote the amplitude of stable and unstable limit cycles respectively. }
\label{fig:subcriticalplot}
\end{figure}

\begin{figure}[h!]
\newcommand{\subwdth}{0.5\textwidth}
\psfrag{time}{\hspace{-0.1cm}  \scriptsize Time}
\psfrag{Rate}{\hspace{-0.7cm} Rate, $R(t)$}
\psfrag{0}{\hspace{-0.04cm}\scriptsize{$0$}}
\psfrag{5}{\hspace{-0.05cm}\scriptsize{$5$}}
\psfrag{10}{\hspace{-0.1cm}\scriptsize{$10$}}
\psfrag{15}{\hspace{-0.1cm}\scriptsize{$15$}}
\psfrag{20}{\hspace{-0.1cm}\scriptsize{$20$}}
\psfrag{25}{\hspace{0cm}\scriptsize{$25$}}
\psfrag{50}{\hspace{-0.1cm}\scriptsize{$50$}}
\psfrag{100}{\hspace{-0.1cm}\scriptsize{$100$}}
\psfrag{200}{\hspace{-0.1cm}\scriptsize{$200$}}
\psfrag{0.5}{\hspace{-0.1cm}\scriptsize{$0.95$}}
\psfrag{1.0}{\hspace{-0.1cm}\scriptsize{$1.0$}}
\psfrag{1.5}{\hspace{-0.1cm}\scriptsize{$1.05$}}
\psfrag{150}{\hspace{-0.1cm}\scriptsize{$150$}}
\psfrag{300}{\hspace{-0.1cm}\scriptsize{$300$}}
\psfrag{400}{\hspace{-0.1cm}\scriptsize{$400$}}
\psfrag{-800}{\hspace{-0.1cm}\scriptsize{$-800$}}
\psfrag{-400}{\hspace{-0.1cm}\scriptsize{$-400$}}
\psfrag{-300}{\hspace{-0.1cm}\scriptsize{$-300$}}
\psfrag{-600}{\hspace{-0.1cm}\scriptsize{$-600$}}
\psfrag{125}{\hspace{-0.1cm}\scriptsize{$125$}}
\psfrag{250}{\hspace{-0.1cm}\scriptsize{$250$}}
\psfrag{60}{\hspace{-0.1cm}\scriptsize{$60$}}
\psfrag{120}{\hspace{-0.1cm}\scriptsize{$120$}}
\psfrag{25000}{\hspace{-0.1cm}\scriptsize{$25000$}}
\psfrag{50000}{\hspace{-0.1cm}\scriptsize{$50000$}}
\psfrag{10000}{\hspace{-0.1cm}\scriptsize{$10000$}}
\psfrag{20000}{\hspace{-0.1cm}\scriptsize{$20000$}}
\begin{tabular}{c}
\subfloat[ $\kappa = 0.95$]{\includegraphics[width=\subwdth]{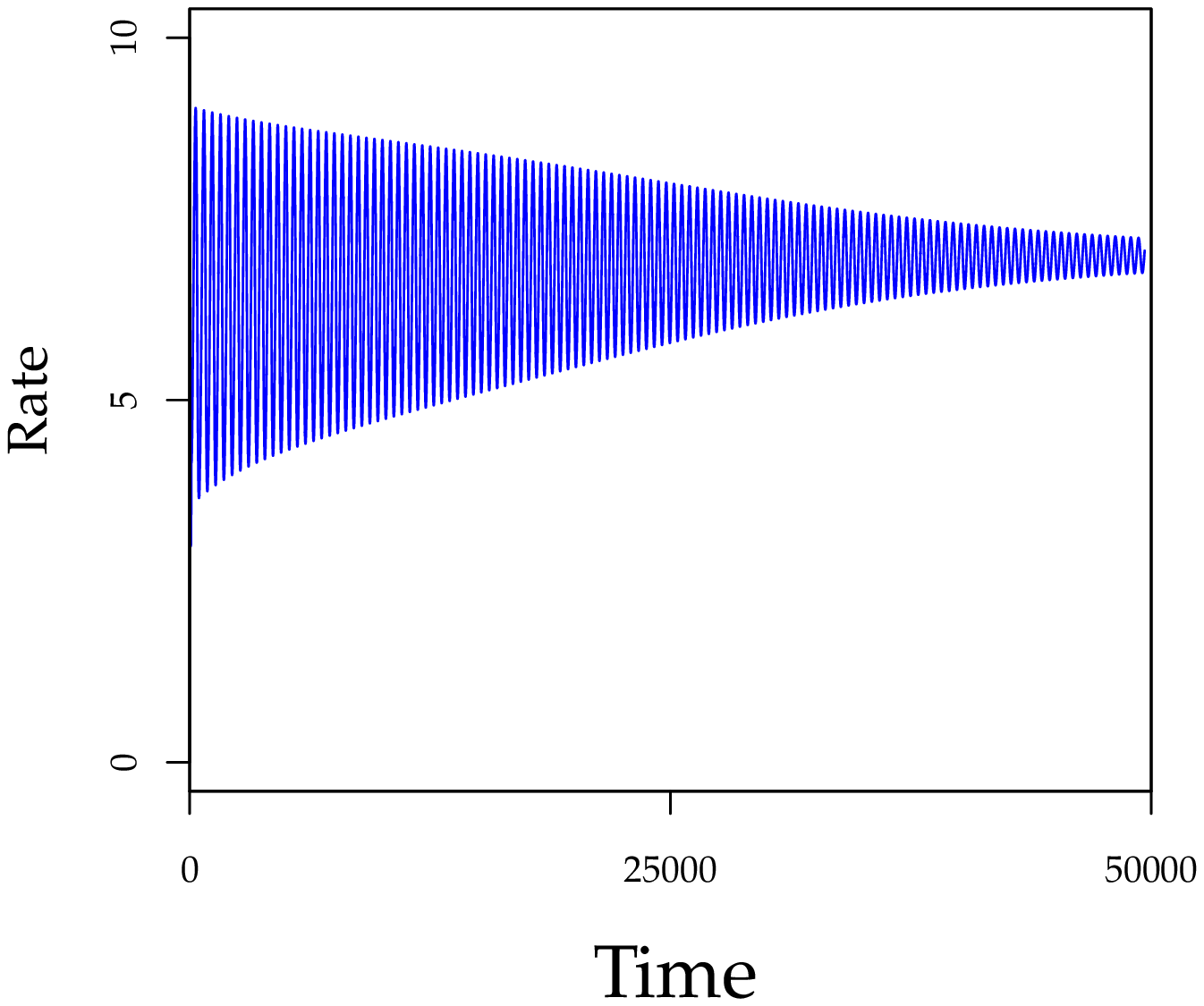}}\hspace{-5mm}
% \subfloat[ $\kappa = 0.97, R_0=2$]{\includegraphics[width=\subwdth]{tompecssub2.eps}}\hspace{-5mm}
\subfloat[ $\kappa = 1.05$]{\includegraphics[width=\subwdth]{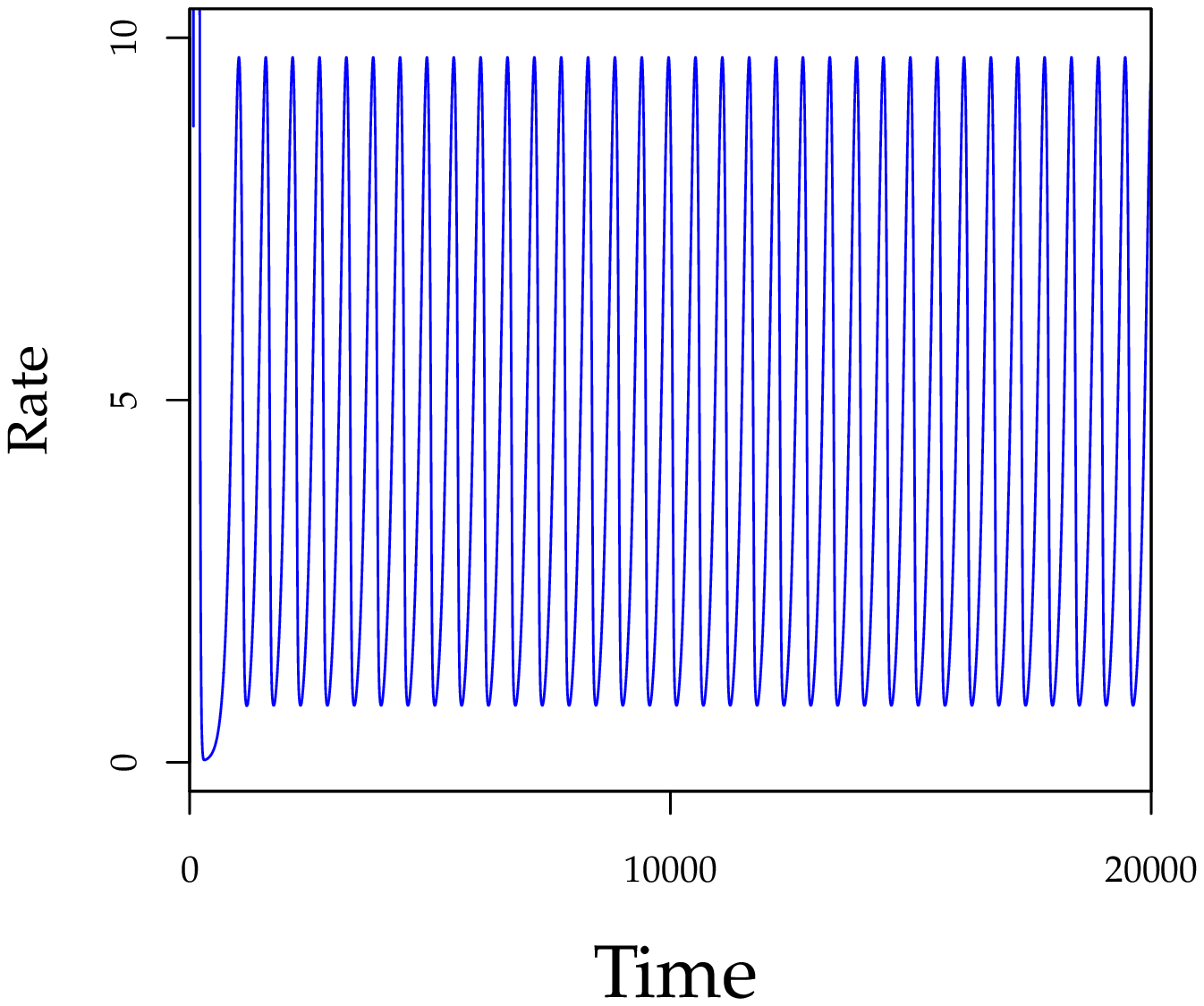}}
\end{tabular}
\caption{Numerical simulations illustrating that the system undergoes a sub-critical Hopf as the bifurcation parameter $\kappa$ increases beyond the critical threshold ($\kappa_c =1$). The values of the parameters used are $a=0.924$, $\tau=100$, $C=10$ and $b=0.257$ ($R^*=7.0$). The initial condition is chosen as $R_0=6.9$.} \label{fig:sub_rcp1d_withq}
\end{figure}

\begin{figure}[h!]
\centering
\psfrag{R}[][][2]{ Amplitude of oscillation}
\psfrag{kappa}[][][2]{ Bifurcation parameter, $\kappa$}
\psfrag{0.95}[][][2]{\scriptsize $0.95$}
\psfrag{0.96}[][][2]{\scriptsize $0.96$}
\psfrag{0.97}[][][2]{\scriptsize $0.97$}
\psfrag{0.98}[][][2]{\scriptsize $0.98$}
\psfrag{0.99}[][][2]{\scriptsize $0.99$}
\psfrag{1.00}[][][2]{\scriptsize $1.00$}
\psfrag{1.01}[][][2]{\scriptsize $1.01$}
\psfrag{1.02}[][][2]{\scriptsize $1.02$}
\psfrag{1.03}[][][2]{\scriptsize $1.03$}
\psfrag{1.04}[][][2]{\scriptsize $1.04$}
\psfrag{1.05}[][][2]{\scriptsize $1.05$}
\psfrag{0}[][][2]{\scriptsize $0$}
\psfrag{10}[][][2]{\scriptsize $10$}
\psfrag{20}[][][2]{$20$}
\psfrag{30}[][][2]{$30$}
\psfrag{40}[][][2]{$40$}
\psfrag{4}[][][2]{\scriptsize $4$}
\psfrag{2}[][][2]{\scriptsize$2$}
\psfrag{5}[][][2]{\scriptsize $5$}
\psfrag{60}[][][2]{$60$}
\psfrag{104}{\small{$\times 10^{-4}$}}
\includegraphics[scale = 0.9,trim=0cm 0cm 0cm 1.7cm, clip=true,width=3.25in,height=2.25in]{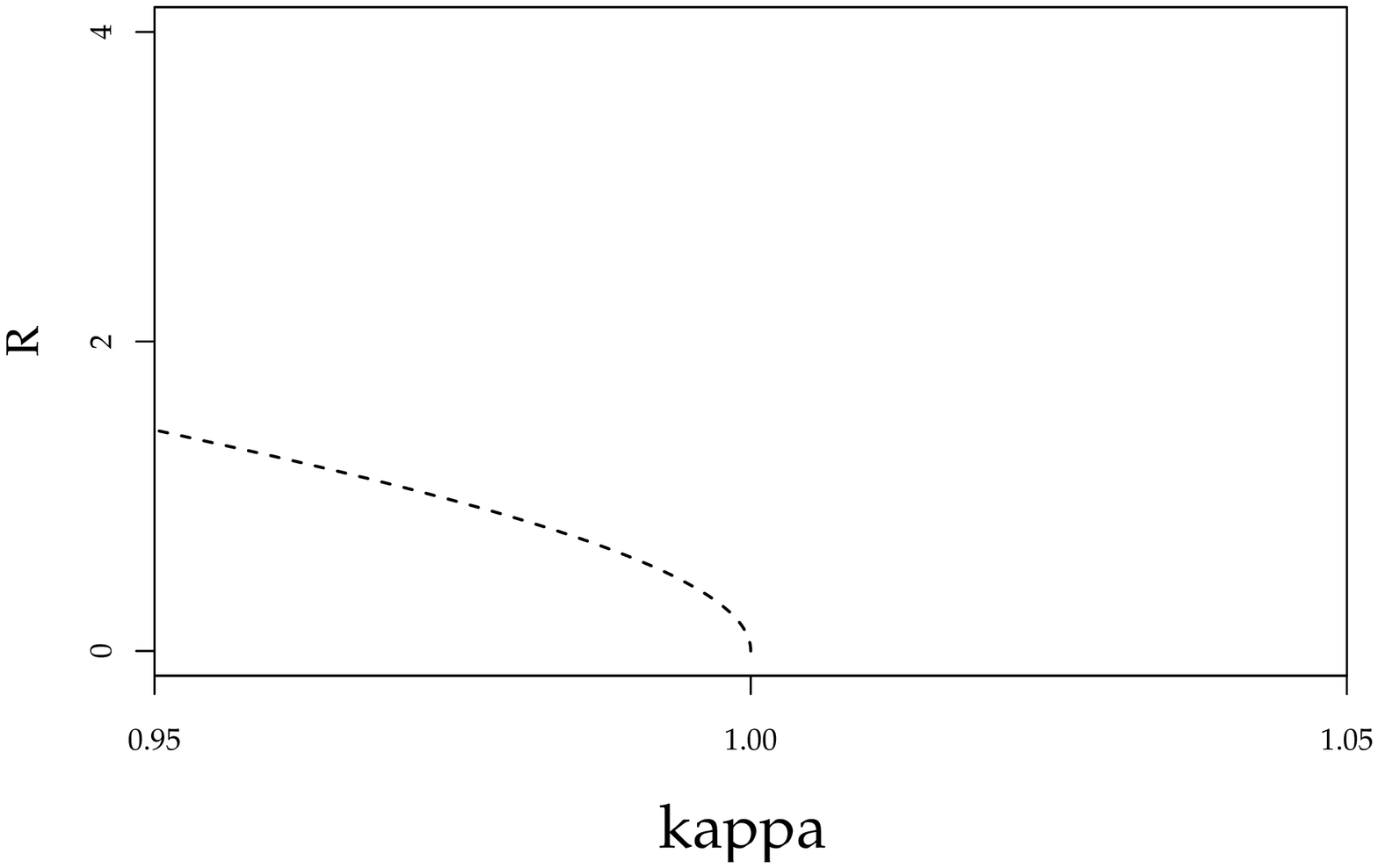}
\vspace{-2mm}
\caption{Bifurcation diagram showing the existence of a sub-critical Hopf for $a=0.827$, $\tau=100$, $C=10$ and $b=0.022$ ($\rho^*=0.90$).}
%\vspace{-5mm}
\label{fig:subcriticalplot1}
\end{figure}

$\textit{Numerical Example 1 (Super-critical)}$: Let us consider the RCP system with $C=10$, $\tau=100$ and $b=0.736$ which corresponds to equilibrium utilization of 55\% of link capacity i.e., $\rho^*=0.55$. For these values, using \eqref{eq:kappa_cval}, we obtain $a=1.01$, for $\kappa_c =1$. Substituting the values in \eqref{eq:myu2rcp2}, we calculate the value of  $ \mu_2$ as $ 2.324 \times 10^{-2} > 0$, implying that the system undergoes a super-critical Hopf bifurcation. The bifurcation diagram drawn using the Matlab package DDE-Biftool \cite{ddetool1,ddetool2} is shown in Figure \ref{fig:supercriticalplot}. As expected, it shows that the system loses local stability via a super-critical Hopf bifurcation, as the bifurcation parameter crosses the critical threshold ($\kappa_c =1$). To validate this, numerical simulations obtained using XPPAUT \cite{xppaut2002} are shown in Figure \ref{fig:super_rcp1d_withq}. For $\kappa=0.95$, and the initial condition as $R_0=5.6$, the system converges to the equilibrium rate, $R^*=5.5$ (see Figure \ref{fig:super_rcp1d_withq}(a)). Whereas, for $\kappa = 1.05 > \kappa_c$ i.e. after the bifurcation, the system leads to the emergence of stable limit cycles.\\
%%%%%%%%%%%%%%%%%%%%%%%%%%%%
\newcommand{\supwdth}{0.5\textwidth}
\begin{figure}[h!]
\psfrag{time}{\hspace{-0.1cm}  \scriptsize Time}
\psfrag{Rate}{\hspace{-0.7cm} Rate, $R(t)$}
\psfrag{0}{\footnotesize{$0$}}
\psfrag{5}{\hspace{-0.05cm}\scriptsize{$5$}}
\psfrag{10}{\hspace{-0.1cm}\scriptsize{$10$}}
\psfrag{15}{\hspace{-0.1cm}\scriptsize{$15$}}
\psfrag{20}{\hspace{-0.1cm}\scriptsize{$20$}}
\psfrag{25000}{\hspace{-0.1cm}\scriptsize{$25000$}}
\psfrag{50000}{\hspace{-0.1cm}\scriptsize{$50000$}}
\psfrag{10000}{\hspace{-0.1cm}\scriptsize{$10000$}}
\psfrag{20000}{\hspace{-0.1cm}\scriptsize{$20000$}}
\psfrag{0.5}{\hspace{-0.1cm}\scriptsize{$0.95$}}
\psfrag{1.0}{\hspace{-0.1cm}\scriptsize{$1.0$}}
\psfrag{1.5}{\hspace{-0.1cm}\scriptsize{$1.05$}}
\psfrag{3}{\hspace{-0.1cm}\scriptsize{$3$}}
\psfrag{1}{\hspace{-0.1cm}\scriptsize{$1$}}
\psfrag{2}{\hspace{-0.1cm}\scriptsize{$2$}}
\begin{tabular}{c}
\subfloat[ $\kappa = 0.95, R_0=5.6$]{\includegraphics[width=\supwdth]{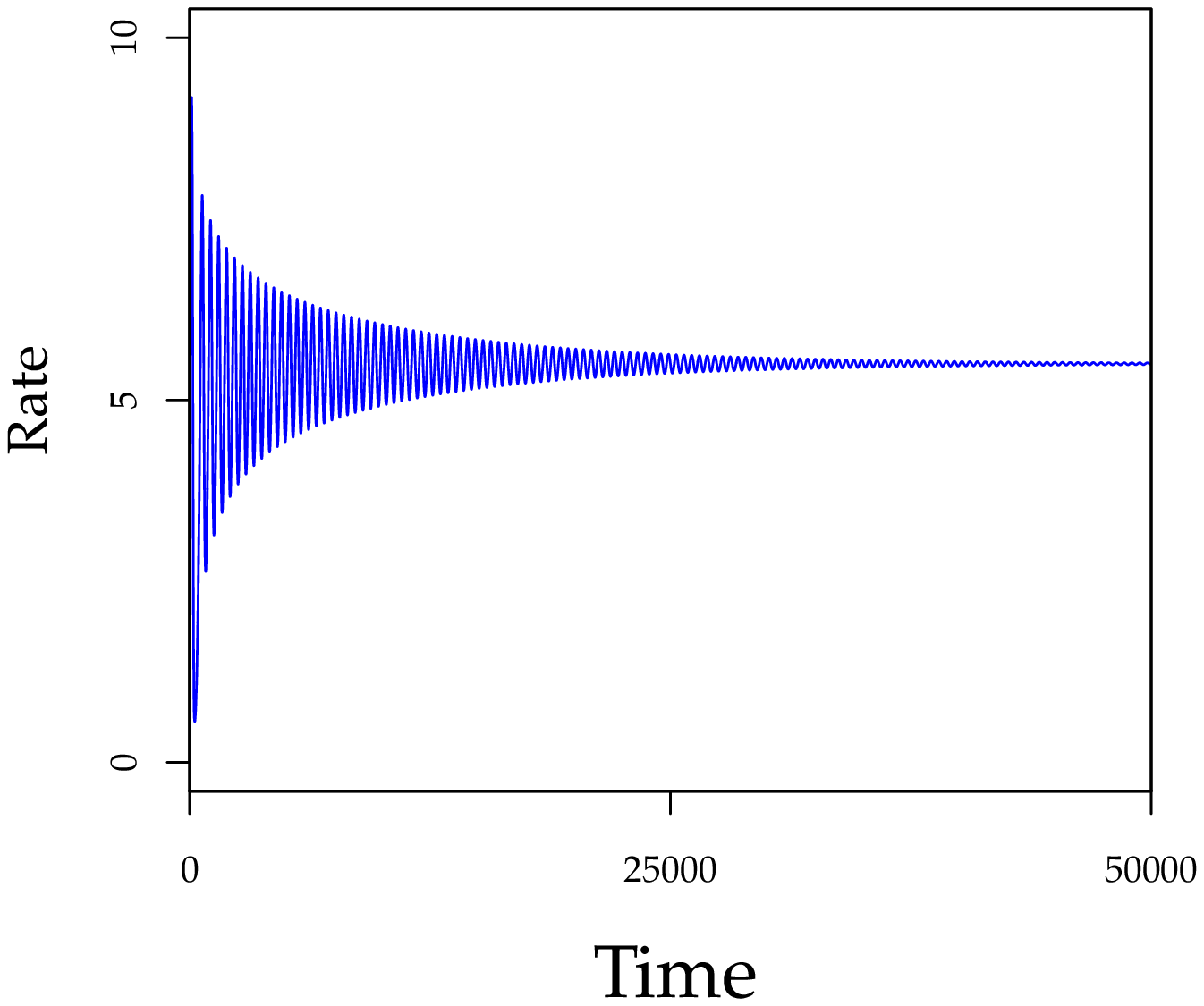}}\hspace{-5mm}
%\subfloat[ $\kappa = 0.97, R_0=2$]{\includegraphics[width=\supwdth]{tompecssuper2.eps}}\hspace{-5mm}
\subfloat[ $\kappa = 1.05, R_0=5.6$]{\includegraphics[width=\supwdth]{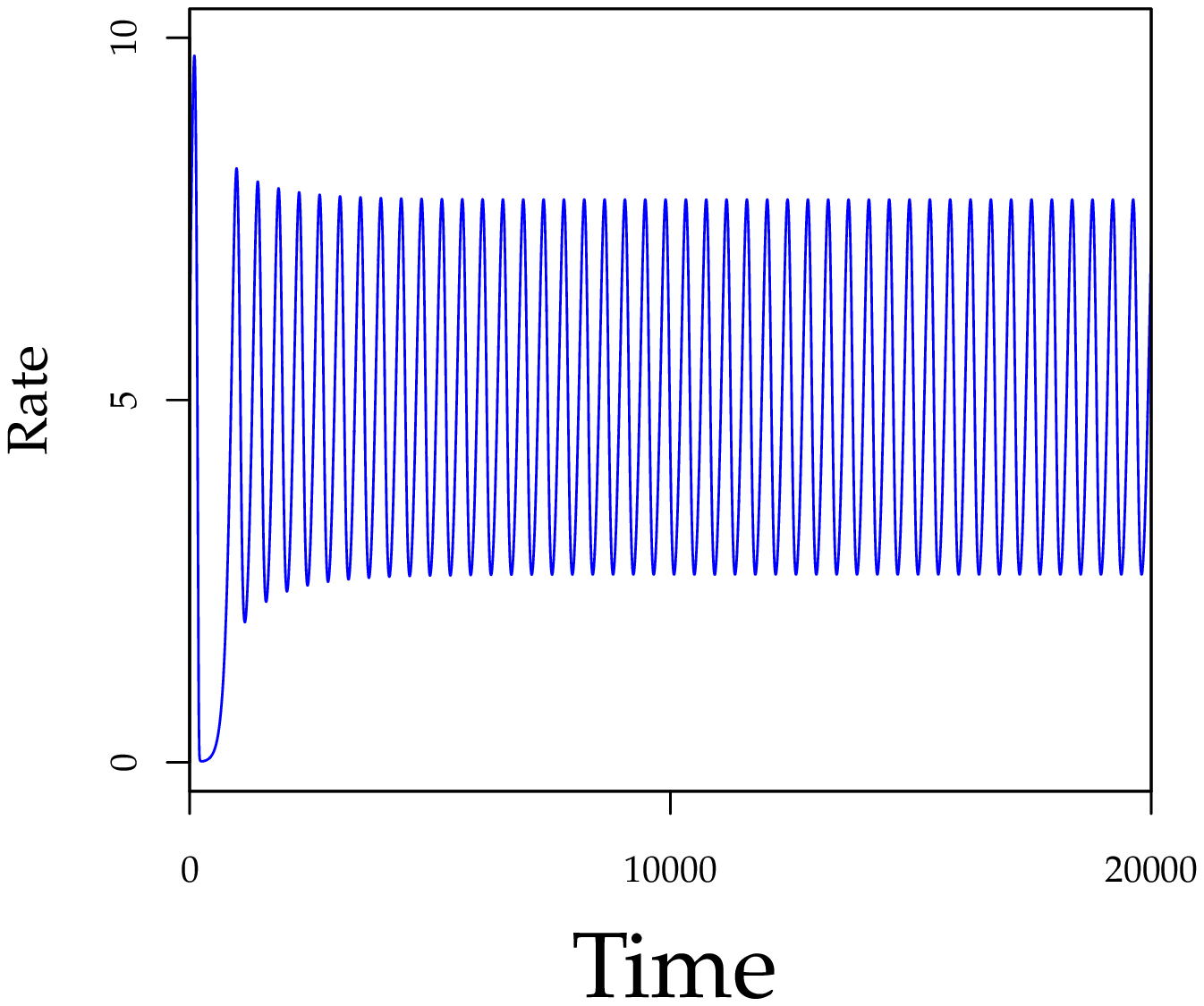}}
\end{tabular}
\caption{Numerical simulations illustrating that the system exhibits a super-critical Hopf bifurcation, as $\kappa$ increases  beyond the critical value. Time series are shown for the cases $\kappa < 1$ and $\kappa > 1$. The parameter values chosen are $a=1.01$, $C = 10$, $\tau = 100$ and $b=0.736$ ($R^*=5.5$).} \label{fig:super_rcp1d_withq}
\end{figure}
\\
$\textit{Numerical Example 2 (Sub-critical)}$: Let $a=0.924$, $C=10$, $\tau=100$ and $b=0.257$ which corresponds to equilibrium utilization of 70\% ($\rho^*=0.70$), the system undergoes a Hopf bifurcation at $\kappa$ = 1. Using \eqref{eq:myu2rcp2}, we calculate $\mu_2  = -3.547 \times 10^{-2}  <0$, implying that the Hopf bifurcation is sub-critical. The bifurcation diagram shown in Figure \ref{fig:subcriticalplot} confirms that the system exhibits a sub-critical Hopf bifurcation, as the bifurcation parameter is varied beyond the critical threshold.
The numerical simulation shown in Figure \ref{fig:sub_rcp1d_withq}(a) illustrates that the system is stable for $\kappa<1$. For $\kappa = 1.05$, as shown in Figure \ref{fig:sub_rcp1d_withq}(b), the system exhibits a limit cycle but now with amplitude larger than that of previous example.
%%%%%%%%%%%%%%%%%%%%%%%%%%%%%%%%%%%%%%%%%%%%%%%%%%%%%
\\
$\textit{Numerical Example 3 (Sub-critical)}$: Consider $a=0.827$, $C=10$, $\tau=100$ and $b=0.022$ ($\rho^*=0.90$), we get $\kappa_c$ = 1. For these values, using \eqref{eq:myu2rcp2}, we obtain $\mu_2  = -5.254 \times 10^{-2}  <0$, implying that the system undergoes a sub-critical Hopf. We can also observe from Figure \ref{fig:subcriticalplot1} that there exists no stable limit cycle in the neighborhood, as the bifurcation parameter is varied beyond the critical threshold. Hence the Hopf bifurcation is sub-critical.
To illustrate the occurrence of a sub-critical Hopf, we present some numerical simulations in Figure \ref{fig:sub1_rcp1d_withq}. Considering the initial condition $R_0=8.9$ and $\kappa=0.95$, the system converges to the stable equilibrium, $R^*=9$ (see Figure \ref{fig:sub1_rcp1d_withq}(a)). Whereas, after the bifurcation i.e. for $\kappa > \kappa_c$, the previously stable fixed point now becomes unstable and also the solution would eventually jump to infinity (Figure \ref{fig:sub1_rcp1d_withq}(b)).
\newcommand{\subwidth}{0.5\textwidth}
\begin{figure}[hbtp!]
\psfrag{time}{\hspace{-0.1cm}  \scriptsize Time}
\psfrag{Rate}{\hspace{-0.7cm}\vspace{2cm} Rate, $R(t)$}
\psfrag{0}{\hspace{-0.04cm}\scriptsize{$0$}}
\psfrag{5}{\hspace{-0.05cm}\scriptsize{$5$}}
\psfrag{10}{\hspace{-0.1cm}\scriptsize{$10$}}
\psfrag{15}{\hspace{-0.1cm}\scriptsize{$15$}}
\psfrag{20}{\hspace{-0.1cm}\scriptsize{$20$}}
\psfrag{25}{\hspace{0cm}\scriptsize{$25$}}
\psfrag{50}{\hspace{-0.1cm}\scriptsize{$50$}}
\psfrag{100}{\hspace{-0.1cm}\scriptsize{$100$}}
\psfrag{200}{\hspace{-0.1cm}\scriptsize{$200$}}
\psfrag{0.5}{\hspace{-0.1cm}\scriptsize{$0.95$}}
\psfrag{1.0}{\hspace{-0.1cm}\scriptsize{$1.0$}}
\psfrag{1.5}{\hspace{-0.1cm}\scriptsize{$1.05$}}
\psfrag{150}{\hspace{-0.1cm}\scriptsize{$150$}}
\psfrag{300}{\hspace{-0.1cm}\scriptsize{$300$}}
\psfrag{400}{\hspace{-0.1cm}\scriptsize{$400$}}
\psfrag{30}{\hspace{-0.1cm}\scriptsize{$30$}}
\psfrag{-400}{\hspace{-0.1cm}\scriptsize{$-400$}}
\psfrag{-300}{\hspace{-0.1cm}\scriptsize{$-300$}}
\psfrag{-600}{\hspace{-0.1cm}\scriptsize{$-600$}}
\psfrag{125}{\hspace{-0.1cm}\scriptsize{$125$}}
\psfrag{250}{\hspace{-0.1cm}\scriptsize{$250$}}
\psfrag{60}{\hspace{-0.1cm}\scriptsize{$60$}}
\psfrag{120}{\hspace{-0.1cm}\scriptsize{$120$}}
\psfrag{25000}{\hspace{-0.1cm}\scriptsize{$25000$}}
\psfrag{50000}{\hspace{-0.1cm}\scriptsize{$50000$}}
\psfrag{10000}{\hspace{-0.1cm}\scriptsize{$10000$}}
\psfrag{20000}{\hspace{-0.1cm}\scriptsize{$20000$}}
\psfrag{5000}{\hspace{-0.1cm}\scriptsize{$5000$}}
\begin{tabular}{c}
\subfloat[ $\kappa = 0.95$]{\includegraphics[width=\subwidth]{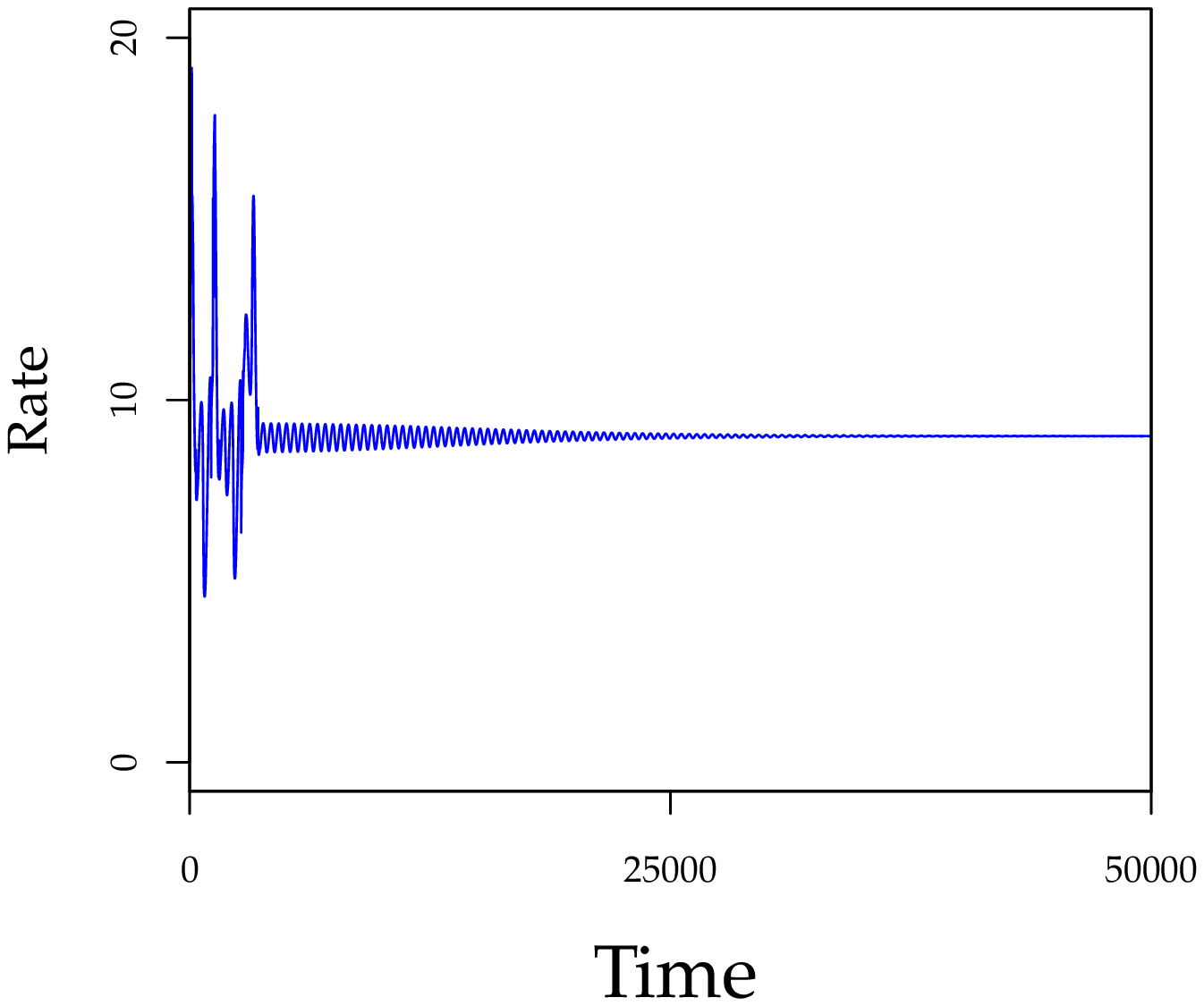}}\hspace{-5mm}
% \subfloat[ $\kappa = 0.97, R_0=2$]{\includegraphics[width=\subwdth]{tompecssub2.eps}}\hspace{-5mm}
\subfloat[ $\kappa = 1.05$]{\includegraphics[width=\subwidth]{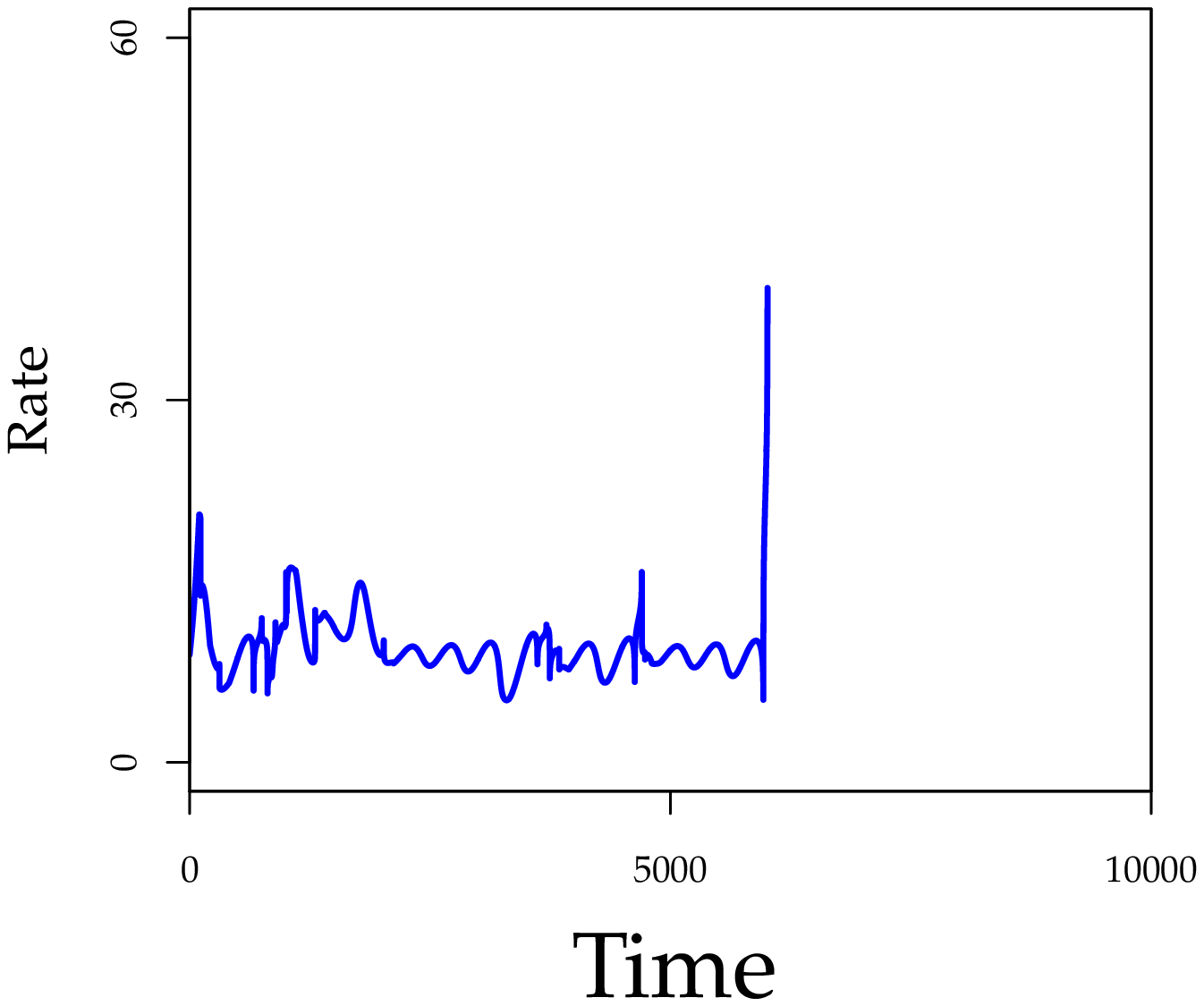}}
\end{tabular}
\caption{Numerical simulations highlighting that the system undergoes a sub-critical Hopf for the parameter values $\kappa_c= 1$, $a=0.827$, $\tau=100$, $C=10$ and $b=0.022$ ($R^*=9.0$).} \label{fig:sub1_rcp1d_withq}
\end{figure}
%%%%%%%%%%%%%%%%%%%%%%%%%%%%%%%%%%%%%%%%%%%%%%%%%%%%%%
% The numerical computations in Fig \ref{fig:supercriticalplot} and \ref{fig:subcriticalplot} highlight the existence of local super-critical Hopf for $\rho^*<\rho^*_b$, and local sub-critical Hopf for $\rho^*>\rho^*_b$.
% \begin{align}
%    \mu_2 =&\,\, \frac{1}{\pi b}\Biggl(  
%    \left(-\frac{a(1+\rho^*)}{C\tau\rho^*}\right)^2\frac{3\pi-2}{5b} \notag \\
%   & + \left( -\frac{a}{C\tau(1-\rho^*)}\right)^2\frac{2(11\pi-4)}{\frac{5a\bigl( 1 +\rho^* \bigr)}{\tau}}      \notag \\
%  & +  \frac{a(1+\rho^*)}{C\tau\rho^*}.\frac{a}{C\tau(1-\rho^*)}\frac{(7\pi-18)}{5\frac{a\bigl( 1 +\rho^* \bigr)}{\tau}} \notag \\
%  & +\frac{2a}{C^2 \tau\rho^*(1-\rho^*)} 
%  - \frac{3\pi a}{C^2 \tau(1-\rho^*)^2} \Biggr) \notag 
%  %
% \end{align}
%%%%%%%%%%%%%%%%%%%%%%%%%%%%%%%%%%%%%%%%%%%%%%%%%%%%%%%%%%%%%%%%%%%%%%%%%%%%%%%%%%%%%%%%%%%%%%%%%%%%%%%%%%%%%%%%%%%%%%%%%%%%%%%%%

In summary, the results of theoretical and numerical analysis reveal that the RCP which uses both rate mismatch and queue size feedback, can undergo a sub-critical Hopf bifurcation, which is undesirable for engineering applications. In fact, in the context of congestion control algorithms, the possibility of occurrence of a sub-critical Hopf has not been extensively studied so far.
%The Hopf bifurcation analysis reveals some key insights into the dynamical properties of RCP. 
The insights from Hopf bifurcation analysis could guide design considerations such that any loss of local stability only occurs via the emergence of small amplitude stable limit cycles. In other words, the nature of Hopf bifurcation and the stability of the bifurcating limit cycles should also be considered while designing congestion control protocols. %Therefore, this study signals the need for additional studies to understand more about this phenomena in . 

\newcommand{\myarrowsize}{0.09cm 5.0}
\begin{figure*}[hbtp!]
\vspace{5mm}
\hspace{-5mm}
\centering
\scalebox{0.55} % Change this value to rescale the drawing.
{\hspace{-15mm}
\begin{pspicture}(0,-2.84083)(18.224112,2.54083)
%\psframe[linewidth=0.028222222,dimen=outer](11.47,0.72328126)(5.41,-0.77671874)
%\includegraphics[viewport= -100 15 -80 180, scale = 2.5]{router}
\includegraphics[viewport= -129 15 -109 60, scale = 1.5]{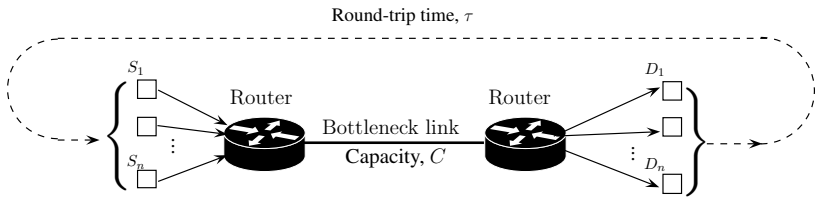}
\includegraphics[viewport= -225 15 -205 60, scale = 1.5]{router}
\psline[linewidth=0.0628222222cm](6.5,-0.047)(10.841,-0.047) % long horizontal lines
\rput(5.516957812,1.05){\begin{Large}Router\end{Large}}
\rput(11.6957812,1.05){\begin{Large}Router\end{Large}}
\rput(8.6957812,0.35){\begin{Large}Bottleneck link \end{Large}}
%\psline[linewidth=0.028222222cm,arrowsize=\myarrowsize,arrowlength=1.4,arrowinset=0.4]{->}(10.05,3.1967187)(10.653681,-.34671876)%arrow2 from src
\usefont{T1}{ptm}{m}{n}
\rput(8.737812,-0.41567187){\begin{Large} Capacity, $C$ \end{Large}}

\psframe[linewidth=0.028222222,dimen=outer](3.015889,0.57917017)(2.544111,0.10739236)
\psline[linewidth=0.028222222cm,arrowsize=\myarrowsize,arrowlength=1.4,arrowinset=0.4]{->}(3.034111,0.36671874)(4.7134111,0.1534671876)
\psframe[linewidth=0.028222222,dimen=outer](3.015889,1.4791701)(2.544111,1.0073924)
\psline[linewidth=0.028222222cm,arrowsize=\myarrowsize,arrowlength=1.4,arrowinset=0.4]{->}(3.05,1.2232813)(4.71,0.37537328124)
%\psframe[linewidth=0.028222222,dimen=outer](3.015889,2.0791702)(2.544111,1.6073923)
%\psline[linewidth=0.028222222cm,arrowsize=\myarrowsize,arrowlength=1.4,arrowinset=0.4]{->}(3.05,1.8232813)(5.43,0.47328126)
%\psframe[linewidth=0.028222222,dimen=outer](3.015889,-0.36082986)(2.544111,-0.8326076)
%\psline[linewidth=0.028222222cm,arrowsize=\myarrowsize,arrowlength=1.4,arrowinset=0.4]{->}(3.05,-0.5967187)(5.43,-0.24671875)
\psframe[linewidth=0.028222222,dimen=outer](3.015889,-0.698082983)(2.544111,-1.14526076)
\psline[linewidth=0.028222222cm,arrowsize=\myarrowsize,arrowlength=1.4,arrowinset=0.4]{->}(3.05,-0.980)(4.71,-0.34671876)
%\psframe[linewidth=0.028222222,dimen=outer](3.015889,-1.8608298)(2.544111,-2.3326077)
%\psline[linewidth=0.028222222cm,arrowsize=\myarrowsize,arrowlength=1.4,arrowinset=0.4]{->}(3.05,-2.1167188)(5.394111,-0.47328126)
%\pscircle[linewidth=0.028222222,dimen=outer](12.08,-0.02671875){0.6}
\psline[linewidth=0.028222222cm,linestyle=dashed,dash=0.16cm 0.16cm](0.65,2.47)(17.61,2.47) % long horizontal lines
%\psline[linewidth=0.028222222cm,linestyle=dashed,dash=0.16cm 0.16cm](0.65,-2.6067189)(17.61,-2.6067189)  % long horizontal lines
% \usefont{T1}{ptm}{m}{n}
% \rput(12.211406,-0.02671875){\begin{Large}$C$ \end{Large}}
% \usefont{T1}{ptm}{m}{n}
% \rput(2.32314062,0.67328125){\begin{Large}$S_1$ \end{Large} }
% \usefont{T1}{ptm}{m}{n}
% \rput(2.32314062,-0.7267187){\begin{Large}$S_2$  \end{Large}}

\usefont{T1}{ptm}{m}{n}
\rput(2.51957812,1.75){\begin{large} $S_1$ \end{large}}
\usefont{T1}{ptm}{m}{n}
\rput(3.4089065,-0.02671875){\Large $\vdots$}
\usefont{T1}{ptm}{m}{n}
\rput(2.51957812,-0.5090567187){\begin{large} $S_n$ \end{large}}

\usefont{T1}{ptm}{m}{n}
\rput(8.957812,3.0){\begin{LARGE}Round-trip time, $\tau$ \end{LARGE}}
%\usefont{T1}{ptm}{m}{n}
%\rput(8.737812,-3.1567187){\begin{LARGE} Round-trip time, $\tau_2$ \end{LARGE}}
\usefont{T1}{ptm}{m}{n}
% \rput(8.6957812,1.5){\begin{LARGE}Router\end{LARGE}}
% \usefont{T1}{ptm}{m}{n}
% \rput(8.737812,-1.751567187){\begin{LARGE} Capacity, $C$ \end{LARGE}}
\psarc[linewidth=0.028222222,linestyle=dashed,dash=0.16cm 0.16cm](0.65,1.25){1.20}{90.0}{270.0}
%\psarc[linewidth=0.028222222,linestyle=dashed,dash=0.16cm 0.16cm](0.65,-1.9767188){0.65}{90.0}{270.0}
% \psline[linewidth=0.028222222cm,linestyle=dashed,dash=0.16cm 0.16cm,arrowsize=\myarrowsize,arrowlength=1.4,arrowinset=0.4]{->}(0.65,-1.3267188)(2.505,-1.3267188) % arrows at source
\psline[linewidth=0.028222222cm,linestyle=dashed,dash=0.16cm 0.16cm,arrowsize=\myarrowsize,arrowlength=1.4,arrowinset=0.4]{->}(0.65,0.0180)(1.6505,0.0180) %arrows at source
% \psline[linewidth=0.028222222cm](8.45,0.72328126)(8.45,-0.77671874)
% \psline[linewidth=0.028222222cm](9.45,0.72328126)(9.45,-0.77671874)
% \psline[linewidth=0.028222222cm](10.45,0.72328126)(10.45,-0.77671874)
\usefont{T1}{ptm}{m}{n}
\rput(14.4089065,-0.202671875){\Large $\vdots$}
% \psline[linewidth=0.028222222cm](6.25,0.72328126)(6.25,-0.77671874)
\usefont{T1}{ptm}{m}{n}
%\rput(5.791406,-1.3467188){\begin{LARGE}$B$ \end{LARGE}}
%\rput{-180.0}(30.648222,-1.1934375){\psframe[linewidth=0.028222222,dimen=outer](15.54,-0.34082985)(15.068222,-0.81260765)}
%\psline[linewidth=0.028222222cm,arrowsize=\myarrowsize,arrowlength=1.4,arrowinset=0.4]{<-}(15.0541115,-0.5732812)(12.534111,-0.31328124)
\rput{-180.0}(30.648222,-2.4334376){\psframe[linewidth=0.028222222,dimen=outer](15.54,-01.15)(15.068222,-1.64)}
\psline[linewidth=0.028222222cm,arrowsize=\myarrowsize,arrowlength=1.4,arrowinset=0.4]{<-}(15.034111,-1.12132813)(12.5115,-0.128671875)
%\rput{-180.0}(30.648222,-4.1934376){\psframe[linewidth=0.028222222,dimen=outer](15.54,-1.8408298)(15.068222,-2.3126075)}
%\psline[linewidth=0.028222222cm,arrowsize=\myarrowsize,arrowlength=1.4,arrowinset=0.4]{<-}(15.074111,-2.1732812)(12.5541115,-0.41328126)
%\rput{-180.0}(30.608223,0.6865625){\psframe[linewidth=0.028222222,dimen=outer](15.52,0.59917015)(15.048223,0.12739237)}
%\psline[linewidth=0.028222222cm,arrowsize=\myarrowsize,arrowlength=1.4,arrowinset=0.4]{<-}(15.0541115,0.32328126)(12.714111,0.10671875)
\rput{-180.0}(30.648222,2.4465625){\psframe[linewidth=0.028222222,dimen=outer](15.54,1.4791701)(15.068222,1.0073924)}
\psline[linewidth=0.028222222cm,arrowsize=\myarrowsize,arrowlength=1.4,arrowinset=0.4]{<-}(15.0541115,1.2867187)(12.511,0.111328125)
\rput{-180.0}(30.648222,2.1986865625){\psframe[linewidth=0.028222222,dimen=outer](15.54,2.0991702)(15.068222,1.6273924)}
\psline[linewidth=0.028222222cm,arrowsize=\myarrowsize,arrowlength=1.4,arrowinset=0.4]{<-}(15.0541115,0.21986865625)(12.571,0.113984375)
\rput{-180.0}(34.92,3.7265625){\psarc[linewidth=0.028222222,linestyle=dashed,dash=0.16cm 0.16cm](17.46,2.5181390){1.283018}{90.0}{270.0}}
\psline[linewidth=0.028222222cm,linestyle=dashed,dash=0.16cm 0.16cm,arrowsize=\myarrowsize,arrowlength=1.4,arrowinset=0.4]{<-}(17.5,-0.0721)(16.171,-0.0721) % arrows at Destination
% \psline[linewidth=0.028222222cm,linestyle=dashed,dash=0.16cm 0.16cm,arrowsize=\myarrowsize,arrowlength=1.4,arrowinset=0.4]{<-}(17.58,-1.3267188)(15.718,-1.3267188)  % arrows at Destination
%\rput{-180.0}(35.12,-3.7134376){\psarc[linewidth=0.028222222,linestyle=dashed,dash=0.16cm 0.16cm](7.61,-1.7567188){0.65}{90.0}{270.0}}
%\usefont{T1}{ptm}{m}{n}
%\rput(3.5962987,0.6717188){\begin{large}$\vdots$ \end{large}}
%\usefont{T1}{ptm}{m}{n}
%\rput(3.5162985,-1.4082812){\begin{large}$\vdots$ \end{large}}
% \usefont{T1}{ptm}{m}{n}
% \rput(14.536299,-1.4082812){\begin{large}$\vdots$ \end{large}}
% \usefont{T1}{ptm}{m}{n}
% \rput(14.556298,0.6717188){\begin{large}$\vdots$ \end{large}}
\usefont{T1}{ptm}{m}{n}
\rput(2.02727048,0.1217187){ \begin{Huge}$\Bigg\{$\end{Huge}}
% \usefont{T1}{ptm}{m}{n}
% \rput(2.2727048,-1.3482812){ \begin{huge}$\Bigg\{$\end{huge}}
% \usefont{T1}{ptm}{m}{n}
% \rput(15.856924,1.0717187){ \begin{huge}$\Bigg\} $\end{huge}}
 \usefont{T1}{ptm}{m}{n}
 \rput(15.876924,-0.013882813){ \begin{Huge}$\Bigg\} $\end{Huge}}
\usefont{T1}{ptm}{m}{n}
\rput(14.98,1.71971171876){\begin{large}$D_1$ \end{large}}
\usefont{T1}{ptm}{m}{n}
\rput(14.98,-0.5698882812){\begin{large}$D_n$  \end{large}}
\end{pspicture} 
}
%\vspace{8mm}
\caption{Simulation setup with a single resource of capacity $C$ and $n$ sources that each producing Poisson traffic. The round-trip times of all the flows are assumed to be same.}
\label{fig:toydiagram1}
\end{figure*}

\section{Packet-level simulations}
In this section, the theoretical insights are validated by investigating if the packet-level simulations of the underlying system exhibits the qualitative properties predicted through the analysis of the fluid model. The packet-level simulations are done using a discrete event RCP simulator (for more details, refer to \cite{krv2009}).

 %%%%%%%%%%%%%%%%%%%%%%%%%%%%%%%%%%%%%%%%%%%%%%%%%%%%%%%%%%%%%%%%%%%%%%%%%%%%%%%%%%%%%%%%%%%%%
\begin{figure}[hbtp!]
 \psfrag{60}{\hspace{-0.2cm}\small{$50000$}}
\psfrag{0}{\small{$0$}} 
 \psfrag{10}{\hspace{0.1cm}\small{$0$}}
 \psfrag{35}{\hspace{-0.2cm}\small{$25000$}}
\psfrag{20}{\hspace{-0.2cm}\small{$10000$}}
\psfrag{15}{\hspace{-0.2cm}\small{$5000$}}
\psfrag{250}{\small{$250$}}
\psfrag{500}{\small{$500$}}
\psfrag{1000}{\small{$1000$}}
\psfrag{1500}{\small{$1500$}}
\psfrag{2000}{\small{$2000$}}
\psfrag{4000}{\small{$4000$}}
\psfrag{10000}{\small{$10000$}}
\psfrag{20000}{\small{$20000$}}
\psfrag{7500}{\small{$7500$}}
\psfrag{15000}{\small{$15000$}}
\newcommand{\hite}{3.16cm}
\centering
\begin{tabular}{cc} 
	\hspace{0.3cm}\small{Queue Size (packets)} & \hspace{0.3cm}\small{Rate (bytes/ms)} \\
% 	\includegraphics[height=\hite, width=0.45\textwidth]{queue_size1.eps} &
% 	%\epsfig{figure=100_100_0.2_0.005_1_0/queue_size.eps, scale =0.25} &
% 	\includegraphics[height=\hite, width=0.45\textwidth]{rate1.eps} \\
% 	%\epsfig{figure=100_100_0.2_0.005_1_0/rate.eps, angle=270, scale =\scl} \\
%  \multicolumn{2}{c}{\small{$a=0.2$}}\\
\includegraphics[height=\hite, width=0.45\textwidth]{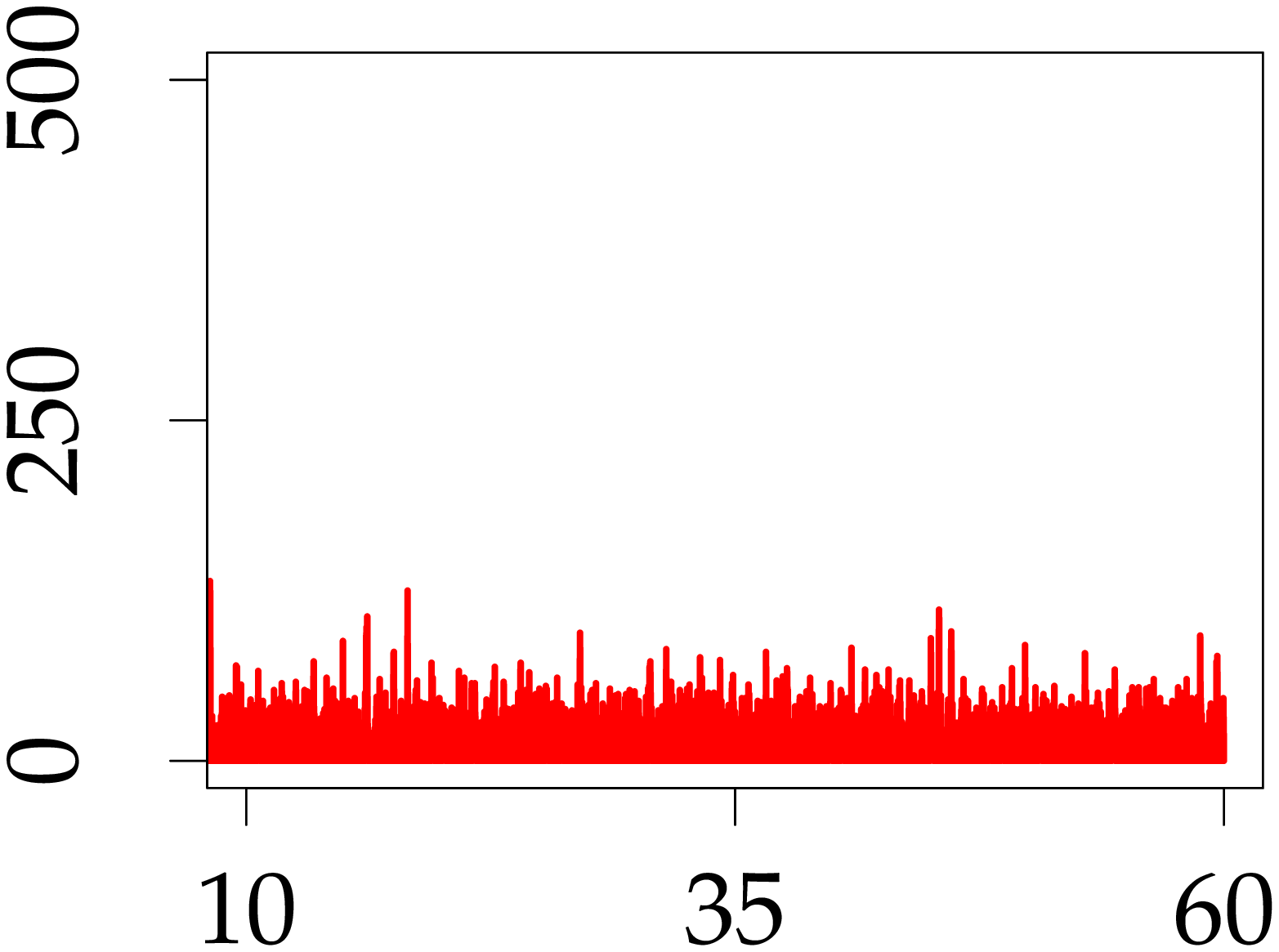} &
\includegraphics[height=\hite, width=0.45\textwidth]{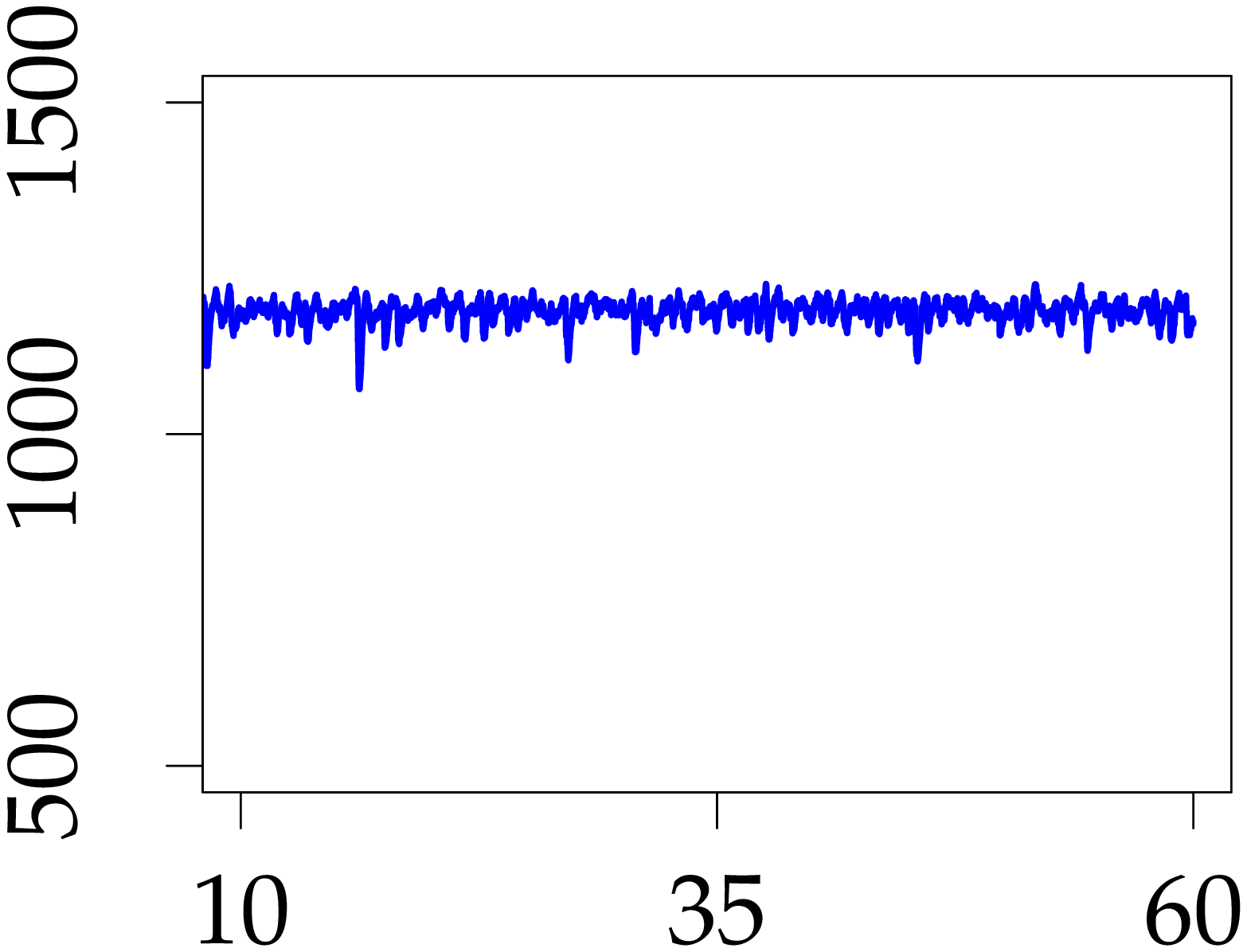} \\
 \multicolumn{2}{c}{\small{$a=0.4$}}\\
% 	\includegraphics[height=\hite,width=0.45\textwidth]{queue_size3.eps} &
% 	\includegraphics[height=\hite,width=0.45\textwidth]{rate3.eps} \\
% \multicolumn{2}{c}{\small{$a=0.6$}}\\
	\includegraphics[height=\hite,width=0.45\textwidth]{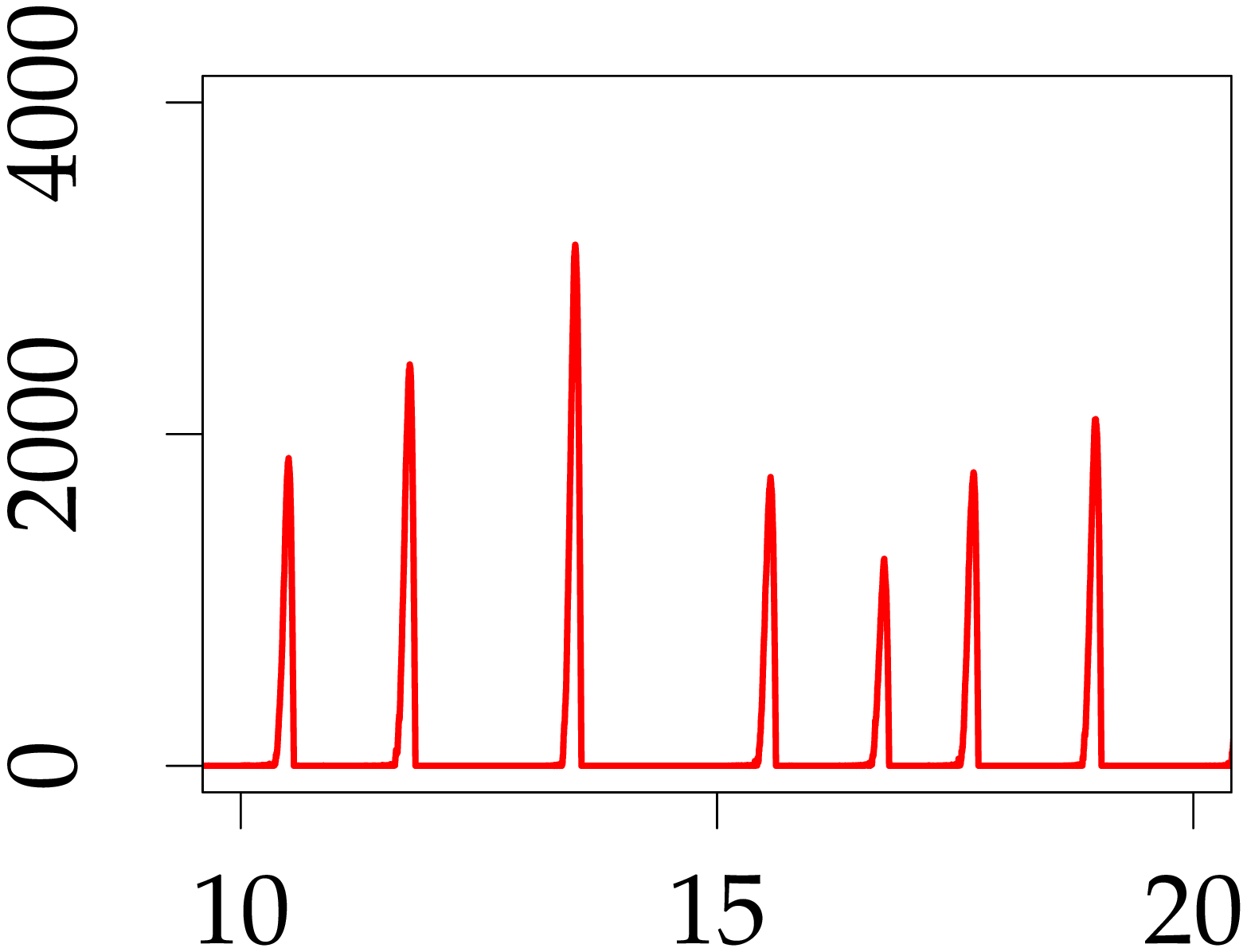} &
	\includegraphics[height=\hite,width=0.45\textwidth]{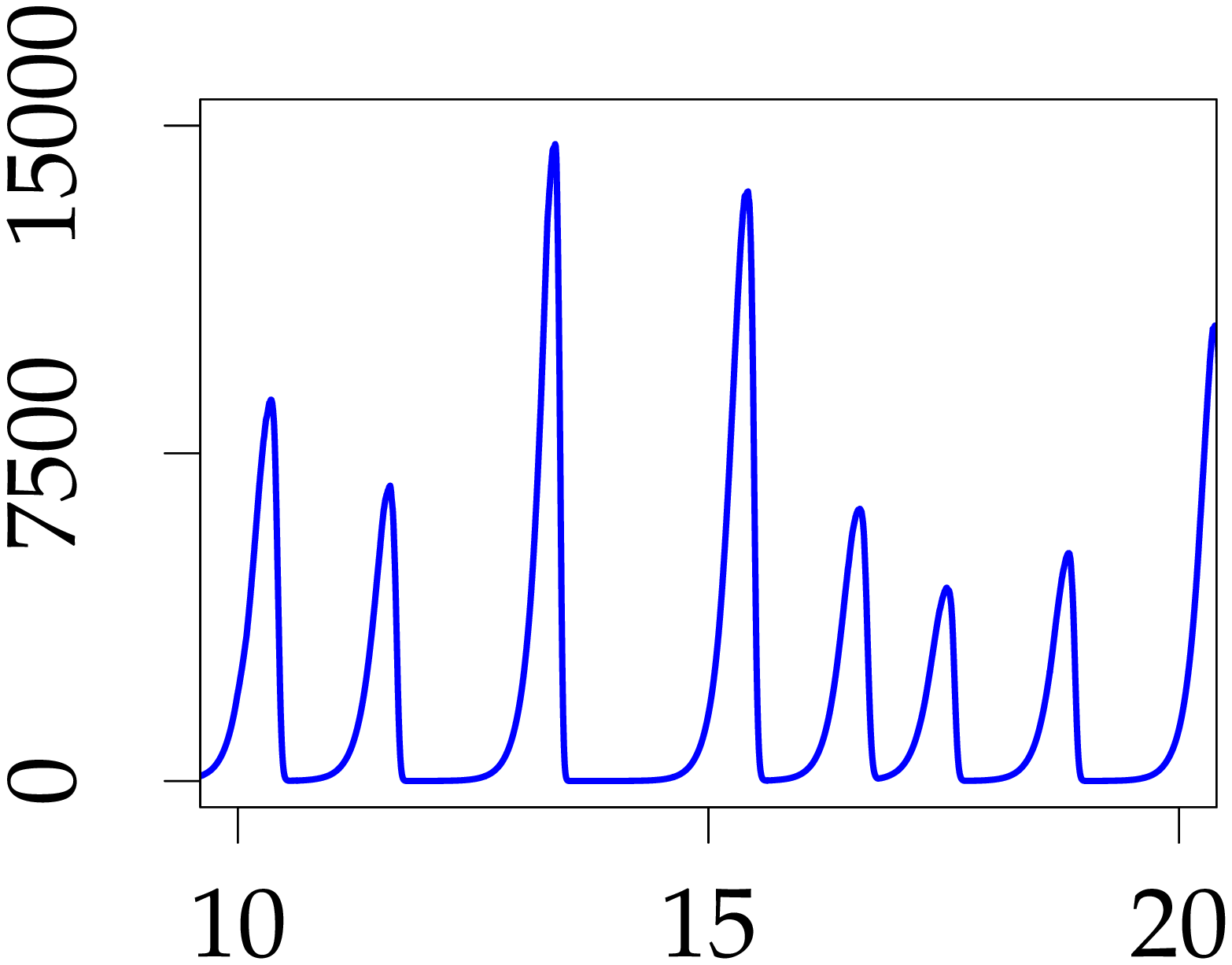} \\
\multicolumn{2}{c}{\small{$a=0.8$}}\\
	\multicolumn{2}{c}{\small{Time (ms)}}\\
\end{tabular}
\caption{\small Traces from a packet-level simulation of RCP which uses both rate mismatch and queue size feedback. Parameter values chosen are: $\tau=100$ ms, $b= 0.005$, $C=1$ Gbps and number of sources = $100$.}
\label{fig:rcp1dhopfplsims_withq}
\end{figure}
 
\begin{figure}[hbtp!]
 \psfrag{60}{\hspace{-0.2cm}\small{$50000$}}
\psfrag{0}{\small{$0$}} 
 \psfrag{10}{\hspace{0.1cm}\small{$0$}}
 \psfrag{35}{\hspace{-0.2cm}\small{$25000$}}
\psfrag{20}{\hspace{-0.2cm}\small{$10000$}}
\psfrag{15}{\hspace{-0.2cm}\small{$5000$}}
\psfrag{250}{\small{$250$}}
\psfrag{500}{\small{$500$}}
\psfrag{1000}{\small{$1000$}}
\psfrag{1500}{\small{$1500$}}
\psfrag{2000}{\small{$2000$}}
\psfrag{4000}{\small{$4000$}}
\psfrag{10000}{\small{$10000$}}
\psfrag{20000}{\small{$20000$}}
\psfrag{7500}{\small{$7500$}}
\psfrag{15000}{\small{$15000$}}
\newcommand{\hite}{3.16cm}
\centering
\begin{tabular}{cc} 
	\hspace{0.3cm}\small{Queue Size (packets)} & \hspace{0.3cm}\small{Rate (bytes/ms)} \\
% 	\includegraphics[height=\hite, width=0.45\textwidth]{queue_size5.eps} &
% 	%\epsfig{figure=100_100_0.2_0.005_1_0/queue_size.eps, scale =0.25} &
% 	\includegraphics[height=\hite, width=0.45\textwidth]{rate5.eps} \\
% 	%\epsfig{figure=100_100_0.2_0.005_1_0/rate.eps, angle=270, scale =\scl} \\
%  \multicolumn{2}{c}{\small{$a=0.4$}}\\
\includegraphics[height=\hite, width=0.45\textwidth]{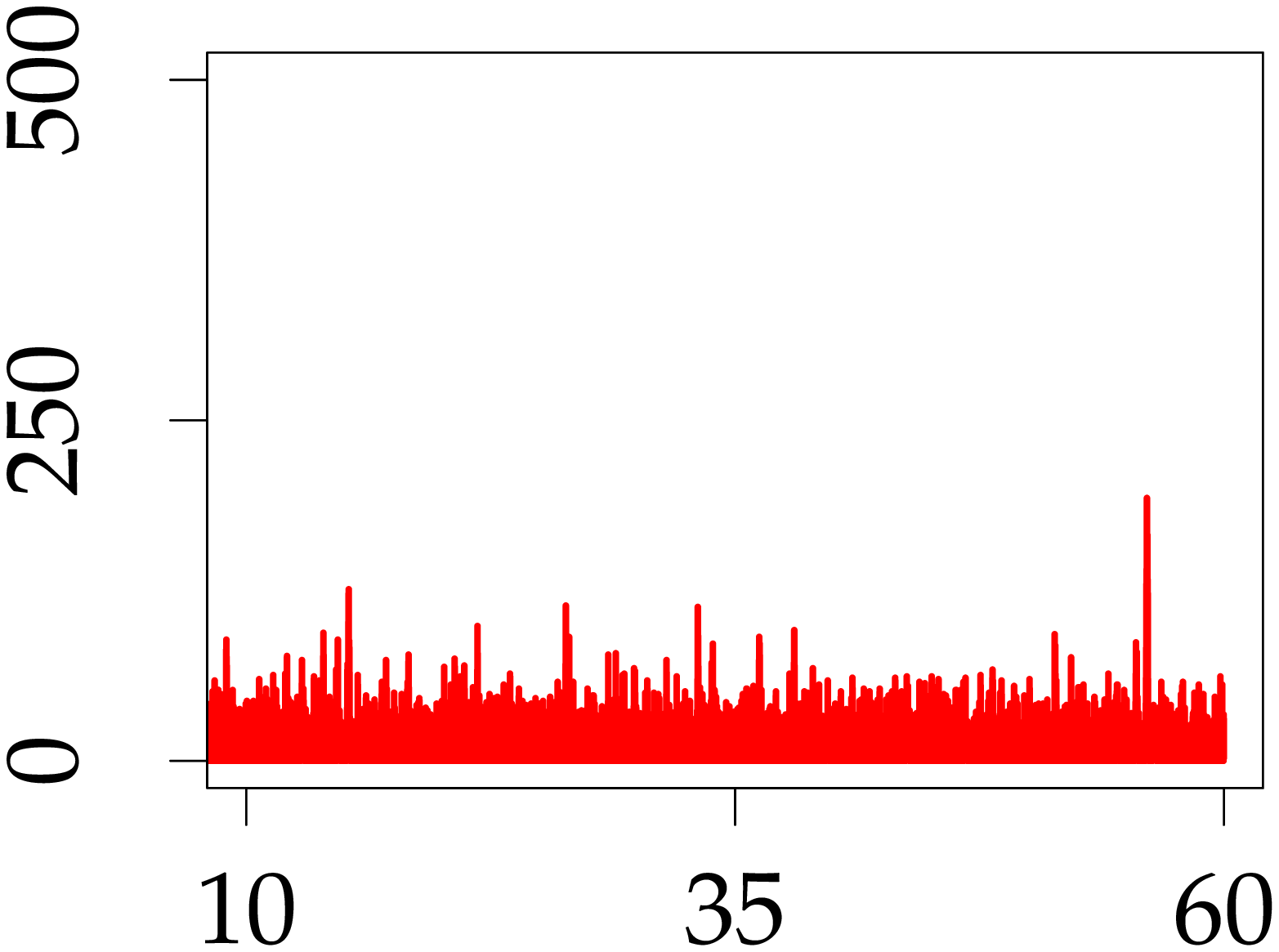} &
\includegraphics[height=\hite, width=0.45\textwidth]{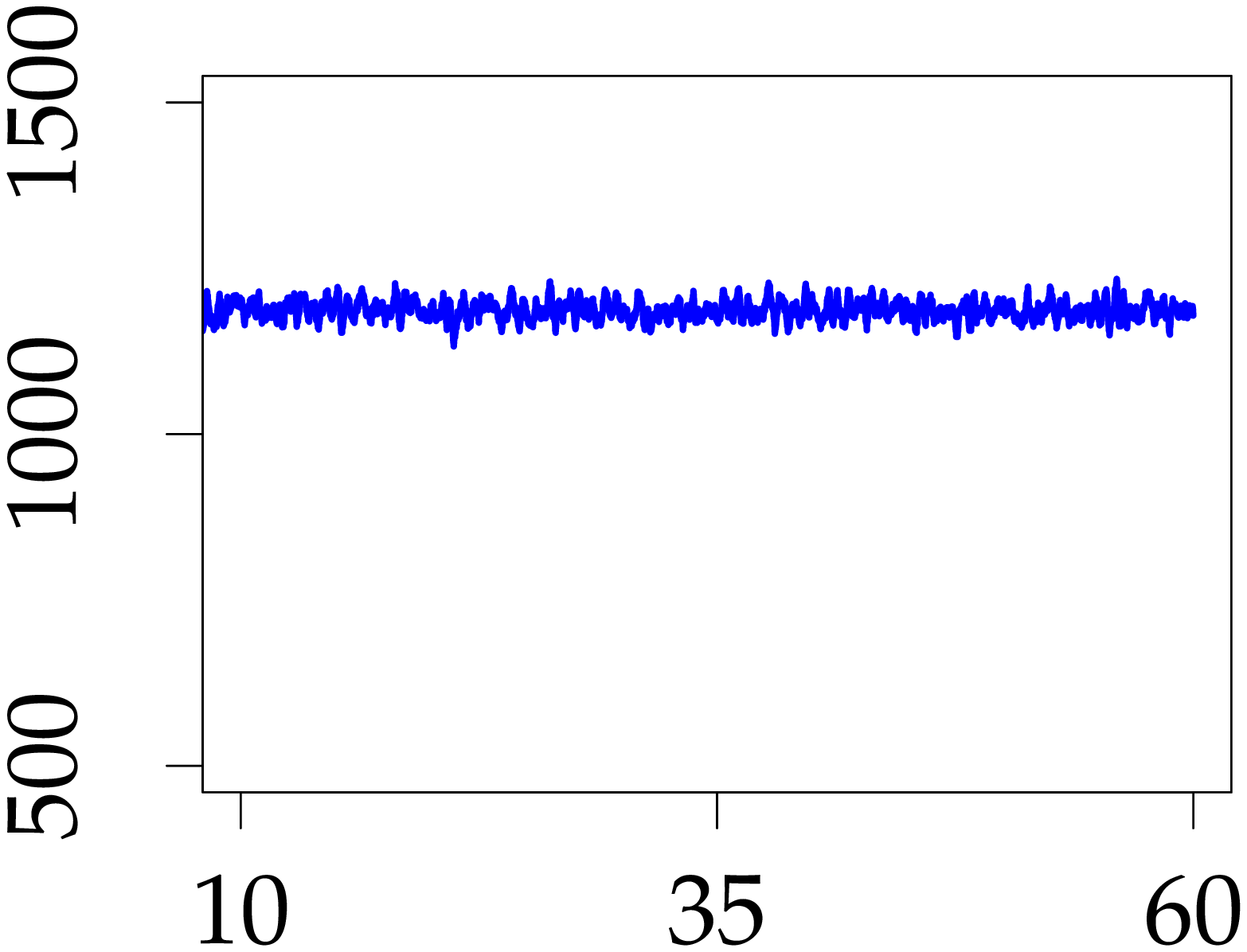} \\
 \multicolumn{2}{c}{\small{$a=0.8$}}\\
% 	\includegraphics[height=\hite,width=0.45\textwidth]{queue_size7.eps} &
% 	\includegraphics[height=\hite,width=0.45\textwidth]{rate7.eps} \\
% \multicolumn{2}{c}{\small{$a=1.2$}}\\
	\includegraphics[height=\hite,width=0.45\textwidth]{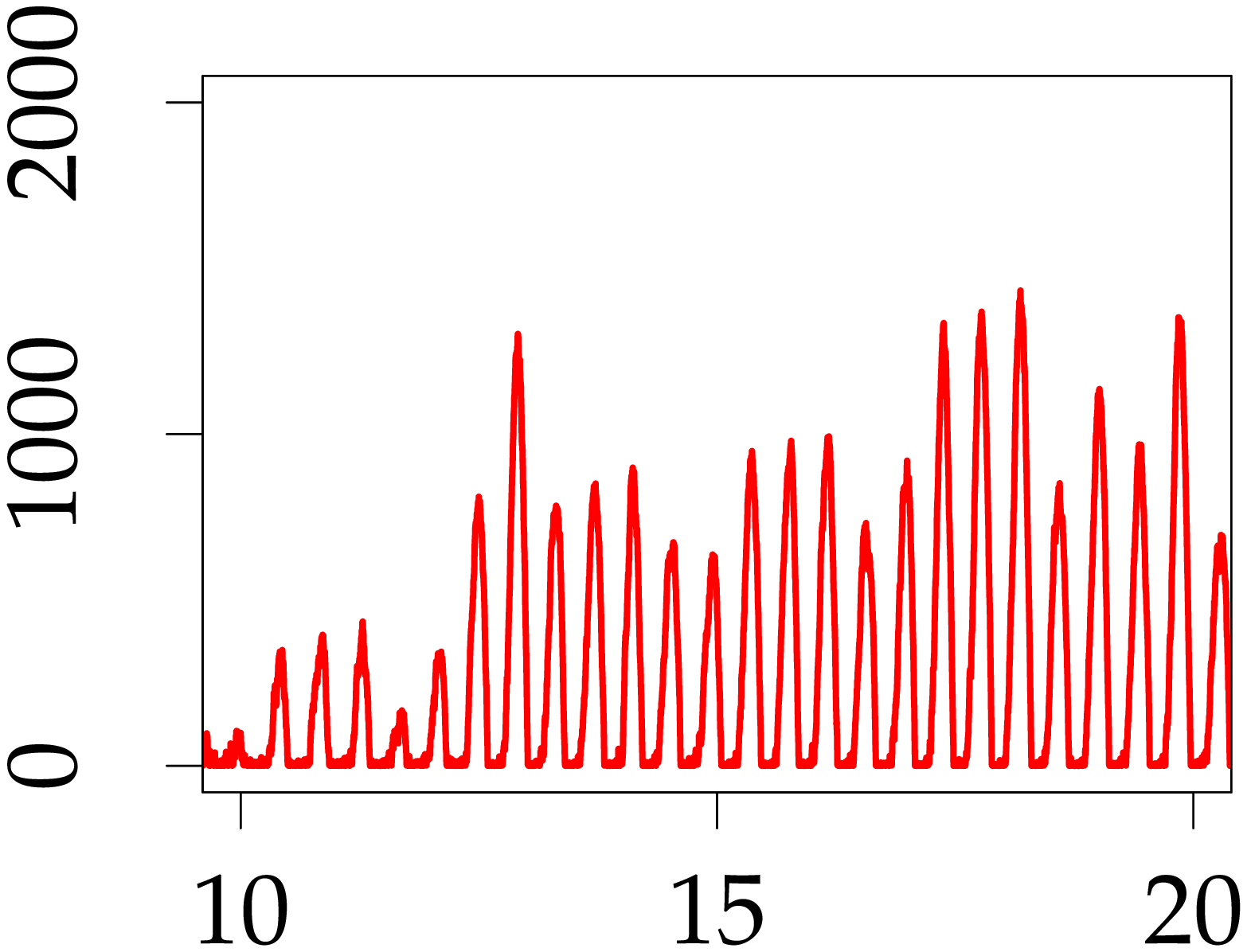} &
	\includegraphics[height=\hite,width=0.45\textwidth]{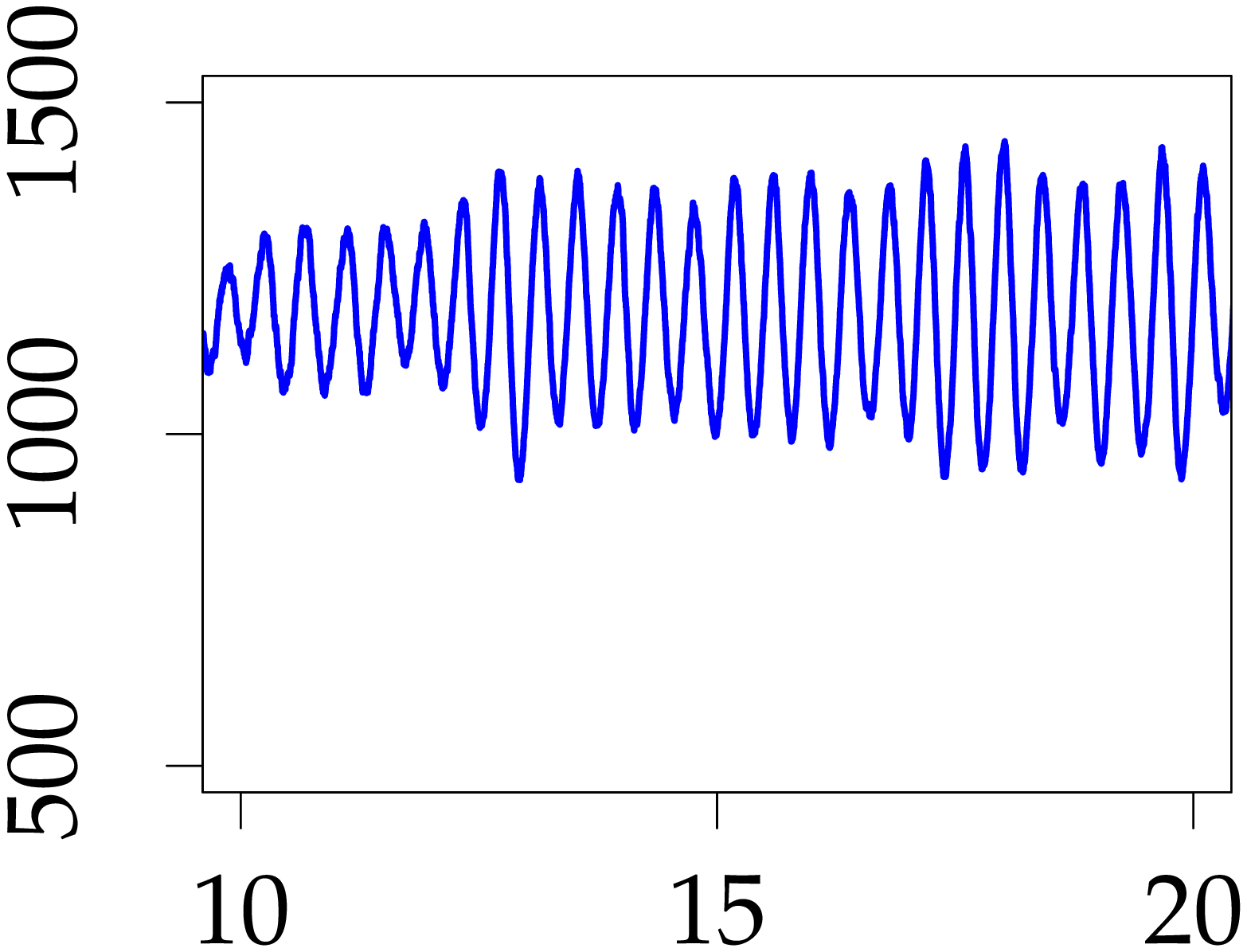} \\
\multicolumn{2}{c}{\small{$a=1.6$}}\\
	\multicolumn{2}{c}{\small{Time (ms)}}\\
\end{tabular}
\caption{\small Simulation traces of RCP which uses only rate mismatch feedback. The parameter values used are $\tau=100$ ms, $b=0$, $\gamma= 0.95$, $C=1$ Gbps and number of sources = $100$.}
\label{fig:rcp1dhopfplsims_withoutq}
\end{figure}
%%%%%%%%%%%%%%%%%%%%%%%%%%%%%%%%%%%%%%%%%%%%%%%%%%%%%%%%%%%%%%%%%%%%%%%%%%%%%%%%%%%%%%%%%%%%%%%
%%%%%%%%%%%%%%%%% PL sims for Bifurcation Analysis %%%%%%%%%%%%%%%%%%%%%%%%%%%%%%%%%%%%%%%

\begin{figure}[hbtp!]
 \psfrag{60}{\hspace{-0.2cm}\small{$50000$}}
\psfrag{0}{\small{$0$}} 
 \psfrag{10}{\hspace{0.1cm}\small{$0$}}
 \psfrag{35}{\hspace{-0.2cm}\small{$25000$}}
\psfrag{20}{\hspace{-0.2cm}\small{$10000$}}
\psfrag{15}{\hspace{-0.2cm}\small{$5000$}}
\psfrag{250}{\small{$250$}}
\psfrag{500}{\small{$500$}}
\psfrag{300}{\small{$300$}}
\psfrag{600}{\small{$600$}}
\psfrag{1000}{\small{$1000$}}
\psfrag{1500}{\small{$1500$}}
\psfrag{2000}{\small{$2000$}}
\psfrag{4000}{\small{$4000$}}
\psfrag{10000}{\small{$10000$}}
\psfrag{20000}{\small{$20000$}}
\psfrag{7500}{\small{$7500$}}
\psfrag{15000}{\small{$15000$}}
\newcommand{\hite}{3.016cm}
\centering
\begin{tabular}{cc} 
	\hspace{0.3cm}\small{Queue Size (packets)} & \hspace{0.3cm}\small{Rate (bytes/ms)} \\
	\includegraphics[height=\hite, width=0.45\textwidth]{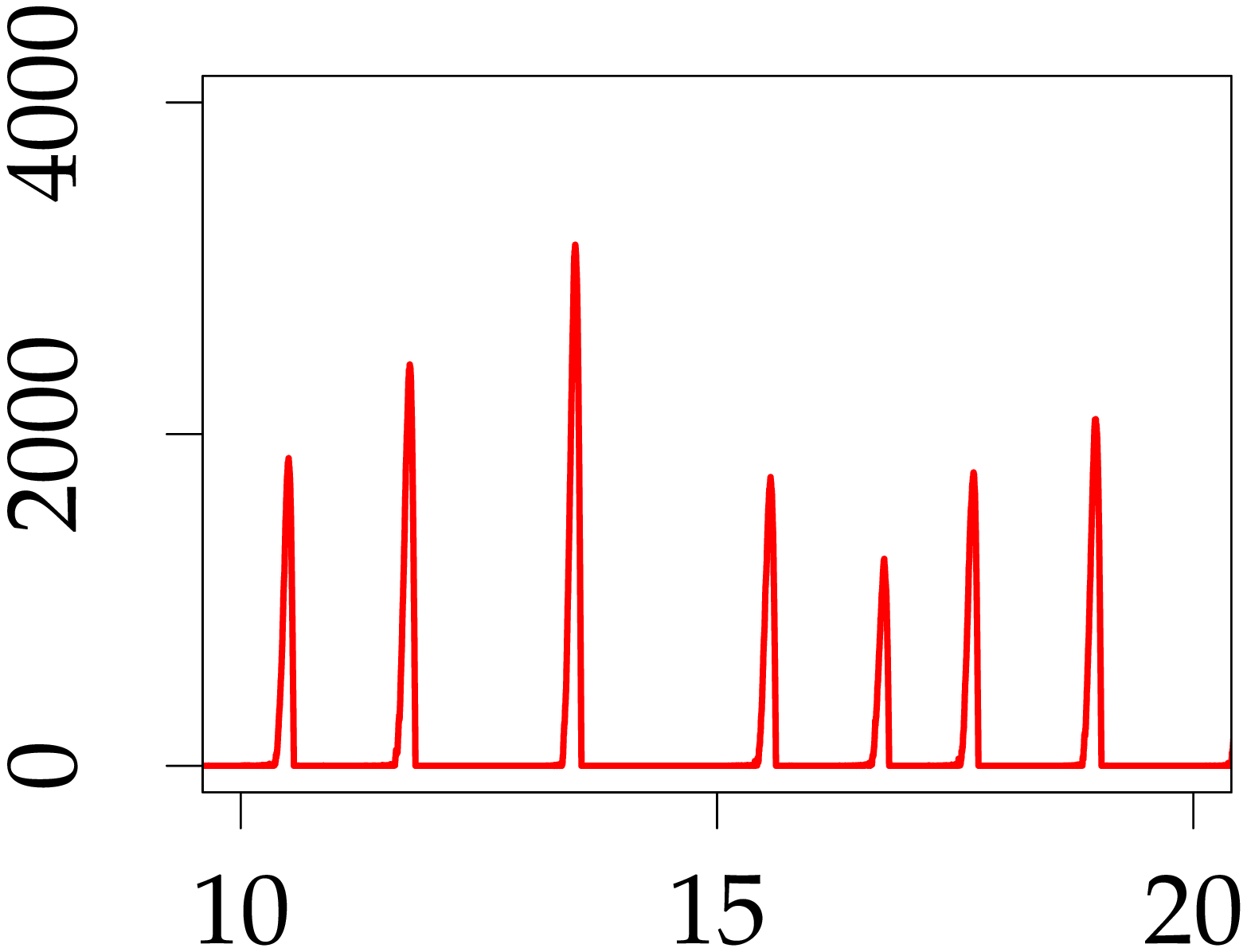} &
	\includegraphics[height=\hite, width=0.45\textwidth]{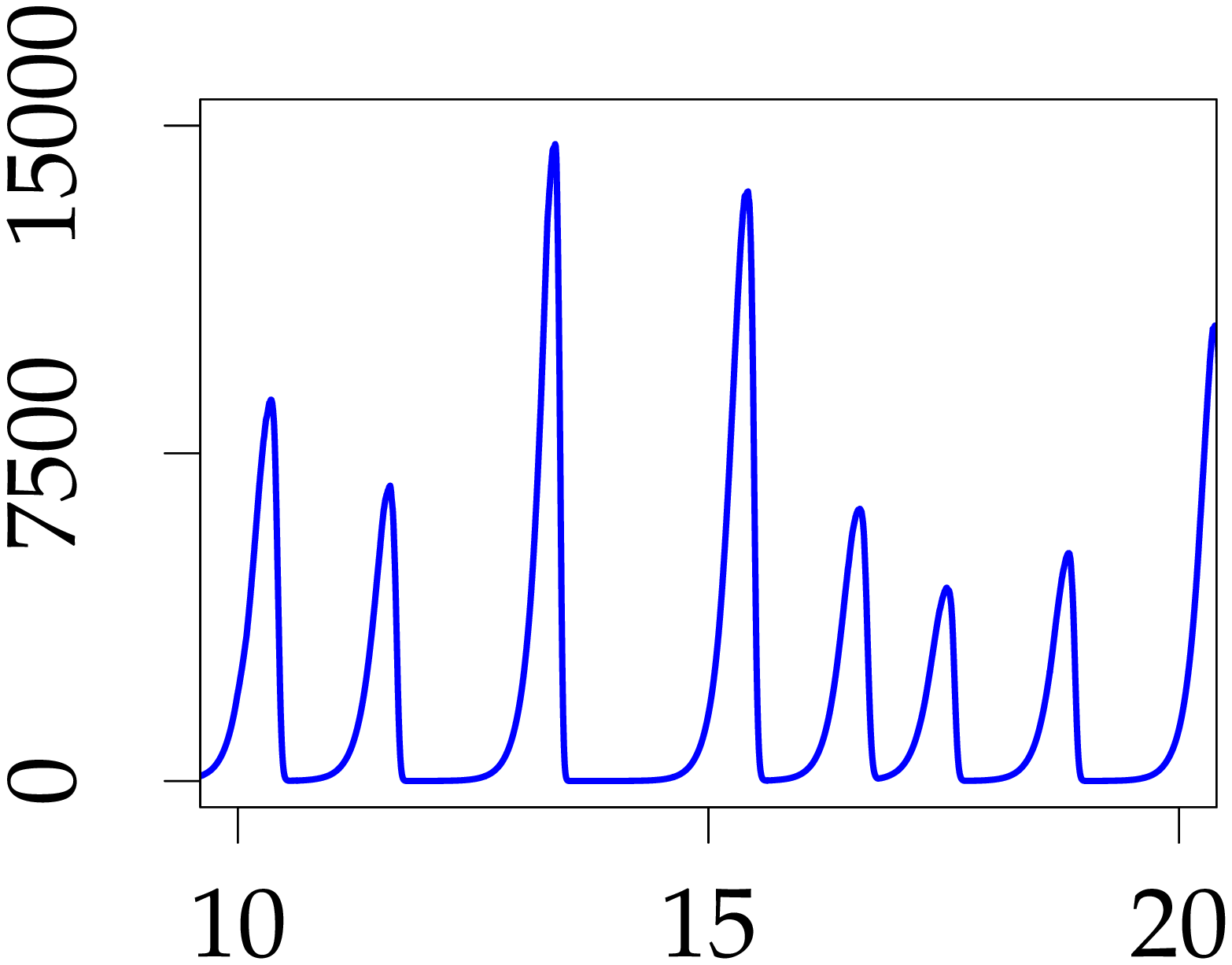} \\
 \multicolumn{2}{c}{\small{With queue feedback: $C=1$ Gbps}}\\
\includegraphics[height=\hite, width=0.45\textwidth]{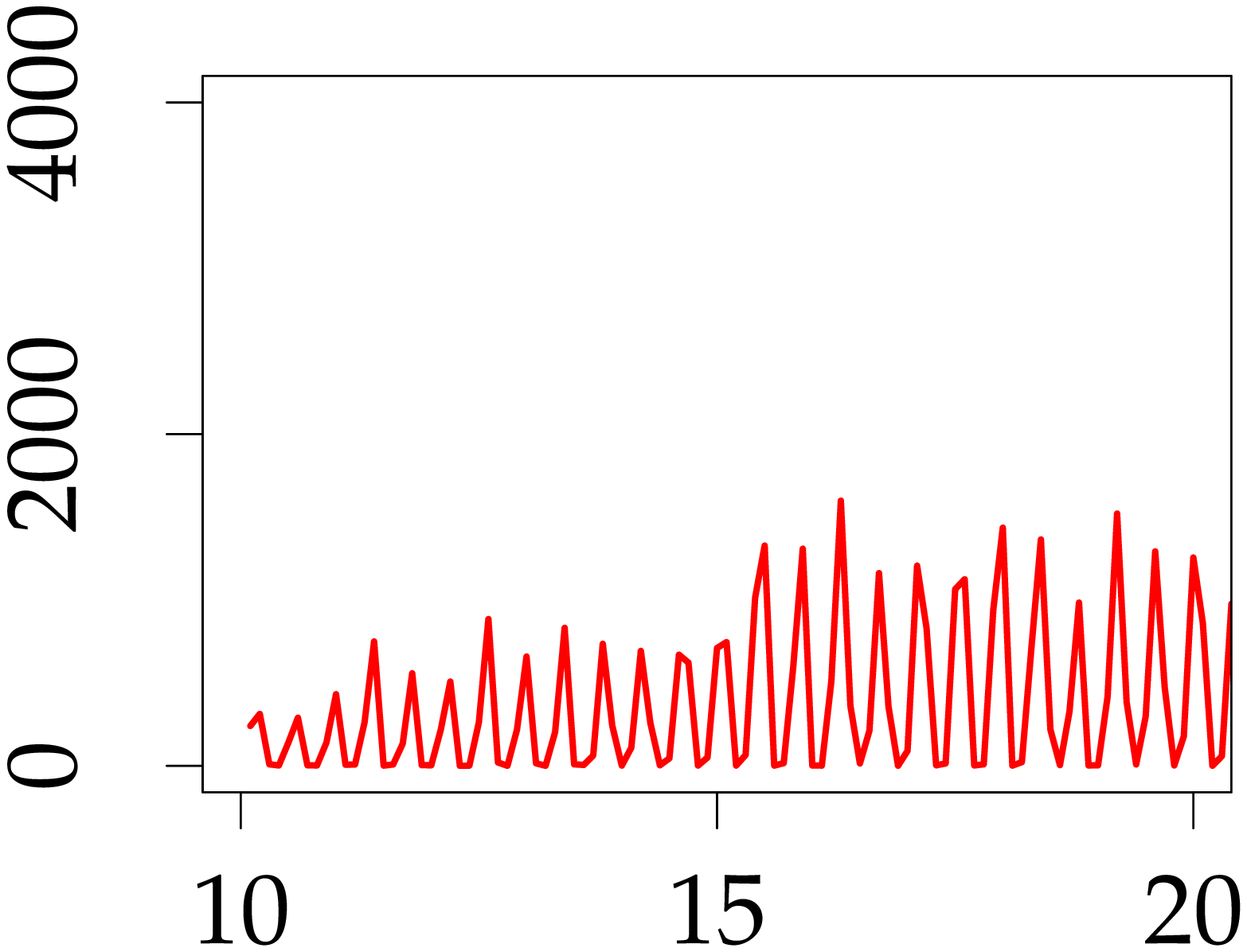} &
\includegraphics[height=\hite, width=0.45\textwidth]{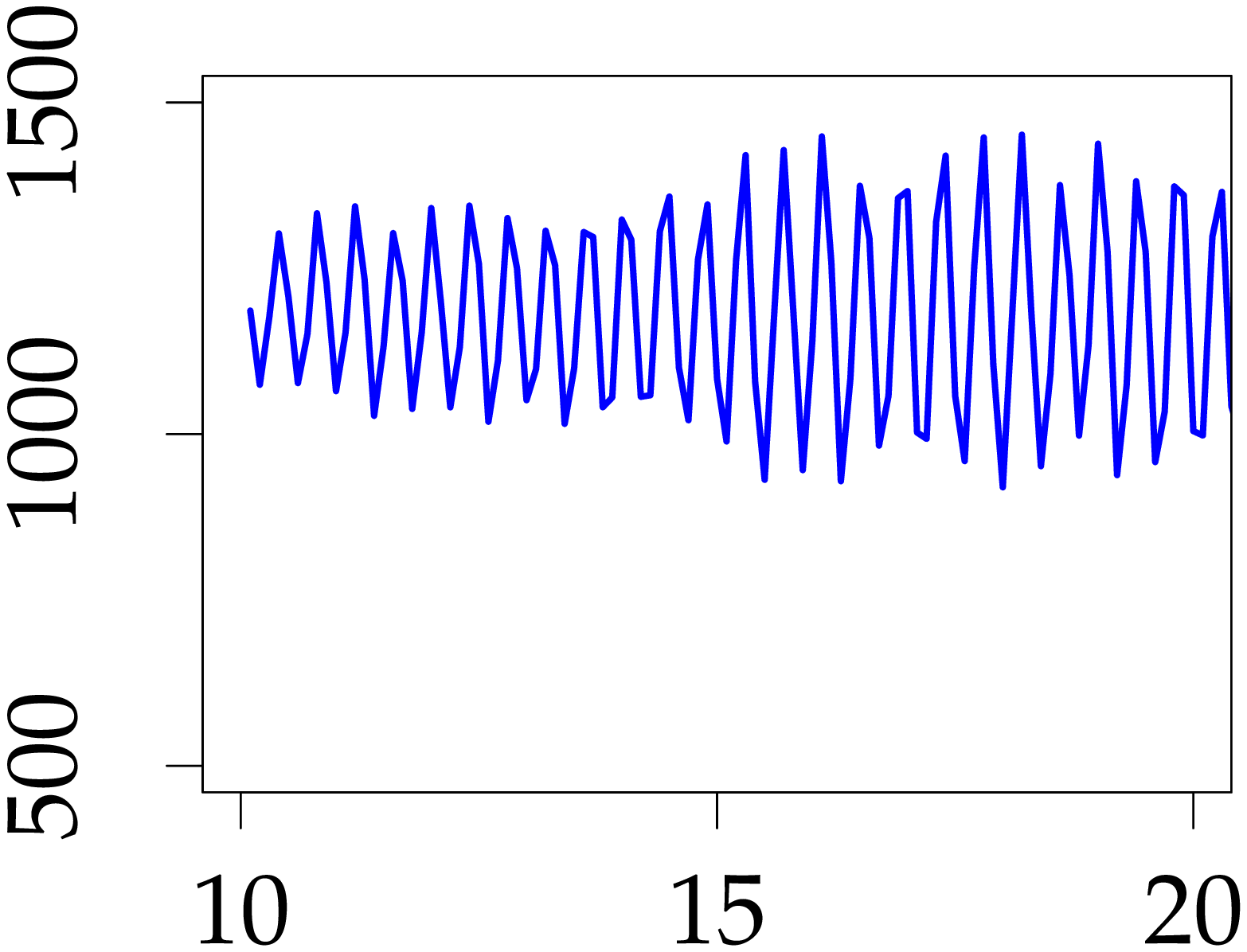} \\
 \multicolumn{2}{c}{\small{Without queue feedback: $C=1$ Gbps}}\\
	\includegraphics[height=\hite,width=0.45\textwidth]{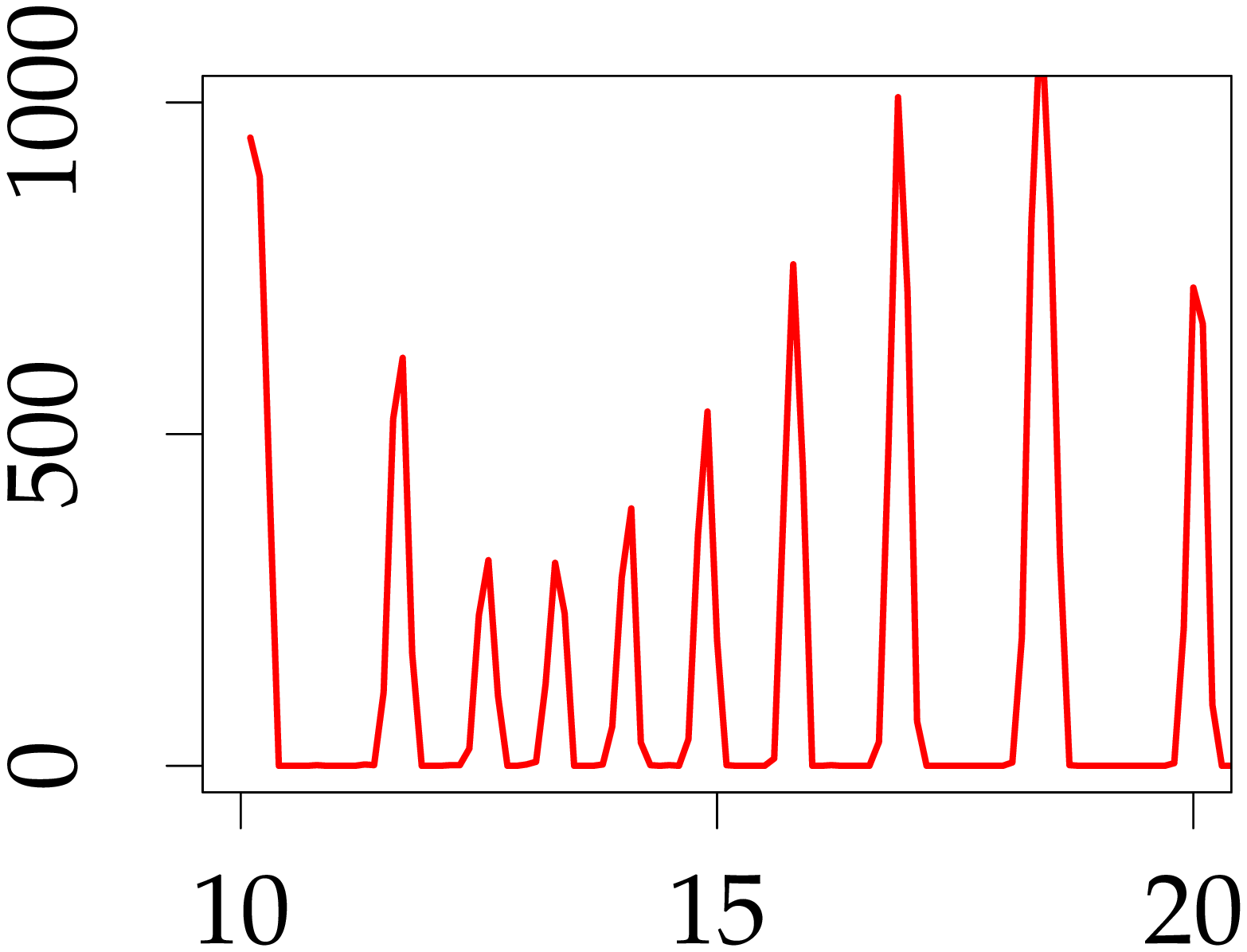} &
	\includegraphics[height=\hite,width=0.45\textwidth]{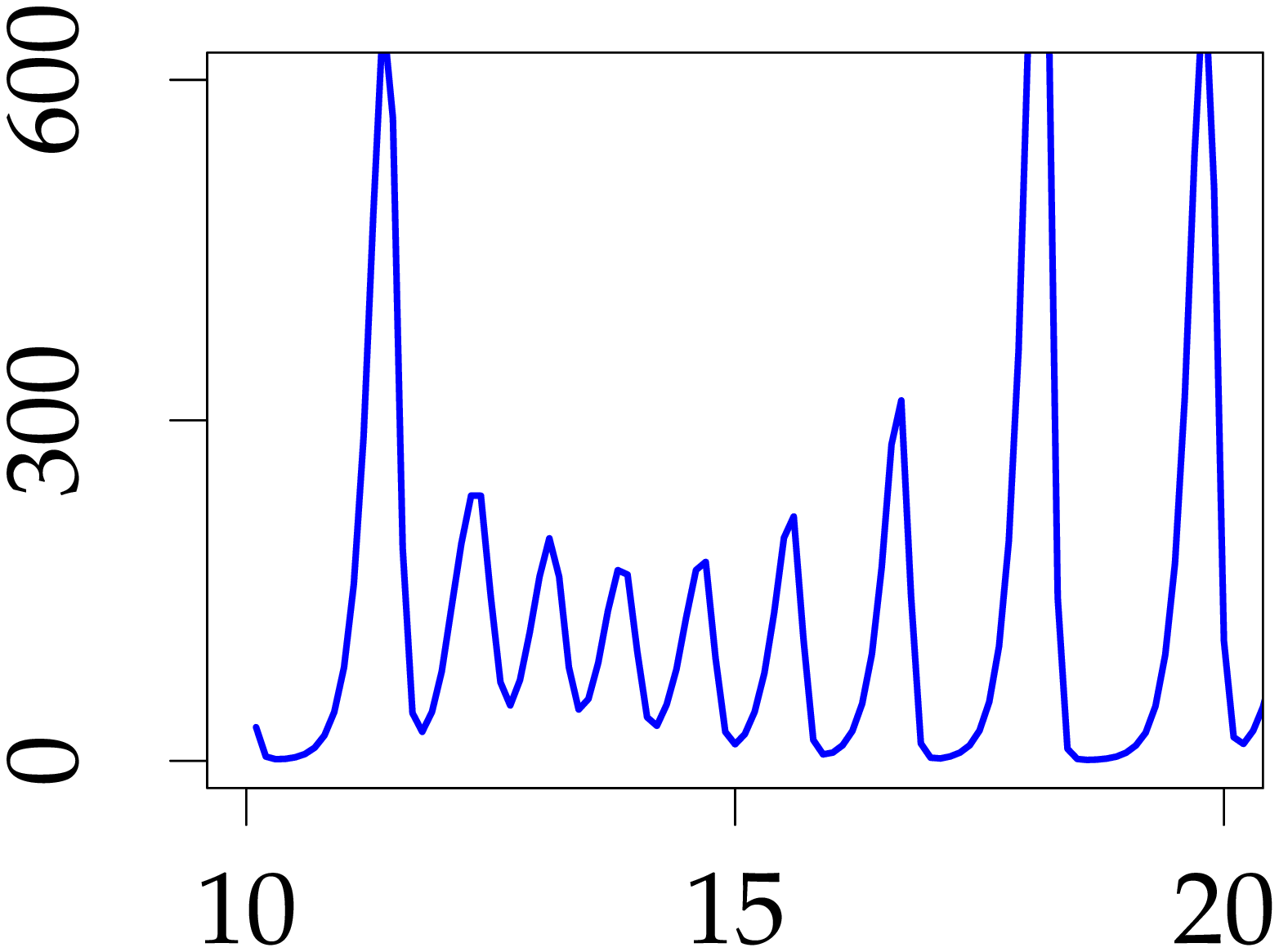} \\
 \multicolumn{2}{c}{\small{With queue feedback: $C=100$ Mbps}}\\
	\includegraphics[height=\hite,width=0.45\textwidth]{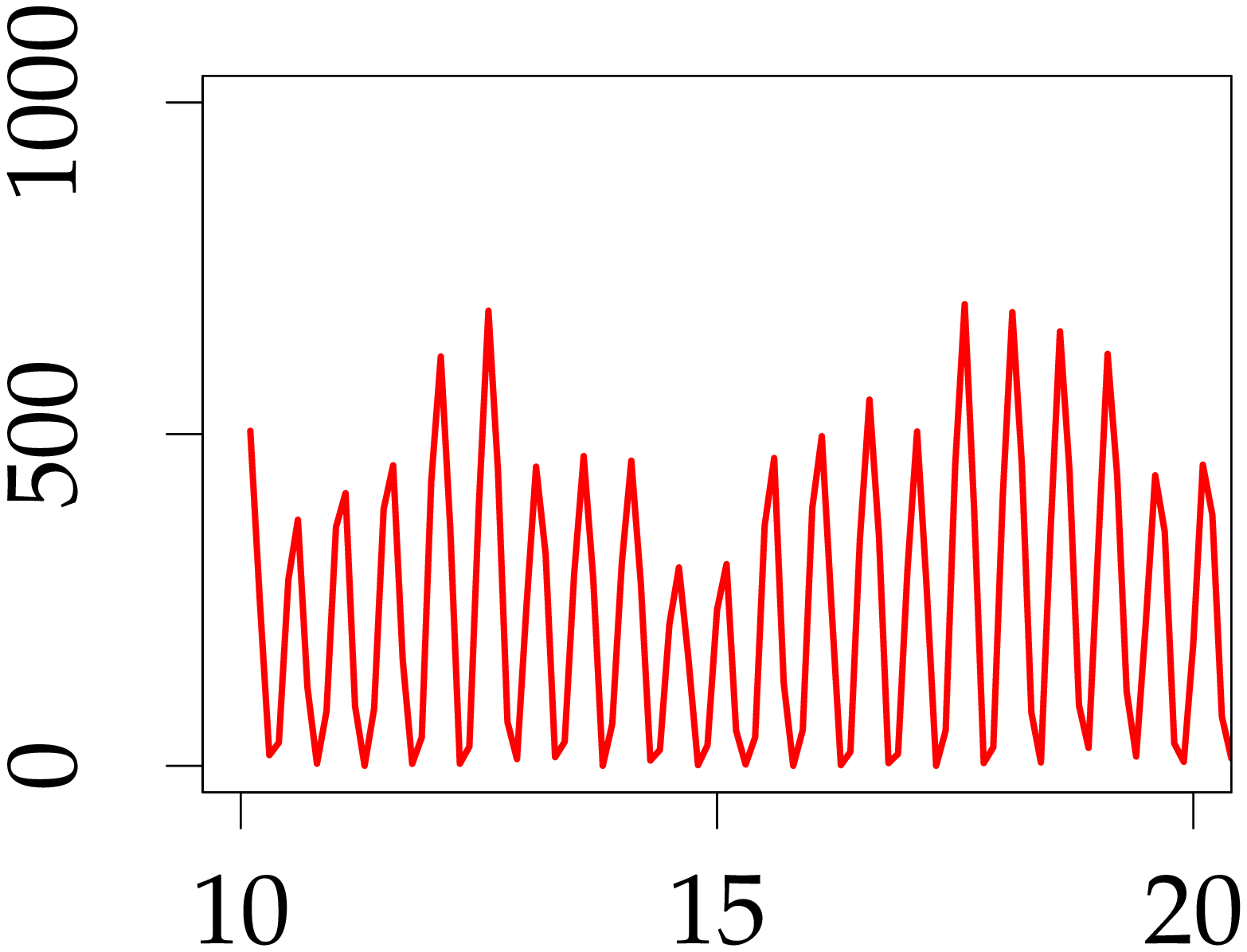} &
	\includegraphics[height=\hite,width=0.45\textwidth]{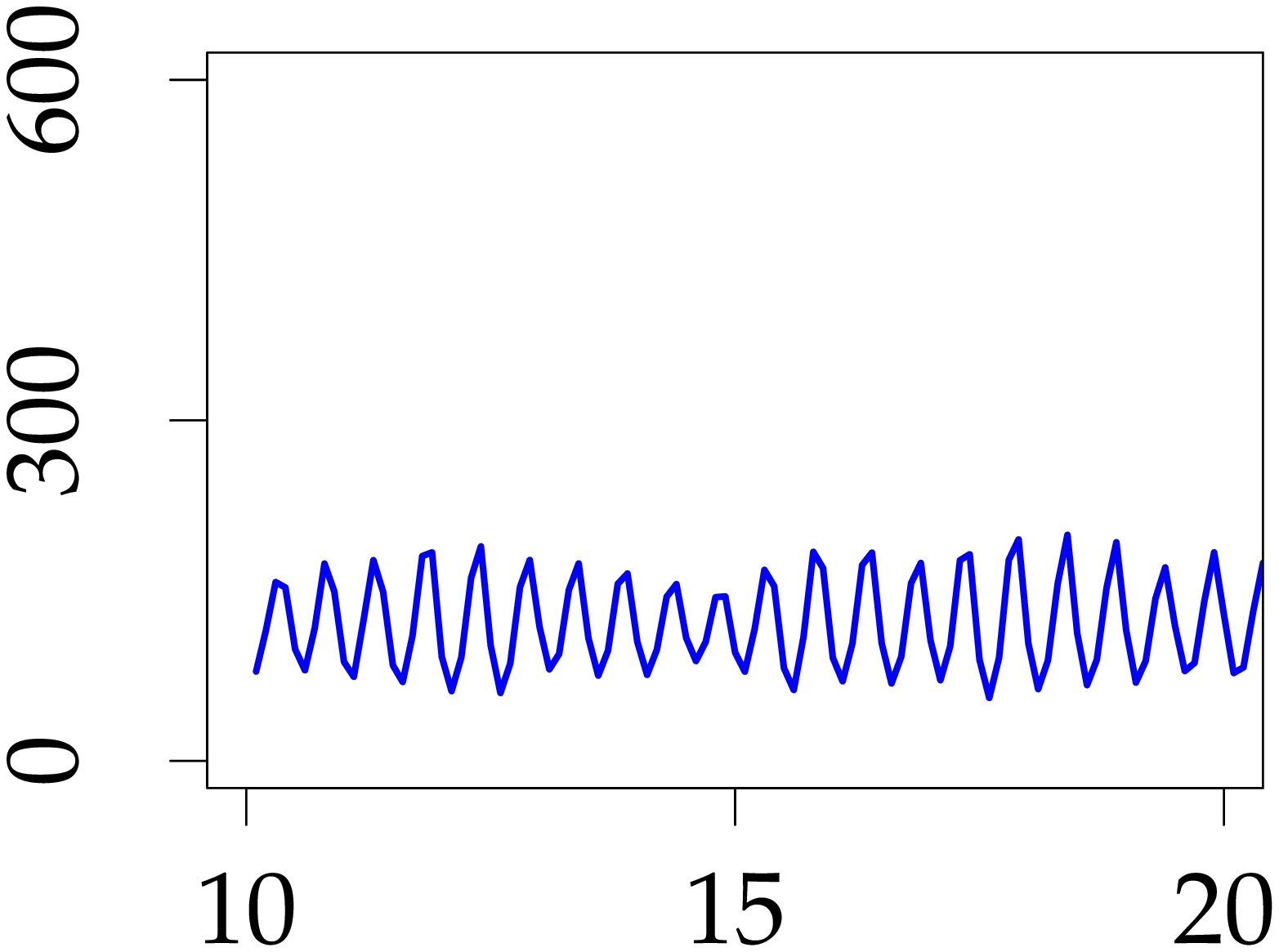} \\
 \multicolumn{2}{c}{\small{Without queue feedback: $C=100$ Mbps}}\\ 
	\multicolumn{2}{c}{\small{Time (ms)}}\\
\end{tabular}
\caption{\small Simulation Traces highlighting that the system which includes queue feedback exhibits limit cycles with amplitude much larger than that of RCP which uses only rate mismatch feedback. The parameter values used are i) with queue feedback: $a=0.8$, $b= 0.005$ and ii) without queue feedback: $a=1.6$, $\gamma = 0.95$. We consider the number of sources as $100$, and each with round-trip time of $100$ ms.}
\label{fig:rcp1dhbaplsims_set1}
\end{figure}
 
\begin{figure}[hbtp!]
 \psfrag{60}{\hspace{-0.2cm}\small{$50000$}}
\psfrag{0}{\small{$0$}} 
 \psfrag{10}{\hspace{0.1cm}\small{$0$}}
\psfrag{19}{\hspace{-0.2cm}\small{$0$}}
\psfrag{24}{\hspace{-0.2cm}\small{$5000$}}
\psfrag{29}{\hspace{-0.2cm}\small{$10000$}}
\psfrag{20}{\hspace{-0.2cm}\small{$10000$}}
\psfrag{15}{\hspace{-0.2cm}\small{$5000$}}
\psfrag{35}{\hspace{0.1cm}\small{$0$}}
\psfrag{90}{\hspace{-0.2cm}\small{$0$}}
\psfrag{95}{\hspace{-0.2cm}\small{$5000$}}
\psfrag{100}{\hspace{-0.2cm}\small{$10000$}}
\psfrag{45}{\hspace{-0.2cm}\small{$10000$}}
\psfrag{40}{\hspace{-0.2cm}\small{$5000$}}
\psfrag{250}{\small{$250$}}
\psfrag{500}{\small{$500$}}
\psfrag{1000}{\small{$1000$}}
\psfrag{1500}{\small{$1500$}}
\psfrag{2000}{\small{$2000$}}
\psfrag{4000}{\small{$4000$}}
\psfrag{10000}{\small{$10000$}}
\psfrag{20000}{\small{$20000$}}
\psfrag{7500}{\small{$7500$}}
\psfrag{15000}{\small{$15000$}}
\newcommand{\hite}{3.16cm}
\centering
\begin{tabular}{cc} 
	\hspace{0.3cm}\small{Queue Size (packets)} & \hspace{0.3cm}\small{Rate (bytes/ms)} \\
	\includegraphics[height=\hite, width=0.45\textwidth]{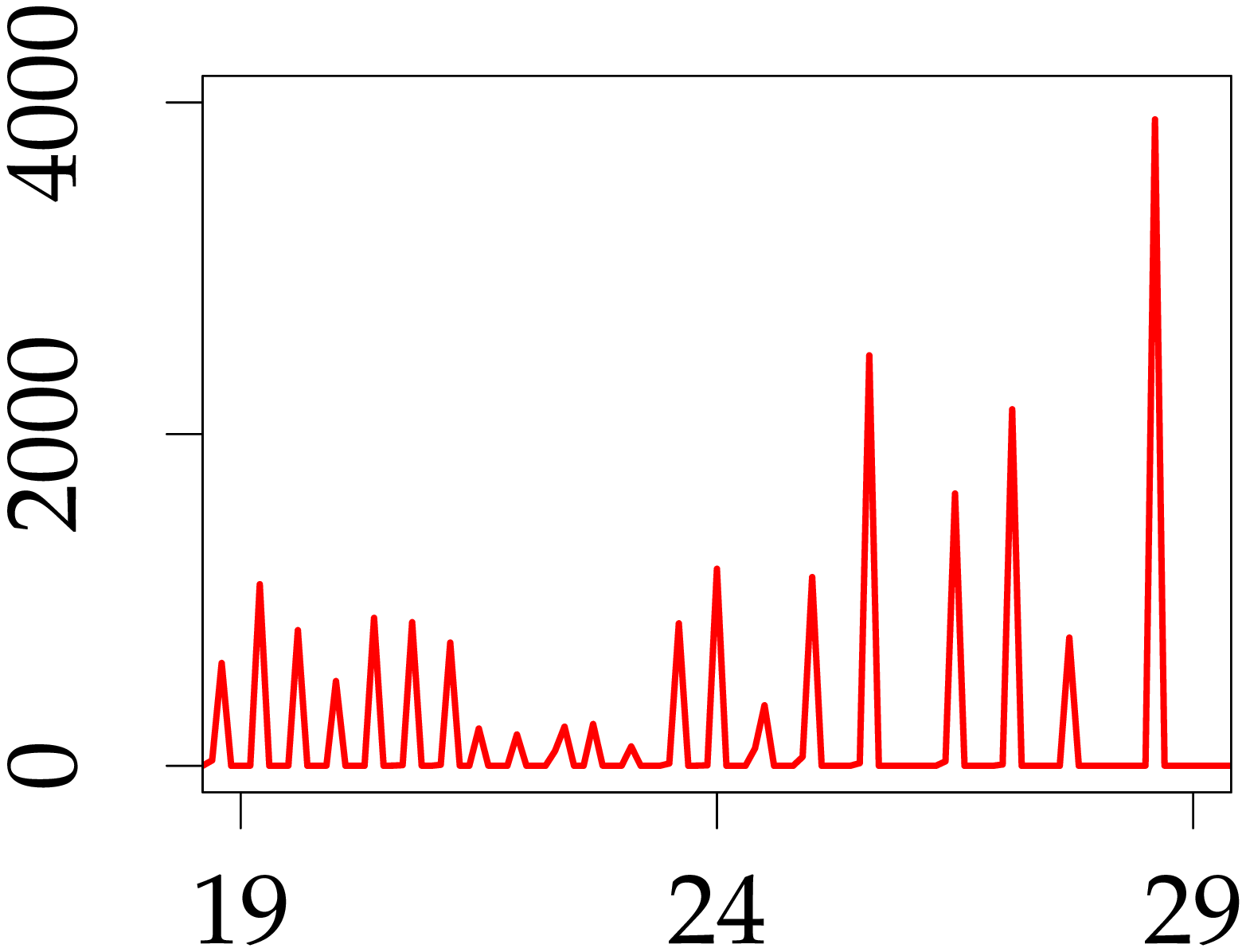} &
	\includegraphics[height=\hite, width=0.45\textwidth]{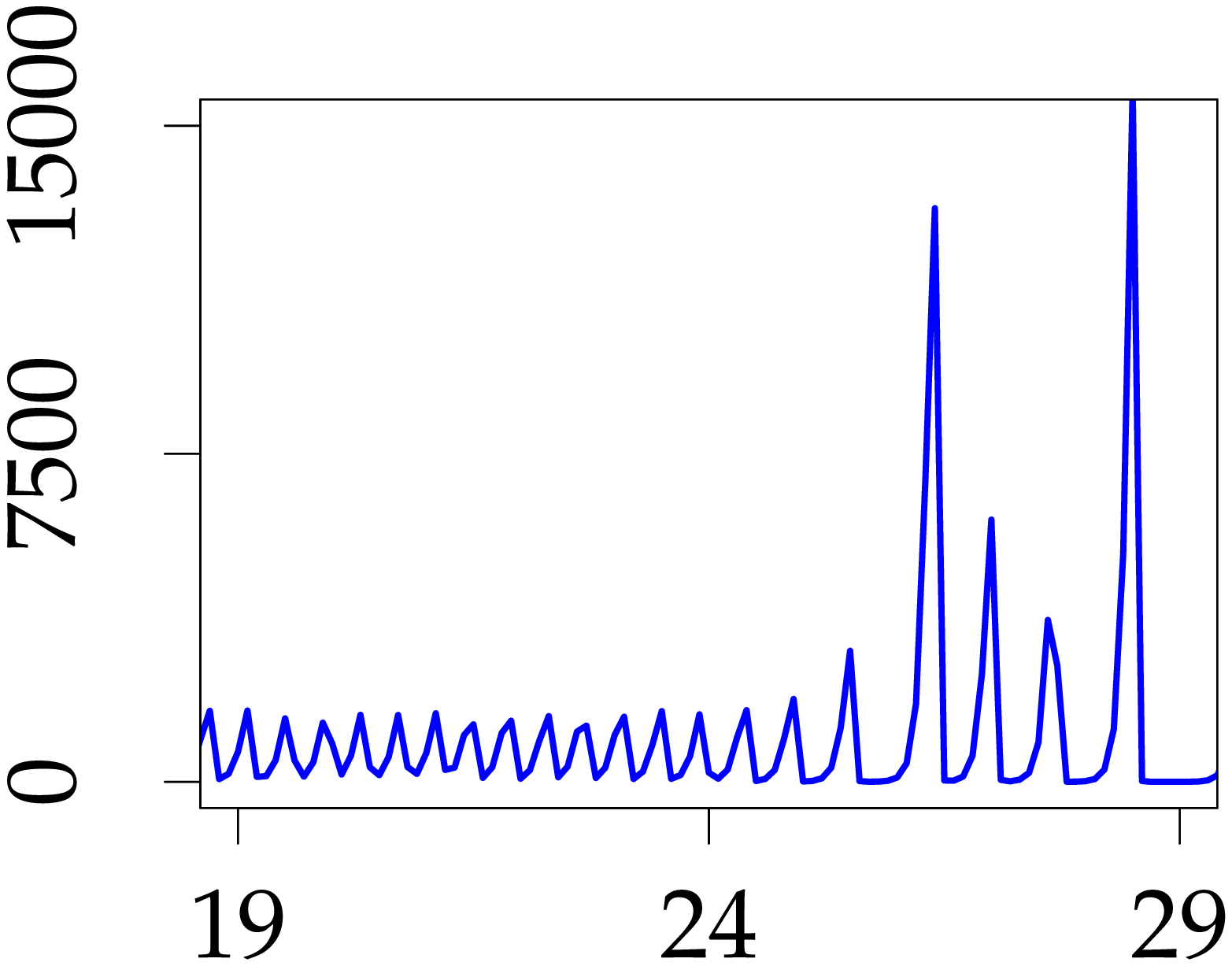} \\
 \multicolumn{2}{c}{\small{With queue feedback: $\tau=50$ ms}}\\
\includegraphics[height=\hite, width=0.45\textwidth]{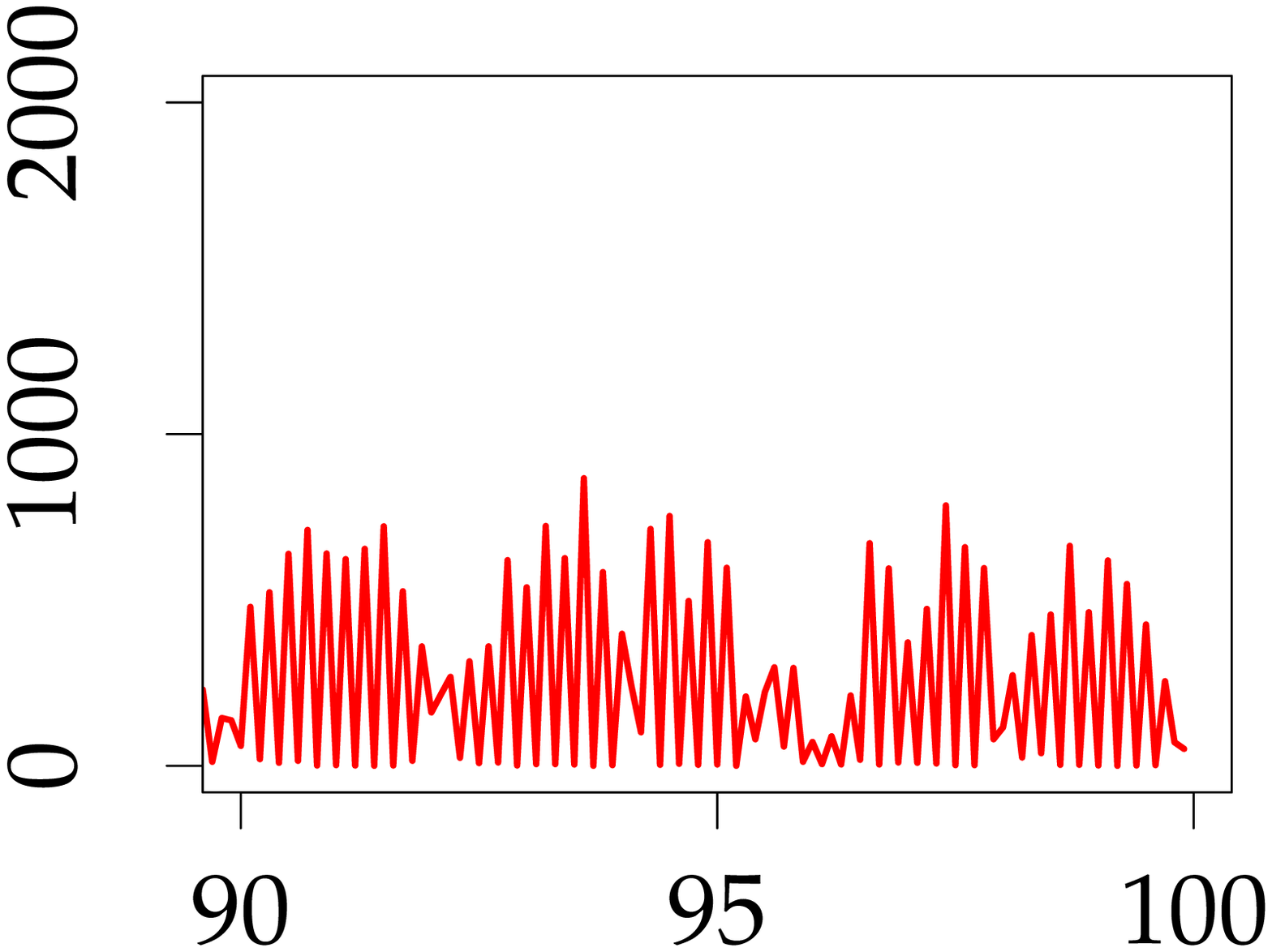} &
\includegraphics[height=\hite, width=0.45\textwidth]{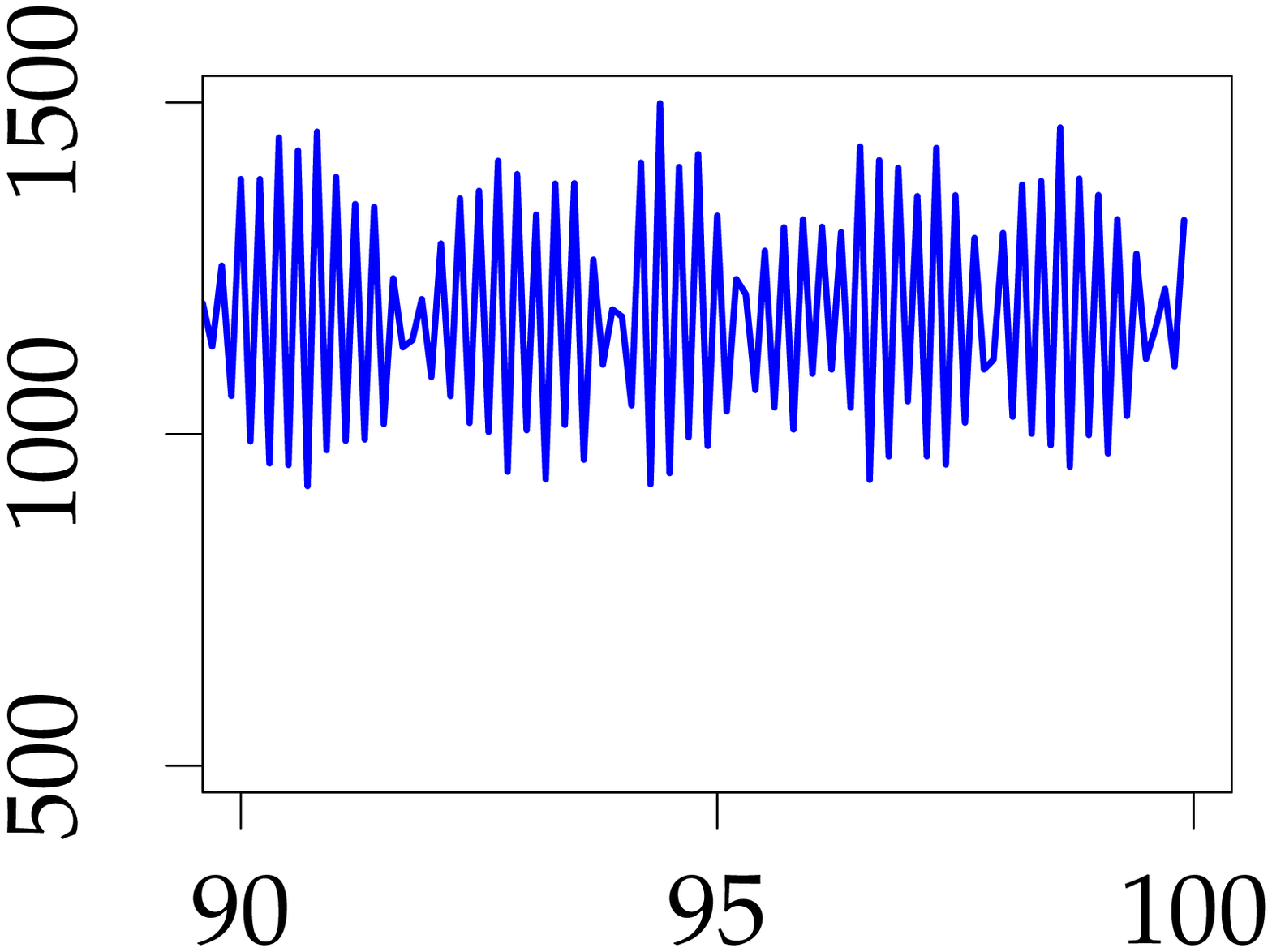} \\
 \multicolumn{2}{c}{\small{Without queue feedback: $\tau=50$ ms}}\\
	\includegraphics[height=\hite,width=0.45\textwidth]{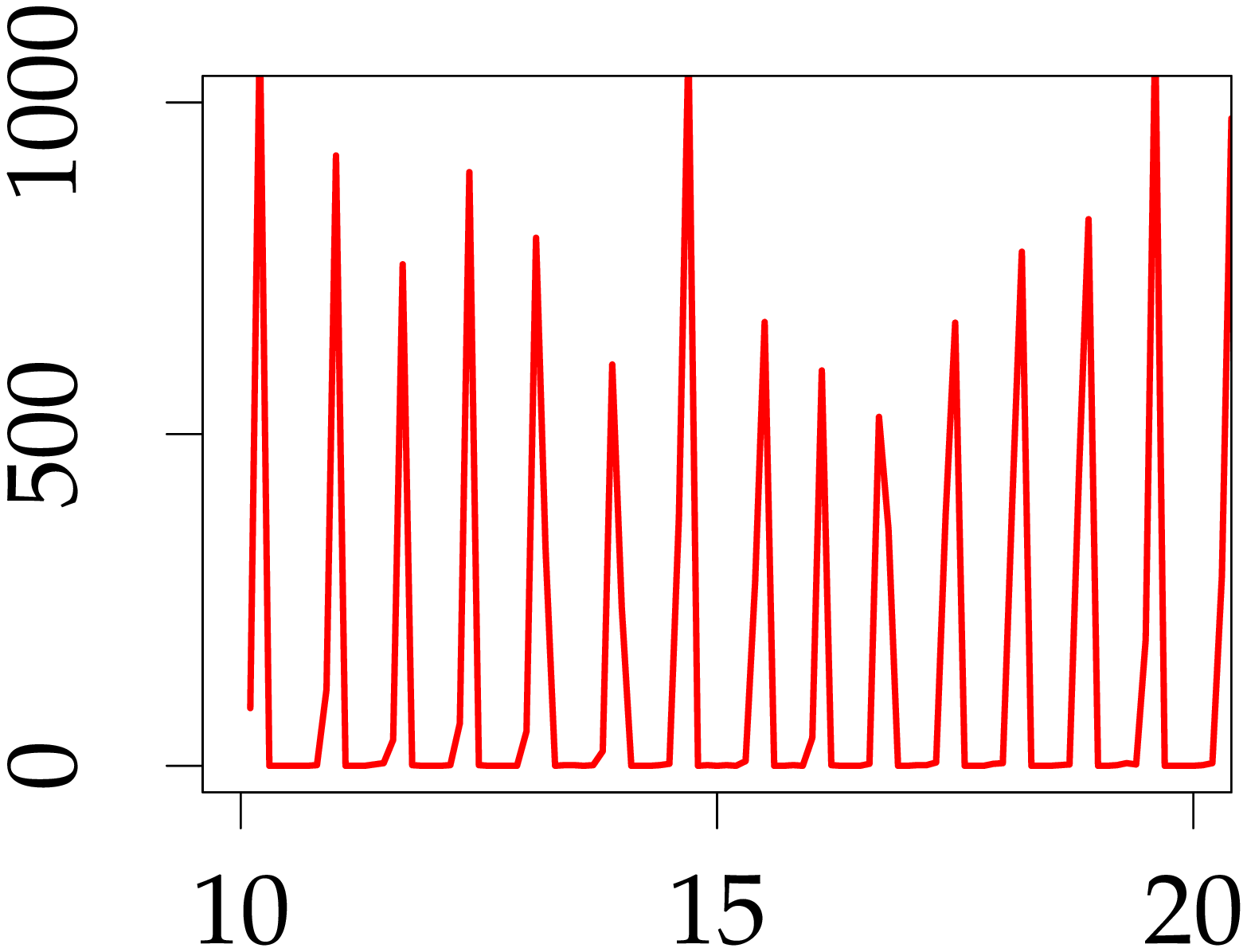} &
	\includegraphics[height=\hite,width=0.45\textwidth]{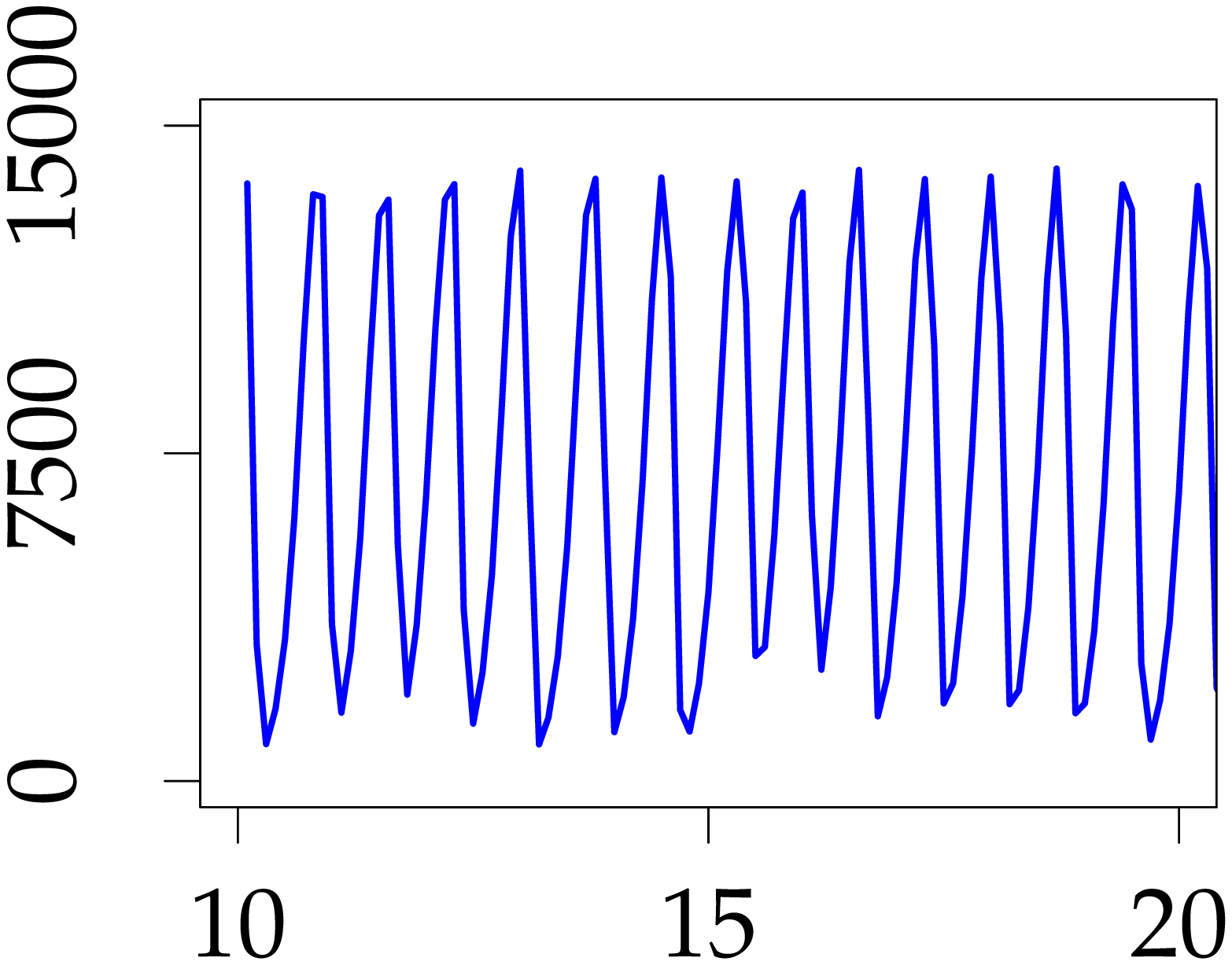} \\
 \multicolumn{2}{c}{\small{With queue feedback: number of sources $=10$}}\\
	\includegraphics[height=\hite,width=0.45\textwidth]{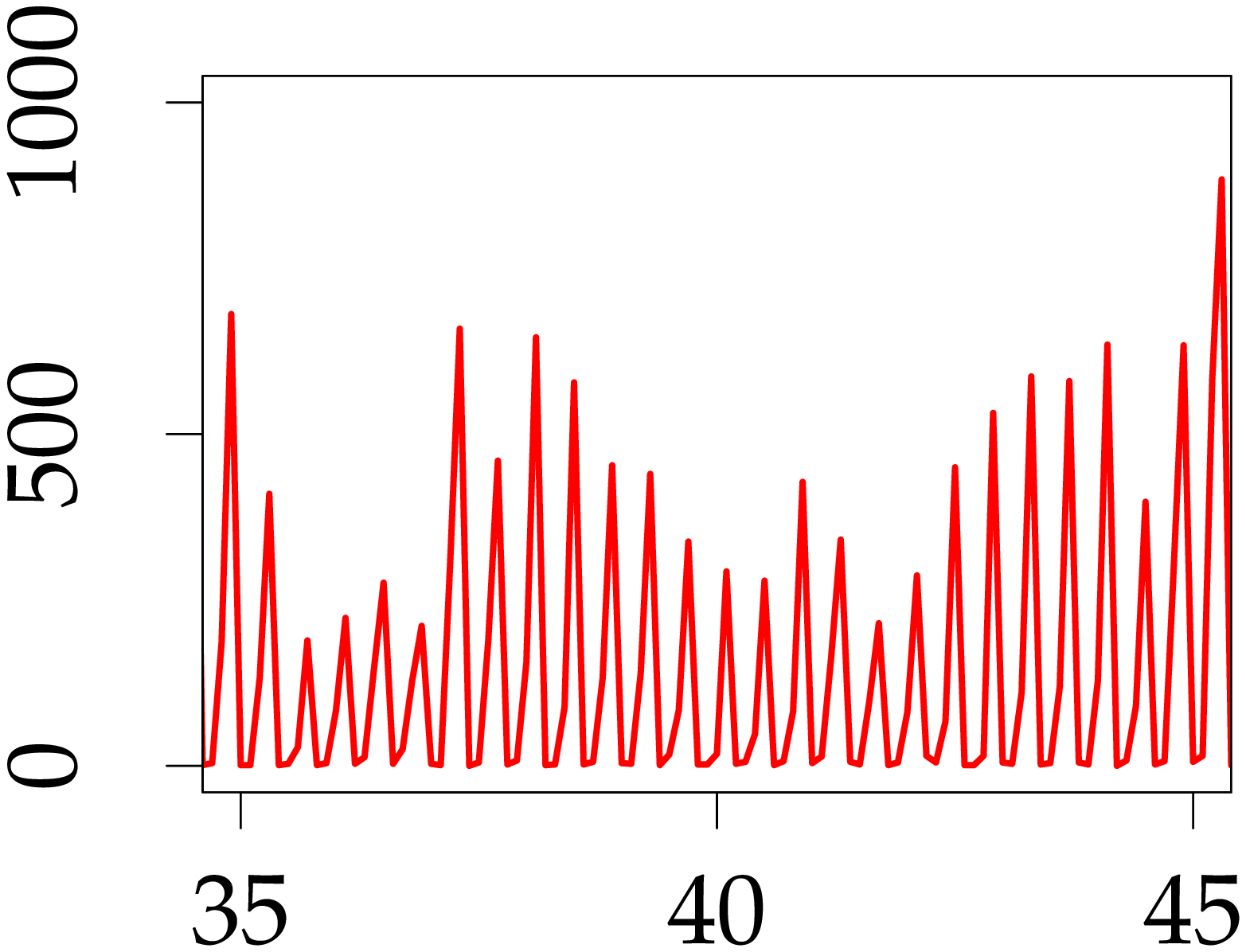} &
	\includegraphics[height=\hite,width=0.45\textwidth]{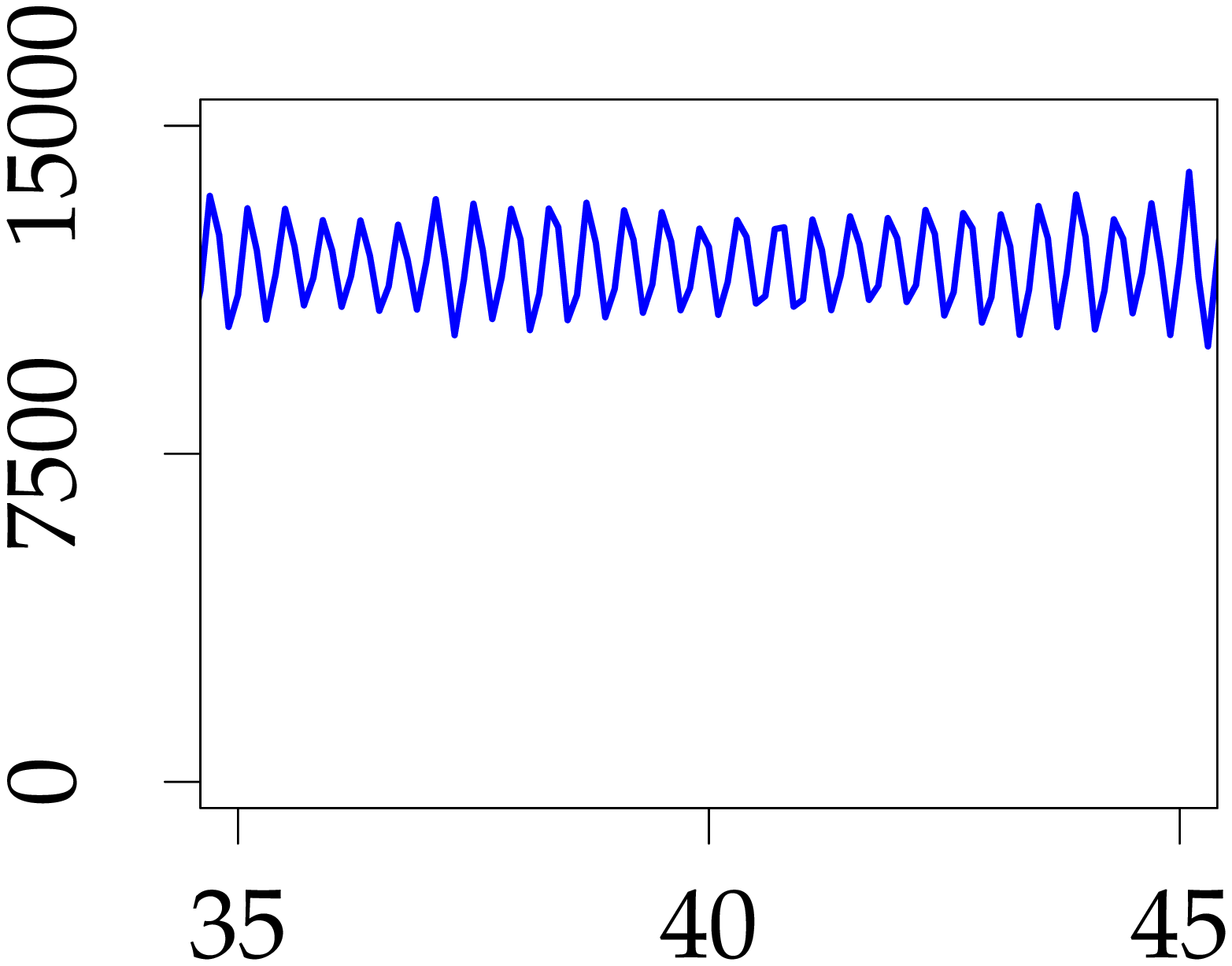} \\
 \multicolumn{2}{c}{\small{Without queue feedback: number of sources $=10$ }}\\
	\multicolumn{2}{c}{\small{Time (ms)}}\\
\end{tabular}
\caption{\small Simulation traces showing that the nature of Hopf bifurcation does not change with $\tau$ and the number of sources. Parameter values chosen are i) with queue feedback: $a=0.8$, $b= 0.005$, $C = 1$ Gbps and ii) without queue feedback: $a=1.6$, $\gamma = 0.95$, $C = 1$ Gbps.}
\label{fig:rcp1dhbaplsims_set2}
\end{figure}
Simulation traces in Figures \ref{fig:rcp1dhopfplsims_withq} and \ref{fig:rcp1dhopfplsims_withoutq} show the evolution of queue size and flow rate for the cases with and without queue size feedback, respectively. The simulated network has a single bottleneck link setup that considers Capacity, $C= 1$ Giga bits per sec (Gbps), number of sources = $100$ and $\tau=100$ ms for all the flows. Here, we set $b=0.005$ which corresponds to equilibrium utilization of 95\% of link capacity. For RCP without queue size feedback, we set $\gamma=0.95$ to achieve the same target link utilization. From these traces, we can observe that, in the presence of queue feedback, the system readily loses stability and leads to the emergence of limit cycles. These observations corroborate the results of our stability analysis which establish that the presence of queue size feedback is associated with a smaller choice of the protocol parameter $a$. 

Similarly, the insights of Hopf bifurcation analysis can be verified from the simulation traces shown in Figure \ref{fig:rcp1dhbaplsims_set1} and \ref{fig:rcp1dhbaplsims_set2}. We can observe that the RCP which uses both rate mismatch and queue feedback exhibits large amplitude limit cycles (due to the occurrence of a sub-critical Hopf). Whereas, in the absence of queue feedback, the system undergoes a super-critical Hopf bifurcation, and leads to the emergence of small amplitude limit cycles. The results of Hopf bifurcation analysis revealed that the type of Hopf bifurcation does not depend on the values of link capacity, round-trip time and the number of flows.  We can verify this by changing the values of these parameters and observe the results in both the cases i.e., with and without queue size feedback (see Figures \ref{fig:rcp1dhbaplsims_set1} and \ref{fig:rcp1dhbaplsims_set2}). 

\section{Conclusions}
RCP estimates the fair rate of flows using feedback based on rate mismatch and queue size. An open design question in RCP is whether it is advantageous to include queue size feedback, given that the protocol already includes feedback based on rate mismatch. To address this question, we linearized the system and analyzed some of the stability and convergence properties for both the design options, i.e., with and without queue size feedback. However, the results of stability and convergence analyses do not provide any design guidelines on whether the queue size feedback is useful or not. This provides motivation for non-linear analysis to study some additional dynamical properties. In particular, we proceeded to analyze the dynamics of both the design choices as conditions for local stability are just violated. We analyzed the type of Hopf bifurcation and the orbital stability of the bifurcating limit cycles. We highlighted that the presence of queue feedback in RCP results in a sub-critical Hopf bifurcation, at high link utilization. A sub-critical Hopf leads to either large amplitude limit cycles or unstable limit cycles, and hence its occurrence should be avoided. Whereas, in the absence of queue feedback, the Hopf bifurcation is always super-critical and leads to the emergence of stable limit cycles of small amplitude. Hence, it is advisable to go with the design choice that uses only rate mismatch feedback. We complemented the analysis with bifurcation diagrams, numerical computations, and packet-level simulations.

Naturally, the work should also extend to consider the cases with multi bottleneck link and heterogeneous delays. It is also important to validate the analytical insights using hardware experiments. 

% we are unable to make any decision on which fairness is desirable.
% The presence of time delayed feedback has implications for stability. Apart from ensuring stability, another important design objective is to make sure that the system converges quickly to a stable equilibrium. Therefore, we explore the impact of queue feedback on the local stability and the convergence properties of RCP.In the stability analysis, we derive stability conditions that enable us to understand the role of various system parameters in ensuring stability. 
% Therefore, based on the linear analysis we are unable to distinguish between the two different design choices.Thus, we proceed to analyze the consequences associated with the loss of stability. 
%\appendices
%\section*{References}

\end{document}